\documentclass[12pt]{article}
\usepackage{natbib}
\usepackage{amsmath}
\usepackage{color}
\usepackage{rotating}
\usepackage{JASA_manu1}
\usepackage{lscape}
\usepackage{graphicx}
\usepackage{bm}
\usepackage{mathrsfs}
\usepackage{graphicx}
\usepackage{booktabs}
\usepackage{caption}
\usepackage{subfig}
\usepackage{amsmath}
\usepackage{rotating}
\renewcommand{\theequation}{\thesection.\arabic{equation}}
\usepackage{slashbox}

\def\be{\begin{equation}}
\def\ee{\end{equation}}
\def\bea{\begin{eqnarray}}
\def\eea{\end{eqnarray}}
\def\bd{\begin{displaymath}}
\def\ed{\end{displaymath}}
\def\bda{\begin{eqnarray*}}
\def\eda{\end{eqnarray*}}
\def\bsm{\begin{small}}
\def\esm{\end{small}}

\def\ha1{\hat \beta_1}

\def\bb0{\delta_\beta}

\def\bsc{\begin{scriptsize}}
\def\esc{\end{scriptsize}}

\input{epsf.sty}
\usepackage{epsfig}

\RequirePackage[OT1]{fontenc}
\RequirePackage{amsthm,amsmath}
\RequirePackage[colorlinks=true,citecolor=blue,urlcolor=blue]{hyperref}

\begin{document}

\title{A Rate Optimal Procedure for Sparse Signal Recovery under Dependence
\footnote{Emails: junli@math.kent.edu, pszhong@stt.msu.edu}}
\author{Jun Li and Ping-Shou Zhong \\
Kent State University and Michigan State University }
\date{}
\maketitle

\begin{center}
\textbf{Abstract}
\end{center}
The paper considers the problem of identifying the sparse different components between two high dimensional means of column-wise dependent random vectors. We show that the dependence can be utilized to lower the  identification boundary for signal recovery. Moreover, an optimal convergence rate for the marginal false non-discovery rate (mFNR) is established under the dependence. The convergence rate is faster than the optimal rate without dependence. To recover the sparse signal bearing dimensions, we propose a Dependence-Assisted Thresholding and Excising (DATE) procedure, 
which is shown to be rate optimal for the mFNR with the marginal false discovery rate (mFDR) controlled at a pre-specified level. Simulation studies and case study are given to demonstrate the performance of the proposed signal identification procedure.   



\setcounter{section}{1} \setcounter{equation}{0}
\section*{\large 1. Introduction}

In genetic studies, one important task is selecting the differentially expressed genes, which can be crucial in identifying novel biomarkers for cancers. 
Motivated by the problem of identifying differentially expressed genes, we consider the high dimensional model
\begin{eqnarray}
X_{ij}=\mu_i+\epsilon_{ij}, \qquad \epsilon_{ij}\stackrel{i.i.d.}{\sim}\mbox{N}(0, \Sigma_i)\, \, \, \mbox{for} \,\,i=1, 2 \,\,\, \mbox{and} \, \,\, 1\le j\le n_i, \label{model}
\end{eqnarray}
where $\mu_i$ is a $p$ dimensional population mean vector and $\Sigma_i$ is a $p\times p$ covariance matrix. If we let $\delta=\mu_1-\mu_2=(\delta_1,\cdots,\delta_p)^{T}$, our interest is to determine which components of $\delta$ are non-zero. 

Due to high dimensionality and relatively small sample sizes in modern statistical data such as the microarray data, we consider $p \gg n_i$. Despite the large number of dimensions, we assume that there are only a small number of signal bearing dimensions. This assumption is thought to be reasonable in many applications. For instance, it is commonly believed that there are only a small number of genes that are significantly differentially expressed between two treatments in a study. Therefore, $\delta$ is sparse in the sense that most of its components are zero but only a small portion of them are non-zero. In order to recover these sparse signals, a commonly used approach is the multiple testing procedure. Each dimension $j \in \{1, \cdots, p\}$ is tested by a $t$-statistic which is expected to have significant value if $\delta_j \ne 0$ and, conversely, to be insignificant if $\delta_j =0$. After all the p-values associated with the $t$-statistics are ranked, the dimensions with p-values smaller than a critical p-value threshold are selected and treated as signal bearing dimensions. 

In the multiple testing procedure, the threshold is chosen to control the false discovery rate (FDR), which is defined as the fraction of false positives among all the rejected hypotheses. For this purpose, Benjamini and Hochberg (1995) introduced a novel procedure (BH procedure) which is shown to be more desirable than other procedures such as the Bonferroni correction that control the familywise error rate (FWER) since the former is less conservative than the latter. 
However, the BH procedure relies on the assumption that the test statistics corresponding to the true null hypotheses are independent.  
In real applications, it is also important to consider the effect of dependence on multiple testing. For example, in genetic studies, genes are actually correlated to achieve certain biological tasks due to the internal structure dictated by the genetic networks of living cells (e.g. gene ontology). It has been shown that the presence of the dependence among test statistics can substantially affect the number of reported non-null hypotheses since the empirical null distribution of dependent p-values can be significantly different from the theoretical null distribution under independent assumption (Efron, 2007). Then the outcome of genetic studies by simply ignoring the intergene correlation is implausible, and a clear strategy to control the false positives in the multiple testing for dependent data is needed (Qiu et al., 2005). 

Some efforts have been made to address the effect of dependence on the multiple testing by assuming some special dependence structures. For example, Benjamini and Yekutieli (2001) showed that when the test statistics corresponding to the true null hypotheses have the positive regression dependence, the BH procedure asymptotically controls the FDR as well as the independence case. Based on a hidden Markov model for the dependence structure, Sun and Cai (2009) proposed an oracle and an asymptotically optimal data-driven procedures which were shown to be able to minimize the false non-discovery rate (FNR) while controlling the FDR at a pre-specified level. Xie, Cai and Li (2011) established a Bayes oracle rule  along with its data adaptive rule based on independent data, which were shown to be optimal in that it minimizes the sum of false negatives and false positives. They also argued that the proposed methods are still valid and remain optimal under short-range dependence. 
 
In this paper, we exploit the nature of dependence differently by investigating its effect on the signal identification boundary without assuming any particular dependence structure. The identification boundary is defined to be a line that separates the plane of signal sparsity and signal strength into two regions. In the region above the line, signals can be recovered individually. But below the line, a successful identification is impossible (Donoho and Jin, 2004, Hall and Jin, 2010, and Ji and Jin, 2012). 
Although the identification boundary for independent data is well established, we are not aware of any existing results exploring the benefits of data dependence in terms of the identification boundary. 
Here we show that the signal identification boundary by incorporating data dependence is lower than that without existence of dependence. More precisely, the explicit expression for the identification boundary is established when dependence is present.


When identifying the signals, people are interested in the procedure that minimizes the FNR while the FDR is controlled at a certain level. However, in the setting of data dependence, the optimal convergence rate for the FNR is still largely unknown. Not only does the paper demonstrate the benefits of data dependence in terms of the identification boundary, but it also establishes the optimal convergence rate for the marginal false non-discovery rate (mFNR) under dependence, which is shown to be faster than the rate with independent data.


To identify the sparse signals, we propose a dependence-assisted thresholding and excising (DATE) procedure. 
The procedure is implemented by first transforming the original data through the matrix
\be
\Omega=(\omega_{kl})=(\frac{n_2}{n_1+n_2}\Sigma_1+\frac{n_1}{n_1+n_2}\Sigma_2)^{-1}. \label{transform_matrix}
\ee
Then, the null components of the transformed data are removed by conducting a marginal thresholding, which is then followed by an additional step to excise the fake signals by maximizing a penalized MLE.  As we will show in Section 4, the proposed procedure attains not only the signal identification boundary under dependence but also the optimal convergence rate for the mFNR with the marginal false discovery rate (mFDR) controlled at a pre-selected level, and thus is superior compared with other methods without taking data dependence into account.

The rest of the paper is organized as follows. In Section 2, we establish two lower bounds: one for the risk function (\ref{risk}) and another for the convergence rate of the mFNR. To show the optimality of these two bounds, we first demonstrate the benefit of transforming the data by the matrix $\Omega$ in (\ref{transform_matrix}) in Section 3. Then a thresholding and excising procedure based on the transformed data is introduced in Section 4. The proposed procedure is shown to be able to achieve two lower bounds established in Section 2 and thus is rate optimal. Section 5 illustrates some numerical studies and Section 6 reports an empirical study to select differentially expressed genes for a human breast cancer data set. Discussion is given in Section 7. All technical details are relegated to the Appendix.

\setcounter{section}{2} \setcounter{equation}{0}
\section*{\large 2. Lower Bounds for Signal Identification under Dependence}

To establish the lower bound of the signal identification boundary in the dependent setting, we start with some notations and definitions. 
Denote $S_{\beta}=\{k: \delta_k \ne 0 \}$ to be a set including the locations of the non-zero $\delta_k$. The number of non-zero elements in $S_{\beta}$ is $p^{1-\beta}$ for $\beta \in (\frac{1}{2}, 1)$. Define $L_p$ to be a slowly varying logarithmic function in the form of $(a\mbox{log}p)^b$. 
Without loss of generality, we assume both $\Sigma_1$ and $\Sigma_2$ are standardized to have unit diagonal elements. With matrix $\Omega=(\omega_{ij})$ defined in (\ref{transform_matrix}), let 
\be 
\underline{\omega}=\underline{\lim}_{p \to \infty}\min\limits_{1\le k \le p}\omega_{kk}, \quad \mbox{and} \quad \bar{\omega}=\overline{\lim}_{p \to \infty}\max\limits_{1\le k \le p}\omega_{kk}. \label{u_l}
\ee 
We model $\delta$ to satisfy the following condition (see Ji and Jin, 2012):\\
(C1). The components of $\delta$ follow a mixture distribution
\[
\delta_k \stackrel{i.i.d.}{\sim} (1-p^{-\beta})h_0+p^{-\beta}\pi_p, \quad k=1, \cdots, p, 
\]
where $h_0$ is a point mass at $0$ and $\pi_p$ is a distribution with the support $[-\sqrt{2r\mbox{log}p/n}, 0) \cup (0, \sqrt{2r\mbox{log}p/n}]$ for $r > 0$ and $n=\frac{n_1 n_2}{n_1+n_2}$. 

In the independent case, the identification boundary that describes the relationship between signal sparsity $\beta$ and signal strength $r$ is defined to be a line $r=\beta$ in the $\beta$-$r$ plane. In the region above the line, it is possible to identify them individually, but it becomes impossible in the region below the line. Since stronger magnitude of signals is needed to discover non-zero components individually, the identification boundary lies above the detection boundary that separates the $\beta$-$r$ plane into the so-called detectable region and undetectable region.

\begin{figure}[t]
\centering
\vspace{-5cm}
\includegraphics[width=\textwidth]{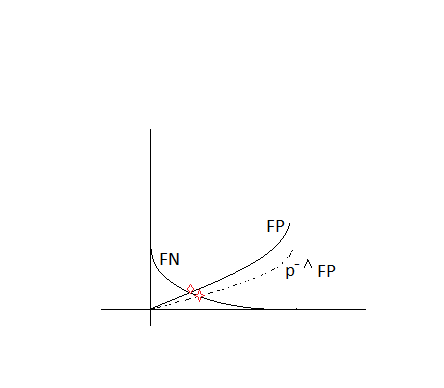}
\vspace{-3cm}
\caption{The horizontal axis represents the number of $\hat\delta_k\neq0$. The diamond is the intersection point of the false positives line (FP) and the false negatives line (FN) where $H(0)$ is minimized and the star is the intersection point where $H(\Lambda)$ is minimized.  }
\label{hlambda-plot}
\end{figure}

Given $\delta_k$ for $1\le k \le p$, $\hat{\delta}_k$  is denoted as an estimate of $\delta_k$. For any signal identification procedure, there are generally two types of error related with the signal estimate $\hat{\delta}_k$:  the false negative meaning that $\delta_k \ne 0$ but $\hat{\delta}_k=0$, and the false positive representing that $\delta_k=0$ but $\hat{\delta}_k\ne 0$. Then the optimal procedure for signal recovery can be defined as the one that minimizes the expected weighted sum of false negatives and false positives:
\be
H(\Lambda)=\mbox{E}\biggl\{\sum_{k \in S_{\beta}}\mbox{I}(\hat{\delta}_k=0)+p^{-\Lambda}\sum_{l\in S_{\beta}^c}\mbox{I}(\hat{\delta}_l \ne 0)\biggr\}, \label{risk}
\ee  
where the weight $p^{-\Lambda}$ with $\Lambda \in [0, \infty)$ is chosen to adjust the level of false positives. If $\Lambda=0$, there is no preference on either the false positives or the false negatives, and the risk (\ref{risk}) becomes the misclassification error adopted by Ji and Jin (2012) for establishing the optimal convergence rate for the variable selection problem in the high-dimensional regression model. On the other hand, choosing a larger value of $\Lambda$ leads to a smaller weight function $p^{-\Lambda}$, which consequently allows the optimal procedure to produce relatively larger false positives when minimizing $H(\Lambda)$. 
The effect of $\Lambda$ on false positives can be demonstrated by Figure \ref{hlambda-plot}. Assume that the minimization of $H(0)$ is achieved at the intersection point {\it diamond} of the false positives line (FP) and the false negatives line (FN). By multiplying FP with $p^{-\Lambda}$ (dash line), the FP becomes less important in $H(\Lambda)$ and $H(\Lambda)$ is minimized at the intersection point {\it star} which is on the right side of the intersection point {\it diamond}. As a result, the expected false positives corresponding to the minimized $H(\Lambda)$ is larger than that corresponding to the minimized $H(0)$. The universal lower bound of the risk function $H(\Lambda)$ at a fixed value $\Lambda$ is established by the following theorem.


\textbf{Theorem 1.} Assume condition (C1) and the model (\ref{model}) for $X_{ij}$. 
As $p \to \infty$, 
\[
H (\Lambda)\ge \left\{ 
  \begin{array}{l l}
    L_p p^{1-\beta-(\bar{\omega}r-\beta+\Lambda)^2/(4\bar{\omega}r )}, & \quad -r<(\Lambda-\beta)/\underline{\omega} <r\\
    p^{1-\beta}, & \quad r< (\beta-\Lambda)/\bar{\omega}\\
    p^{1-\Lambda}, &\quad r< (\Lambda-\beta)/\bar{\omega}
  \end{array} \right. 
\] 
where $\underline{\omega}$ and $\overline{\omega}$ are defined in (\ref{u_l}), and $L_p$ is a slowly varying logarithmic function.

The universal lower bound varies with different values of $r$, $\beta$ for each fixed value of $\Lambda$. If we choose $\Lambda=0$, the misclassification error has the lower bound
\[
H (0)\ge \left\{ 
  \begin{array}{l l}
    L_p p^{1-\beta-(\bar{\omega}r-\beta)^2/(4\bar{\omega}r )}, & \quad r>\beta/\underline{\omega}\\
    p^{1-\beta}. & \quad r<\beta/\bar{\omega}
  \end{array} \right. 
\] 
Some key observations are as follows.  First, if the signal strength $r<\beta/\bar{\omega}$, the misclassification error is no less than $p^{1-\beta}$, the number of non-zero $\delta_k$, which implies that there exists no successful signal identification procedure. The area $r<\beta/\bar{\omega}$ in $r-\beta$ plane is thereafter called the region of no recovery. On the other hand, if the signal strength attains $r \ge (1+\sqrt{1-\beta})^2/\underline{\omega}$, the misclassification error asymptotically converges to zero and all the signals can be successfully recovered. The corresponding region is called the region of full recovery. The area sandwiched between the no recovery region and the full recovery region satisfies $\beta/\bar{\omega}<r <(1+\sqrt{1-\beta})^2/\underline{\omega}$, having the misclassification error less than the number of signals and greater than zero. This region is called region of partial recovery. Most importantly, since $\bar{\omega} \ge \underline{\omega} > 1$ under data dependency shown by Lemma 1 in Appendix, the partial recovery boundary $r=\beta/\bar{\omega}$ and full recovery boundary $r=(1+\sqrt{1-\beta})^2/\underline{\omega}$ used to separate three regions are lower than those without existence of data dependence.

\begin{figure}[t]
\begin{minipage}[b]{0.5\linewidth}
\centering
\includegraphics[width=\textwidth]{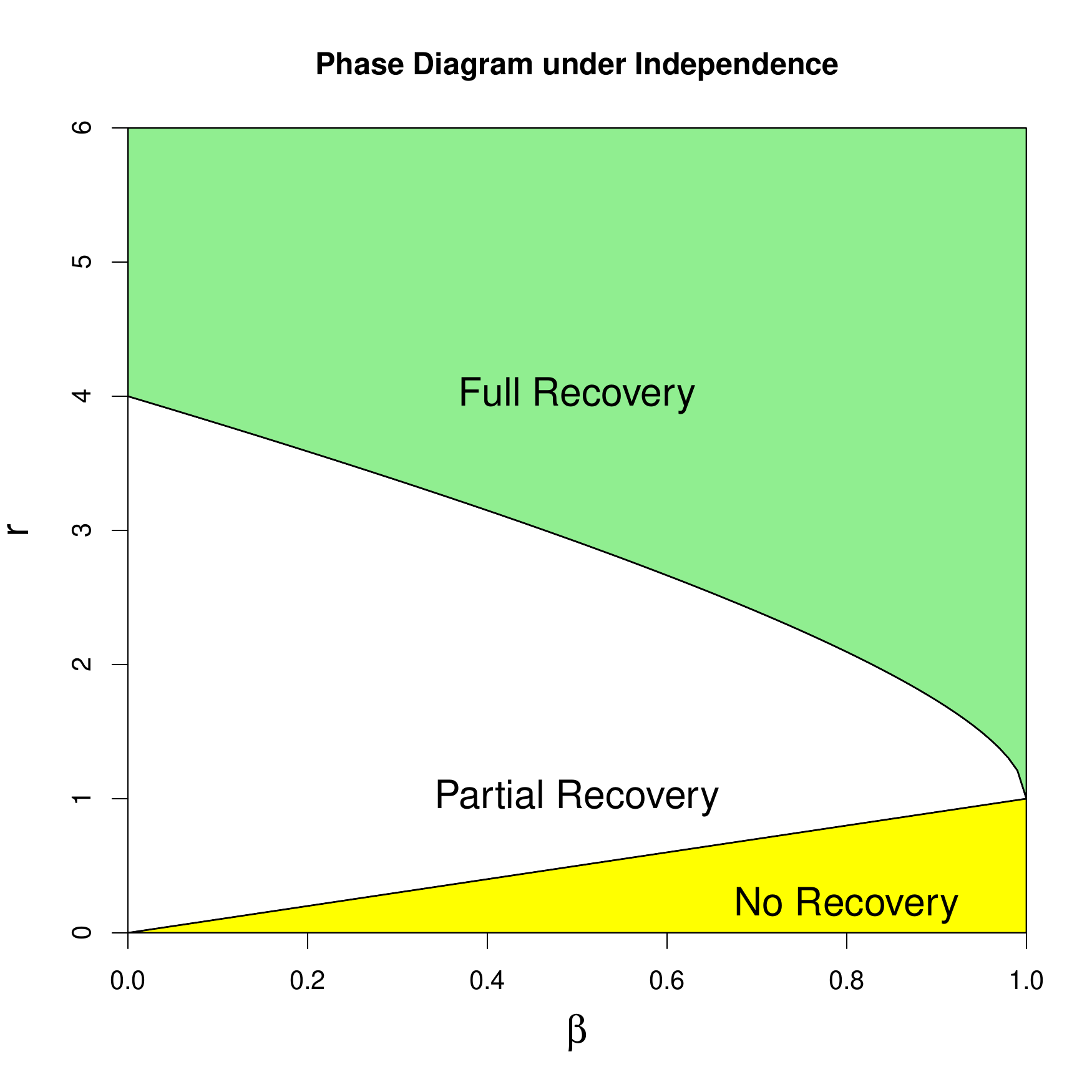}
\end{minipage}
\hspace{0.5cm}
\begin{minipage}[b]{0.5\linewidth}
\centering
\includegraphics[width=\textwidth]{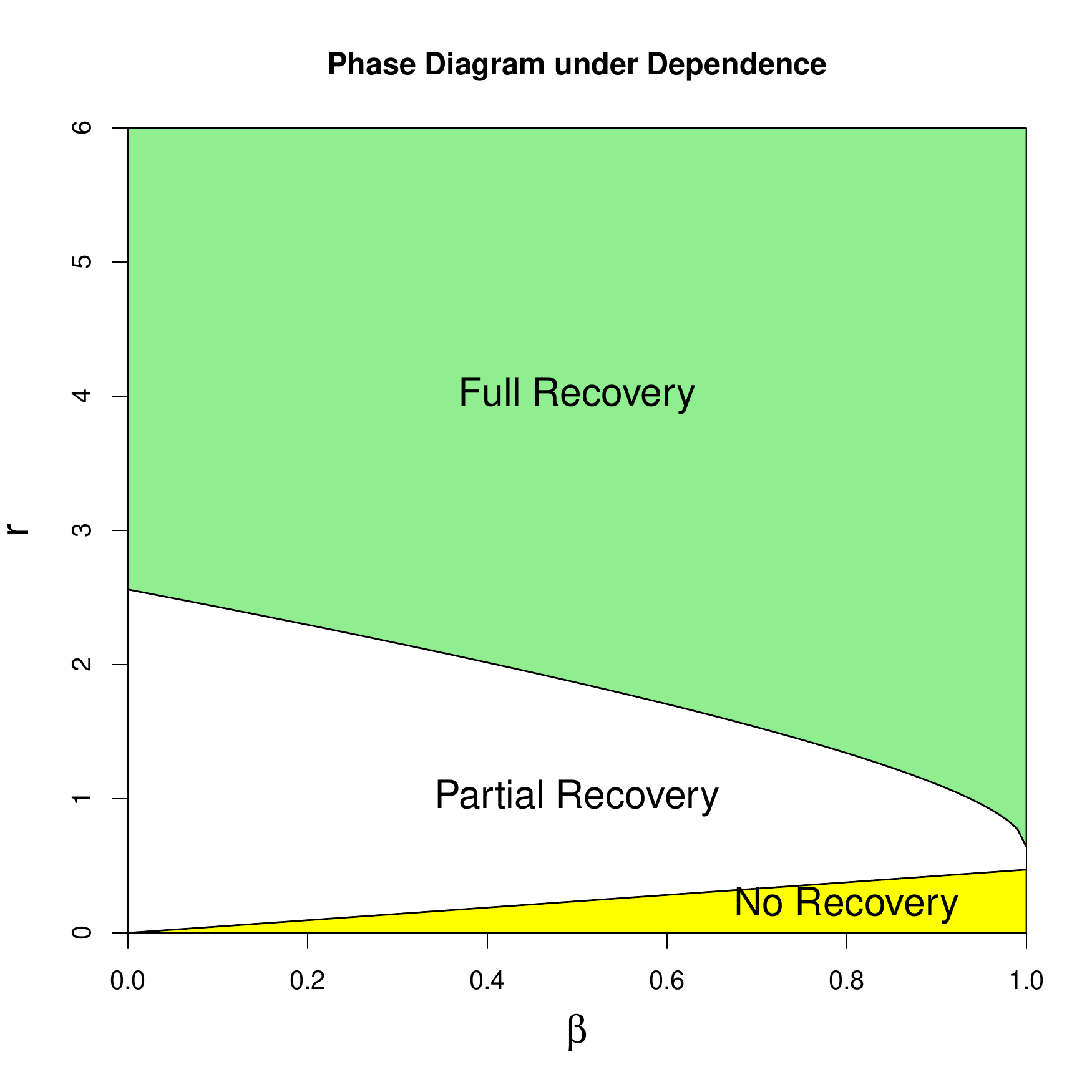}
\end{minipage}
\caption{Left: phase diagram for signal recovery without data dependence. Right: phase diagram for signal recovery with $\Sigma_1=\Sigma_2=(0.6^{|i-j|})$ for $1\le i,j \le p$. }
\label{phase.diagram}
\end{figure}

To demonstrate the observations above, we consider $\Sigma_1=\Sigma_2=(\rho^{|i-j|})$ for $1\le i,j \le p$ in model (\ref{model}) such that the data dependence is exhibited by the value of $\rho$. If $\rho=0$, $\bar{\omega}= \underline{\omega}=1$ since there is no data dependence. On the other hand, if $\rho=0.6$, we obtain $\underline{\omega}= 1.5625$ and $\bar{\omega}= 2.125$. The corresponding phase diagrams with and without data dependence are displayed in Figure \ref{phase.diagram} in which the partial signal identification boundary and the full recovery boundary with $\rho=0.6$ are lower than those with $\rho=0$ due to the fact that $\underline{\omega}>1$ and $\bar{\omega}>1$. As a result, even though the signals with $r < \beta$ are unable to be identified by any procedure if there exists no data dependence, some of them can be recovered as long as the signal strength $r>\beta/2.125$ with the existence of data dependence. The benefit to the full signal identification with the existence of dependence can be seen based on the similar derivation.

There is a close connection between the signal recovery and the weighted risk function $H(\Lambda)$. It has been shown that by properly choosing $\Lambda$, the decision rule that minimizes the weighted risk function $H(\Lambda)$ is also the optimal procedure that controls the marginal FDR at level $\alpha$ and minimizes the marginal FNR (mFNR) in the multiple testing problem ( Sun and Cai, 2007, Sun and Cai, 2009, and Xie, Cai, Maris and Li, 2011). Let FP= false positives, TP=true positives, FN= false negatives and TN= true negatives. The mFDR and mFNR are defined as
\[
\mbox{mFDR}= \biggl \{ \frac{\mbox{E}(\mbox{FP})}{\mbox{E}(\mbox{FP})+\mbox{E}(\mbox{TP})}\biggr\} \quad \mbox{and} \quad \mbox{mFNR}= \biggl \{ \frac{\mbox{E}(\mbox{FN})}{\mbox{E}(\mbox{FN})+\mbox{E}(\mbox{TN})}\biggr\}. 
\]
Genovese and Wasserman (2002) showed that mFDR and mFNR are asymptotically equivalent to FDR and FNR under weak conditions. In general, the connection between $\Lambda$ and $\alpha$ is complicated. The following theorem provides a solution for choosing a proper $\Lambda(\alpha)$ such that the mFDR is controlled at the level of $\alpha<1$. Moreover, it establishes a lower bound for the mFNR subject to the constraint that $\mbox{mFDR}\le \alpha$.


\textbf{Theorem 2.} Assume condition (C1) and (\ref{model}) for $X_{ij}$. 
If we choose 
\[
\Lambda(\alpha)=\underline{\omega}r+\beta-2\sqrt{\underline{\omega}r\beta\bigg(1-\frac{g(\alpha,p)}{\beta}\biggr)}, \quad \mbox{where} \,\, g(\alpha,p)=\frac{\mbox{log}\{\frac{\alpha}{(1-\alpha)}\sqrt{4\pi \beta \mbox{log}p}\}}{\mbox{log}p},
\]
then as $p \to \infty$, 
\[
\mbox{mFNR} \ge L_p p^{-\beta-\bigl\{\sqrt{\bar{\omega}r}-\sqrt{\beta-g(\alpha, p)}\bigr\}^2} \quad \mbox{and} \quad \mbox{mFDR} \le \alpha <1.
\] 

Similar to the weighted risk function, the lower bound for the mFDR is accelerated with existence of dependence since $\bar{\omega}>1$. 
In order to show that the lower bounds in Theorems 1 and 2 are tight, we need to search for a signal identification procedure that is able to attain the universal lower bounds. As we will see in next section, the key for this procedure is to take the data dependence into account, which can be done by transforming the data via the matrix $\Omega$ defined in (\ref{transform_matrix}).

\setcounter{section}{3} \setcounter{equation}{0}
\section*{\large 3. Data Transformation}

Some additional assumptions are needed to establish the theoretical performance of the procedure we will introduce in this and next sections. \\
(C2). The eigenvalues of $\Sigma_i$ for $i=1, 2$ satisfy $C_0^{-1} \le \lambda_{\mbox{min}}(\Sigma_i) \le \lambda_{\mbox{max}}(\Sigma_i) \le C_0$ for some constant $C_0>0$.\\
(C3). The matrix $\Omega$ in (\ref{transform_matrix}) is presumably sparse and belongs to the class
\[
\mathcal{V}(c_p, M_p)=\biggl\{\Omega: ||\Omega||_{L_1} \le M, \max \limits_{1 \le j \le p} \sum_{i=1}^p |\omega_{ij}|^{\zeta} \le c \quad \mbox{for} \quad 0 < \zeta <1 \biggr\},
\]
where $M$ and $c$ are fixed constants.\\
(C4). As $n\to \infty$, $p \to \infty$ and 
$\mbox{log}p=c\,n^{\theta}$ for $c>0$ and $\theta <\frac{1-\zeta}{2-\zeta}$ where $\zeta$ is defined in (C3).

Conditions (C2) and (C3) define a class of matrices with sparse structures, which is originally proposed by Bickel and Levina (2008b). 
Condition (C4) specifies the exponential growth of dimension $p$ with $n$. All of these conditions are commonly assumed in the literature.

For signal identification, we need to define a statistic to estimate the magnitude of the signal. To this end, we let $\bar{X}_i^{(k)}=\sum_{j=1}^{n_i} X_{ij}^{(k)}/n_i$ for $i=1, 2$ 
where $X_{ij}^{(k)}$ is the $k$th component of $X_{ij}$. Then a measure of $n \delta_k^2$ is defined by   
\be
L_k={n\{\bar{X}_1^{(k)}-\bar{X}_2^{(k)}\}^2}, \quad k=1,\cdots, p. \label{stat_z} 
\ee
Since the marginal variances of $L_k$ are the same, the probability of the non-null component being identified depends on the value of $L_k$ or essentially its signal strength $\delta_k$. The magnitude of $\delta_k$ can be enhanced by transforming $X_{ij}$ into $Z_{ij}=\Omega X_{ij}$ where $\Omega=(\omega_{kl})$ for $1 \le k, l \le p$ is defined in (\ref{transform_matrix}). The similar transformation was also considered in Hall and Jin (2010) for their innovated higher criticism test, and Cai, Liu and Xia (2014) for testing the equality of two sample mean vectors. 

To appreciate signal enhancement induced by the transformation, we let $\delta_{\Omega}$ be the difference in two population mean vectors after the transformation. Then the following relationship holds between $\delta_{\Omega}$ and the original signal strength $\delta$:
\be
\delta_{\Omega,k}=\omega_{kk} \delta_k+\sum_{l \ne k \in S_{\beta}}\omega_{kl} \delta_l, \quad \mbox{for} \quad k=1, \cdots, p. \nonumber 
\ee
Lemma 2 in Appendix shows that for sparse signals and sparse $\Omega$ assumed in (C3), $\delta_{\Omega,k} = \omega_{kk} \delta_k+o(n^{-1/2})$, 
which implies that if $k \in S_{\beta}$, 
\be
\frac{\delta_{\Omega,k}}{\sqrt{\omega_{kk}}}=\sqrt{\omega_{kk}}\delta_k+o(n^{-1/2}).  \label{signal_transform1}
\ee
This, together with $\omega_{kk}\ge 1$ in Lemma 1, leads to 
\[
 \frac{\delta_{\Omega,k}}{\sqrt{\omega_{kk}}} \ge {\delta_k}.
\]
Therefore, for signal identification, we propose the following test statistic
\[ 
{T}_k=\frac{n\{\bar{{Z}}_1^{(k)}-\bar{{Z}}_2^{(k)}\}^2}{{\omega}_{kk}},  \quad k=1,\cdots, p, 
\] 
which is constructed based on the transformed data and has the standardized  signal strength $\sqrt{n}\delta_{\Omega,k}/\sqrt{\omega_{kk}}$ greater than standardized signal strength $\sqrt{n}\delta_k$ of the test statistic (\ref{stat_z}). 

In real applications, $\Omega$ is unknown and needs to be estimated. When $\Sigma_1$ and $\Sigma_2$ are bandable, $\Omega$ can be estimated through the Cholesky decomposition proposed by Bickel and Levina (2008a). 
Yuan and Lin (2007) considered an $L_1$ penalized normal likelihood estimator for the sparse precision matrix. More can be found in Friedman, Hastie and Tibshirani (2008). 
Cai, Liu and Luo (2011) introduced an CLIME estimator based on the constrained $L_1$ minimization approach for precision matrix estimate. With estimated $\hat{\Omega}$, the transformed signal for $k \in S_{\beta}$ is $\hat{\delta}_{\Omega,k}=\sum_{l \in S_{\beta}}\hat{\omega}_{kl}\delta_l$. Similar to $\delta_{\Omega,k}$ when $\Omega$ is known, Lemmas 1 and 2 show that under some mild conditions, with probability equal to 1,  
\[
\frac{\hat{\delta}_{\Omega,k}}{\sqrt{\hat{\omega}_{kk}}} \ge {\delta_k},
\]
Therefore, we consider the following test statistics based on the transformed data $\hat{Z}_{ij}=\hat{\Omega} X_{ij}$ as the starting point of the proposed signal identification procedure:
\be 
\hat{T}_k=\frac{n\{\bar{\hat{Z}}_1^{(k)}-\bar{\hat{Z}}_2^{(k)}\}^2}{\hat{\omega}_{kk}},  \quad k=1,\cdots, p. \label{stat1_est}
\ee 

The advantage of the statistics in (\ref{stat1_est}) relative to (\ref{stat_z}) is that the standardized signal strength has been enhanced by incorporating the dependence, which potentially increases the probability of weak signals being identified by the signal recovery procedure. However, since $\delta_{\Omega,k}=\sum_{l \in S_{\beta}} \omega_{kl}\delta_l$, a side effect of the transformation is that it generates some fake signals, i.e., $\delta_k=0$ but $\delta_{\Omega,k} \ne 0$ if $\omega_{kl}\ne 0$ for some $l \in S_{\beta}$. Therefore, a successful signal recovery procedure benefited by data transformation requires to remove these fake signals. As we will discuss in next section, fake signals can be successfully excised by a penalized method with $L_0$ penalty.  As revealed by Ji and Jin (2012), this approach is very effective in cleaning fake signals but suffers the computational intensity if dimension $p$ is large. To reduce the complexity of the original signal selection problem, we first need a dimension reduction procedure, which is fulfilled by a thresholding step as we will discuss in next section.

\setcounter{section}{4} \setcounter{equation}{0}
\section*{\large 4. DATE procedure to recover signals}

To introduce our signal identification procedure, we first focus on most interesting case where $\underline{\omega}r <(\sqrt{1-\Lambda}+\sqrt{1- \beta})^2$. According to Theorem 1, this case indicates that the weighted risk $H(\Lambda)$ does not converge to zero but is less than $p^{1-\beta}$. The corresponding region on $r-\beta$ plane is the partial recovery under a fixed value $\Lambda$.  The case $\underline{\omega}r \ge (\sqrt{1-\Lambda}+\sqrt{1- \beta})^2$ corresponding to the full recovery region is an easier problem due to the relatively larger signal strength. We will discuss it at the end of this section. 

As we have discussed in the previous section, after data transforming, $p$ coordinates consist of the signals, fake signals and noise. As the first step of the proposed method for signal recovery, a thresholding is conducted to remove the noise. 
After all the $p$ dimensions are checked by a threshold function $2s\mbox{log}p$, we set $\hat{\delta}_k=0$ for $k \in \{1,\cdots, p\}$ if and only if 
\be
\hat{T}_k<2s \mbox{log}p, \label{threshold}
\ee
where $s>0$ is chosen to control the level of the threshold, and the decision on other coordinates with $\hat{T}_k \ge 2s \mbox{log}p$ will be made in another step following the thresholding step. Although imposing the threshold is to prevent noise, it can potentially screen out signals and thus produce the false negatives. 
The following Lemma establishes the upper bound of the expected false negatives generated in the thresholding step (\ref{threshold}). 

\textbf{Lemma 3.} Assume conditions (C1), (C3) and (C4). Let $s \in (0, \frac{(\underline{\omega}r+\beta-\Lambda)^2}{4\underline{\omega}r})$ and $\beta-\Lambda <\underline{\omega}r <(\sqrt{1-\Lambda}+\sqrt{1- \beta})^2$. As $p \to \infty$,
\[
\mbox{E}\biggl\{\sum_{k=1}^p (\hat{\delta}_k=0, \delta_k\ne 0)\biggr\}
\le L_p p^{1-\beta-{(\underline{\omega}r-\beta+\Lambda)^2}/{(4\underline{\omega}r)}}.
\] 

Since the error above is no more than the error rate established in Theorem 1 provided that $\underline{\omega}=\bar{\omega}$, it does not affect the rate optimality of the whole identification procedure as long as the error made in the following excising step is under control. 

The fake signals generated by the transformation are able to survive from the thresholding if 
\[
\hat{T}_k \ge 2s \mbox{log}p, \quad k \notin S_{\beta}.
\]
To excise these fake signals, we implement an $L_0$ penalization approach, which is originally designed for the regression problem.
For the purpose of variable selection, this approach directly penalizes the number of non-zero parameters but is hampered by high dimensionality since it requires an exclusive search of all $2^{p}$ submodels and is computationally intensive. However, as we will show in the following, this NP hard problem can be circumvented thanks to an important consequence of conducting the thresholding. To see it, we let $\mathcal{U}(s)$ be a set including all components survived from the thresholding
\be
\mathcal{U}(s)=\{k: \hat{T}_k\ge 2s\mbox{log}p, 1\le k \le p\}. \label{signal.set}
\ee
We define $V_0=\{1, \cdots, p\}$ to be a set of notes and 
\be
\Omega^*(i,j)=\hat{\Omega}(i,j)\mbox{I}_{\{|\hat{\Omega}(i,j)|\ge log^{-1}p\}} \label{reg_omega}
\ee
to be regularized $\hat{\Omega}$. Then according to the Gaussian graph theory, given the precision matrix $\Omega^*$, any $i\ne j \in V_0$ are connected if and only if $\Omega^*(i,j)\ne 0$. The following Lemma 4 summarizes the consequence after conducting the thresholding. 

\textbf{Lemma 4.} Assume the conditions (C1)-(C4). Let $s \in (0, \frac{(\underline{\omega}r+\beta-\Lambda)^2}{4\underline{\omega}r})$ and $\beta-\Lambda <\underline{\omega}r <(\sqrt{1-\Lambda}+\sqrt{1- \beta})^2$. With probability $1- L_p p^{-\beta-{(\underline{\omega}r-\beta+\Lambda))^2}/{(4\underline{\omega}r)}}$, $\mathcal{U}(s)$ are split into disconnected clusters of size no more than a positive integer $K$ with respect to $(V_0, \Omega^*)$.

According to Lemma 4, the $L_0$ penalization approach can be effectively adopted to each of self-connected subsets with relatively small size. 
Let $I_0=\{i_1,\cdots,i_m\}$ be one of the self-connected subsets with size $m \le K$, and $\hat{A}=\hat{\Omega}^{I_0,I_0}$ be an $m \times m$ matrix with $\hat{\Omega}^{I_0,I_0}(k,l)=\hat{\Omega}(i_k,i_l)$. To excise the fake signals in $I_0$, we find an $m$-dimensional vector $\hat{\delta}(I_0)$ each component of which is equal to either $0$ or $\delta^{date}$ or $-\delta^{date}$ to minimize the following function:
\be
n\biggl\{(\bar{\hat{Z}}_1-\bar{\hat{Z}}_2)^{I_0}-\hat{A}\delta\biggr\}^{\prime}\hat{A}^{-1}\biggl\{(\bar{\hat{Z}}_1-\bar{\hat{Z}}_2)^{I_0}-\hat{A}\delta \biggr\}+(\lambda^{date})^2||\delta||_0, \label{pmle_e}
\ee 
where $\lambda^{date}$ and $\delta^{date}$ are two tuning parameters. 
After we apply the $L_0$ penalization approach to all the self-connected subsets, each of $\delta_k$ for $k=1, \cdots, p$ is eventually determined by  the proposed DATE procedure which can be summarized by the following algorithm.
\begin{itemize}
\item[(1).] Transform data $X_{ij}$ to obtain $\hat{Z}_{ij}=\hat{\Omega} X_{ij}$ where $\hat{\Omega}$ is estimated $\Omega$;
\item[(2).] Conduct the thresholding described by (\ref{threshold}) such that the coordinates $k=1,\cdots,p$ are assigned to either $\mathcal{U}(s)$ or its complement $\mathcal{U}^c(s)$ where $\mathcal{U}(s)$ is defined in (\ref{signal.set}). For all $k \in \mathcal{U}^c(s)$, we set $\hat{\delta}_k=0$;
\item[(3).] Allocate $l \in \mathcal{U}(s)$ into different self-connected subsets $\{I_0^{(1)}, I_0^{(2)}, \cdots, I_0^{(h)}\}$ with respect to $(V_0, \Omega^*)$. For $I_0^{(1)}$, $\delta(I^{(1)}_0)$ is equal to $\hat{\delta}(I^{(1)}_0)$ each component of which is chosen to be either $0$ or $\delta^{date}$ or $-\delta^{date}$ in order to minimize the penalized function (\ref{pmle_e}). Repeat the same procedure to other $I^{(j)}_0$ where $j \in \{2,\cdots, h\}$ to determine $\delta_l$ for $l \in \mathcal{U}(s)$.      
\end{itemize}

To easily measure the performance of the proposed DATE procedure, we further assume the following condition which is analogous to (C1) but requires a slightly stronger signal strength than (C1). A similar strategy was also taken in Ji and Jin (2012) to measure the performance of an UPS procedure for variable selection in the high dimensional regression problem.
     
$(\mbox{C1})^{\prime}$. Similar to (C1), the components of $\delta$ follow the mixture distribution with 
$\pi_p$ being a distribution on the support $[-(1+\eta)\sqrt{2r\mbox{log}p/n}, -\sqrt{2r\mbox{log}p/n}] \cup [\sqrt{2r\mbox{log}p/n}, (1+\eta)\sqrt{2r\mbox{log}p/n}]$ where $\eta \le \frac{\beta-\Lambda}{\sqrt{C_0}r} \frac{\sqrt{\beta r}}{\sqrt{(\underline{\omega}r-\beta+\Lambda)^2+4\underline{\omega}r\beta}}$ and the constant $C_0$ is defined in (C2).   

The following theorem establishes the upper bound of the risk (\ref{risk}) for  the proposed DATE procedure.

\textbf{Theorem 3.} Assume conditions (C2)-(C4) and $(\mbox{C1})^{\prime}$. Choose $s \in (0, \frac{(\underline{\omega}r+\beta-\Lambda)^2}{4\underline{\omega}r})$ and $\beta-\Lambda <\underline{\omega}r <(\sqrt{1-\Lambda}+\sqrt{1- \beta})^2$, and set the tuning parameters in (\ref{pmle_e}) to be 
\[
\quad \lambda^{date}=\sqrt{2(\beta-\Lambda) \mbox{log}p}, \quad \delta^{date}=\sqrt{2r\mbox{log}p/n}. 
\]
As $p \to \infty$, the weighted risk (\ref{risk}) for the DATE satisfies 
\[
H(\Lambda)\le L_p  p^{1-\beta-(\underline{\omega}r-\beta+\Lambda)^2/(4\underline{\omega}r )}. 
\]

Since $(\underline{\omega}r-\beta+\Lambda)^2/(4\underline{\omega}r )\le (\bar{\omega}r-\beta+\Lambda)^2/(4\bar{\omega}r )$, the lower bound in Theorem 1 is no greater than the upper bound in Theorem 3. Specially, these two bounds match each other if $\bar{\omega}=\underline{\omega}$, which implies both bounds are tight and thus the DATE procedure is rate optimal in terms of the risk (\ref{risk}). 

Our ultimate goal is to apply the DATE procedure to signal identification. So we need to ensure that it can successfully control the FDR  at any desired level $\alpha <1$. By carefully reviewing the whole procedure, we see that the thresholding step (\ref{threshold}) is designated to control the false negatives and the success of the FDR control is determined only by the excising step (\ref{pmle_e}) where the role is played by the tuning parameter $\lambda^{date}$. Due to the adoption of $L_0$ penalty, smaller value of $\lambda^{date}$ allows more toleration for the false positives and thus leads to greater FDR. It turns out that if we subtract an additional term from the $\lambda^{date}$ in Theorem 3, the mFDR can be successfully controlled at $\alpha <1$ and the rate of the mFNR is accordingly established by Theorem 4.   

\textbf{Theorem 4.} Assume conditions (C2)-(C4) and $(\mbox{C1})^{\prime}$. Choose $s \in (0, \beta)$, $\beta-\Lambda <\underline{\omega}r <(\sqrt{1-\Lambda}+\sqrt{1- \beta})^2$ and $\Lambda=(\sqrt{\underline{\omega}r}-\sqrt{\beta})^2$. As $p \to \infty$, by setting the tuning parameters of the DATE as
\[
\quad \lambda^{date}=\sqrt{2(\beta-\Lambda) \mbox{log}p-\Upsilon}, \quad \delta^{date}=\sqrt{2r\mbox{log}p/n}, 
\]
where
\[
\Upsilon=\frac{4\underline{\omega}r}{\underline{\omega}r+\beta-\Lambda}\biggl(\frac{1}{2}\mbox{log}\mbox{log}p+\mbox{log}\biggl\{\frac{\alpha\sqrt{\pi}(\underline{\omega}r+\beta-\Lambda)}{2\sqrt{\underline{\omega}r}(1-\alpha)} \biggr\} \biggr).
\]
Then, 
\[
\mbox{mFDR}\le \alpha \quad \mbox{and} \quad \mbox{mFNR} \le L_p  p^{-\beta-(\sqrt{\underline{\omega}r}-\sqrt{\beta})^2}. 
\]

Since $\bar{\omega}r \ge \underline{\omega}r > \beta$, the optimal rate of the mFNR in Theorem 2 is not faster than the rate in Theorem 4 and two rates are equal to each other asymptotically if $\bar{\omega}=\underline{\omega}$. This, combining with the fact that $\mbox{mFDR} \le \alpha <1$, shows that the proposed DATE procedure is optimal in that it minimizes the mFNR subject to the constraint that mFDR is controlled at the desired level $\alpha <1$.

There are three tuning parameters needed to estimated in the proposed signal identification procedure: the level of threshold $s$ in (\ref{threshold}), two tuning parameters $\delta^{data}$ and $\lambda^{date}$ in (\ref{pmle_e}). To select tuning parameters $\lambda^{data}$ and $\delta^{date}$, we estimate the sparsity $\beta$, the signal magnitude $r$ and $\underline{\omega}$ by the following estimators:
\[
\hat{\beta}=-\frac{\mbox{log}\biggl\{\frac{1}{p}\sum_{k=1}^p \mbox{I}(\hat{T}_k > 2q \mbox{log}p)\biggr\}}{\mbox{log}p}, \hat{r}=\frac{1}{{2p^{1-\hat{\beta}}\,\mbox{log}p}}\,{\sum_{k=1}^p \frac{\hat{T}_k-1}{\hat{\omega}_{kk}} \,\mbox{I}(\hat{T}_k > 2q \mbox{log}p)}, 
\]
and
\begin{eqnarray}
\hat{\underline{\omega}}=\min\limits_{1\le k \le p}\hat{\omega}_{kk}, \label{tuning_e}
\end{eqnarray}
where $q$ is another threshold level controlling the accuracy of estimate in $\beta$ and $r$ and the question of properly choosing both $s$ and $q$ is addressed in Theorem 5.   
With two tuning parameters $\lambda^{data}$ and $\mu^{date}$ estimated by plugging the $\hat{\beta}, \hat{r}, \hat{\underline{\omega}}$ into the expressions defined in Theorem 4, 
the following theorem shows that the performance of the DATE procedure with estimated parameters (\ref{tuning_e}) is asymptotically equivalent to the DATE in Theorem 4.

\textbf{Theorem 5.} Assume conditions (C2)-(C4) and $(\mbox{C1})^{\prime}$. As $p \to \infty$, by setting $s \in (0, \beta)$ in (\ref{threshold}), $q \in (\beta, \underline{\omega}r)$ in (\ref{tuning_e}) and estimating the tuning parameters as 
\[
\hat{\lambda}=2\hat{s}\mbox{log}p, \quad \hat{\lambda}^{date}=\sqrt{2(\hat{\beta}-\hat{\Lambda}) \mbox{log}p-\hat{\Upsilon}}, \quad \hat{\delta}^{date}=\sqrt{2\hat{r}\mbox{log}p/n}, 
\]
where
\[
\hat{\Lambda}=(\sqrt{\underline{\hat{\omega}}\hat{r}}-\sqrt{\hat{\beta}})^2, 
\]
\[
\hat{\Upsilon}=\frac{4\underline{\hat{\omega}}\hat{r}}{\underline{\hat{\omega}}\hat{r}+\hat{\beta}-\hat{\Lambda}}\biggl(\frac{1}{2}\mbox{log}\mbox{log}p+\mbox{log}\biggl\{\frac{\alpha\sqrt{\pi}(\underline{\hat{\omega}}\hat{r}+\hat{\beta}-\hat{\Lambda})}{2\sqrt{\underline{\hat{\omega}}\hat{r}}(1-\alpha)} \biggr\} \biggr),  \quad \mbox{and}
\]
$\hat{\beta}$, $\hat{r}$ and $\hat{\underline{\omega}}$ are given by (\ref{tuning_e}), 
then, 
\[
\mbox{mFDR}\le \alpha \quad \mbox{and} \quad \mbox{mFNR} \le L_p  p^{-\beta-(\sqrt{\underline{\omega}r}-\sqrt{\beta})^2}. 
\]

Although two threshold levels $s$ and $q$ are not explicitly specified, simulation studies show that the performance of the proposed procedure is insensitive to $(s, q)$ as long as they are properly chosen from two intervals separated by $\beta \in (0,1)$.   

The optimality of the proposed DATE is established for the signal in the partial recovery region with $\underline{\omega}r <(\sqrt{1-\Lambda}+\sqrt{1- \beta})^2$. If $\underline{\omega}r \ge (\sqrt{1-\Lambda}+\sqrt{1- \beta})^2$, the region is the full recovery region. The lower bounds of the weighted risk $H(\Lambda)$ and the mFNR corresponding to this region converge to zero as $r$ tends to infinity at each fixed large value of $p$ as shown in Theorems 1 and 2. However, even when $\underline{\omega}r \ge (\sqrt{1-\Lambda}+\sqrt{1- \beta})^2$, the upper bounds for these two rates corresponding to the full recovery region will not vanish, since the proposed DATE procedure involves data transformation, precision matrix and tuning parameters estimation each of which contributes non-negligible error at the order of $o(p^{-1})$. Although this error is very small, it becomes prominent and dominant as $r$ is big enough to make two upper bounds established in Theorems 3, 4 and 5 smaller order of $o(p^{-1})$, and consequently the upper bounds of the weighted risk $H(\Lambda)$ and the mFNR will be at the rate of $o(p^{-1})$.

\setcounter{section}{5} \setcounter{equation}{0}
\section*{\large 5. Simulation Study}

Simulation studies were conducted to demonstrate the performance of the proposed  procedure for signals recovery under different combinations of signal sparsity controlled by $\beta$, signal strength $r$ and data dependence. The proposed procedure is denoted by $\mbox{DATE}_{\Omega}$ if $\Omega$ is known and $\mbox{DATE}_{\hat{\Omega}}$ if $\Omega$ is unknown. For comparison, the BH procedure was also implemented as follows: each of $p$ coordinates is tested by the two-sample t test to obtain the ordered $p$-values $P_{(1)}< \cdots < P_{(p )}$. Based on the cutoff value   
$m=\mbox{max}\{1\le k \le p: P_{(k)} \le k\alpha/p \}$, 
the coordinates with $P_i \le P_{(m)}$ are treated as signal bearing dimensions. 

The random samples $\{X_{ij}\}$ were generated from $\mbox{N}(\mu_i, \Sigma)$ for $i=1, 2$.
Without loss of generality, $\mu_1=0$ and $\mu_2$ had $[p^{1-\beta}]$ nonzero coordinates which were uniformly and randomly drawn from $\{1,\cdots, p\}$. The magnitude of each nonzero entry of $\mu_2$ was randomly drawn from the interval $[\sqrt{r\mbox{log}p/n}, \sqrt{3r\mbox{log}p/n}]$ and then multiplied by a random sign. Four models were considered for the covariance matrix $\Sigma=(\sigma_{ij})$:
\begin{itemize} 
\item[ ] (a). AR(1) model:  $\sigma_{ij}=\rho^{|i-j|}$ for $1\le i, j\le p$.
\item[ ] (b). Block diagonal model: $\sigma_{ii}=1$ for $i=1,\cdots,p$, and $\sigma_{ij}=0.6$ for $2(k-1)+1\le i\ne j \le 2k$ where $k=1, \cdots, [p/2]$.
\item[ ] (c). Penta-diagonal model: $\sigma_{ii}=1$ for $i=1,\cdots,p$, $\sigma_{ij}=0.5$ for $|i-j|=1$ and $\sigma_{ij}=0.2$ for $|i-j|=2$.
\item[ ] (d). Random sparse matrix model: first generate a $p \times p$ matrix $\Gamma$ each row of which has only one non-zero element that is randomly chosen from $\{1, \cdots, p\}$ with magnitude generated from Unif(1, 2) multiplied by a random sign. $\Sigma$ is then obtained by standardizing $\Gamma \Gamma^{T}+\bm{\mbox{I}}$ to have unit diagonal elements. 
\end{itemize}

To apply the $\mbox{DATE}_{\hat{\Omega}}$, we need to estimate $\Omega$. For models $(a)-(c )$, the Cholesky decomposition approach (Bickel and Levina, 2008a) was implemented. 
Recall that the precision matrix $\Omega$ can be decomposed as   
$\Omega=(I-A)^{\prime} D^{-1} (I-A)$ where $A$ is a lower triangular matrix with zero diagonals and $D$ is a diagonal matrix. The elements below the diagonal element on the $k$th row of $A$ can be thought as the regression coefficients of the $k$th component on its predecessors, and the $k$th diagonal element of $D$ is the corresponding residual variance. 
Let $A_{\tau}$ be the $\tau$-banded lower triangular matrix of $A$ and $D_{\tau}$ be the corresponding residual variances on the diagonals. The $\tau$-banded precision matrix $\Omega_{\tau}=(I-A_{\tau})^{\prime} D^{-1}_{\tau} (I-A_{\tau})$. Given a sample, $A_{\tau}$ and $D_{\tau}$ can be estimated by  the least square estimation, which leads to
\[
\hat{\Omega}_{\tau}=(I-\hat{A}_{\tau})^{\prime}\hat{D}_{\tau}^{-1}(I-\hat{A}_{\tau}),
\] 
where the banding width parameter  $\tau$  in the estimation of $\Omega$
was chosen according to the data-driven procedure proposed by Bickel
and Levina (2008a).  For a given data
set, we divided it into two subsamples by repeated ($N=50$ times)
random data split. For the $l$-th split,  $l \in \{1, \cdots,
N\}$, we let
$\hat{{\Sigma}}_{\tau}^{(l)}=\{(I-\hat{A}_{\tau}^{(l)})^{\prime}\}^{-1}
\hat{D}_{\tau}^{(l)} (I-\hat{A}_{\tau}^{(l)})^{-1}$ be the Cholesky
decomposition of ${\Sigma}$ obtained from the first subsample by
taking the same approach described in previous section for
$\hat{A}_{\tau}^{(l)}$ and $\hat{D}_{\tau}^{(l)}$. Also we let
${S}_n^{(l)}$ be the sample covariance matrix obtained from the second
subsample. Then the banding parameter $\tau$ is selected as 
\be
\hat{\tau}=\min \limits_{\tau} \frac{1}{N}\sum_{l=1}^{N}
||\hat{{\Sigma}}_{\tau}^{(l)}-{S}_n^{(l)} ||_{F}, \label{risk2} 
\ee 
where
$||\cdot||_F$ denotes the Frobenius norm.

In model (d), $\Sigma$ is first estimated by applying the thresholding operator $T_m$ to the sample covariance matrix $S_n$ such that
\[
T_m(S_n)=[s_{ij}\mbox{I}(|s_{ij}|\ge m)].
\]
As shown by Bickel and Levina (2008b), $||T_m(S_n)-\Sigma||=o_p(1)$ under the condition (C4) where $||\cdot||$ is the spectral norm. The threshold $m$ can be selected by the cross-validation method. Given a data of size $n$, we split it into two sub-samples with sizes of $n_1=n\{1-\frac{1}{\mbox{log}(n)}\}$ and $n_2=\frac{n}{\mbox{log}(n)}$ for $N=50$ times. For the $l$-th split, let $S_{1,l}$ and $S_{2,l}$ be the sample covariance matrices based on the $n_1$ and $n_2$ observations. The threshold $\hat{m}$ is chosen to minimize
\be
R_s= \frac{1}{N}\sum_{l=1}^{N}
||T_m(S_{1,l})-S_{2,l}||_{F}^2. \label{risk3} 
\ee      
Then $\Omega$ can be estimated by $T^{-1}_{\hat{m}}(S_n)$ since $T_{\hat{m}}(S_n)$ is positive definite provided that $||T_{\hat{m}}-T_0|| \le \epsilon$ and $\lambda_{min} > \epsilon$ (Bickel and Levina, 2008b).

\begin{sidewaysfigure}
\begin{minipage}[b]{0.32\linewidth}
\centering
\includegraphics[width=\textwidth]{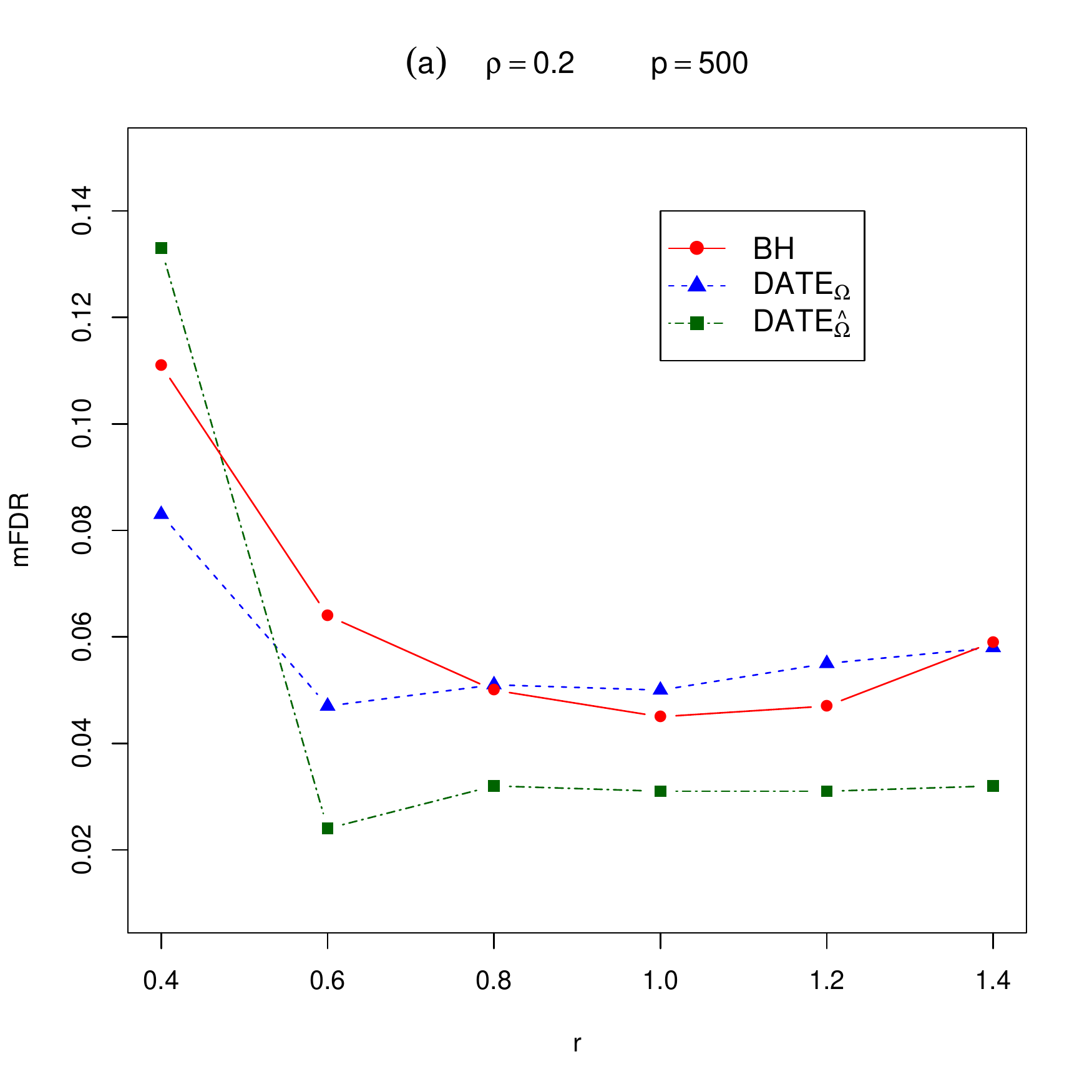}
\end{minipage}
\hspace{0.25cm}
\begin{minipage}[b]{0.32\linewidth}
\centering
\includegraphics[width=\textwidth]{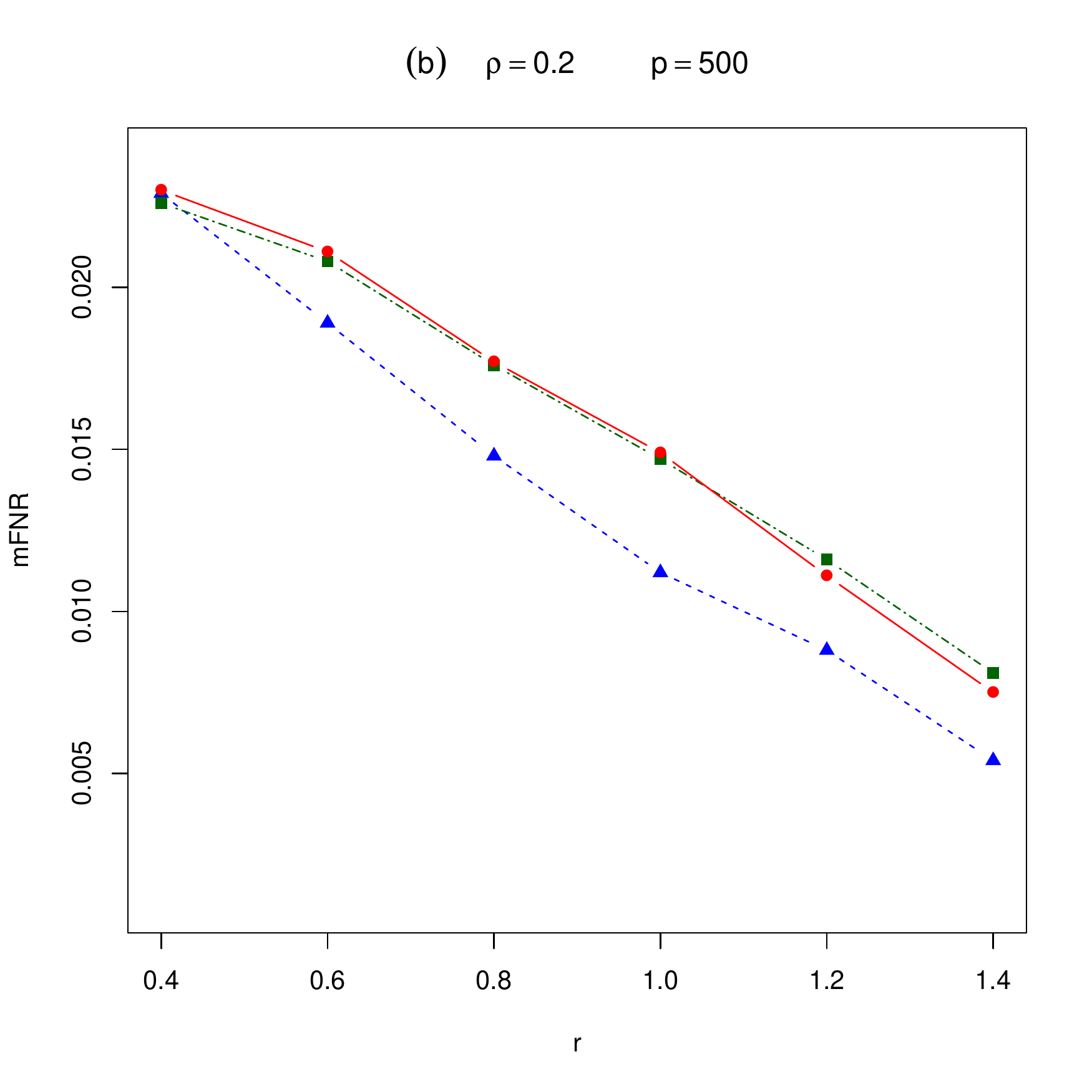}
\end{minipage}
\hspace{0.25cm}
\begin{minipage}[b]{0.32\linewidth}
\centering
\includegraphics[width=\textwidth]{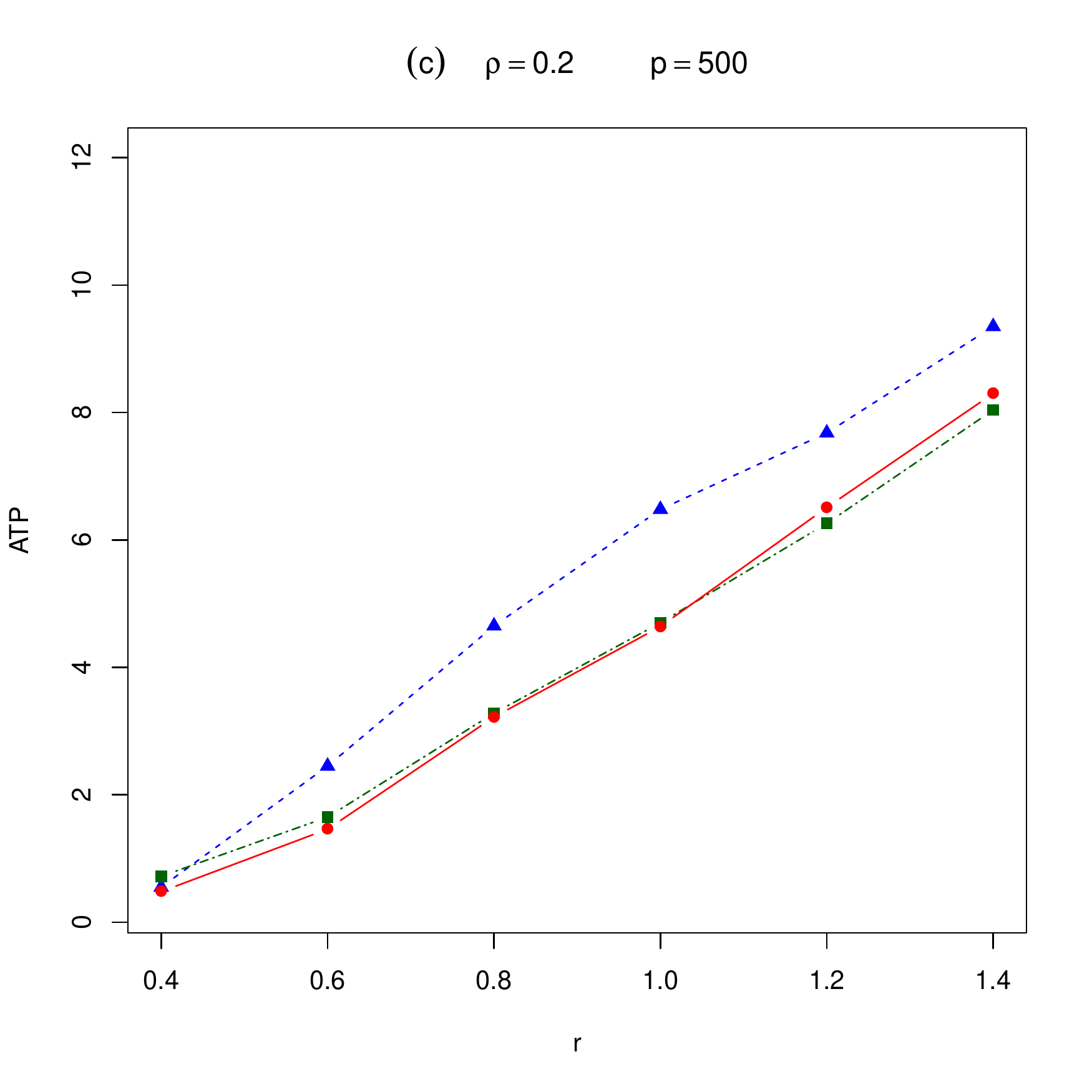}
\end{minipage}
\begin{minipage}[b]{0.32\linewidth}
\centering
\includegraphics[width=\textwidth]{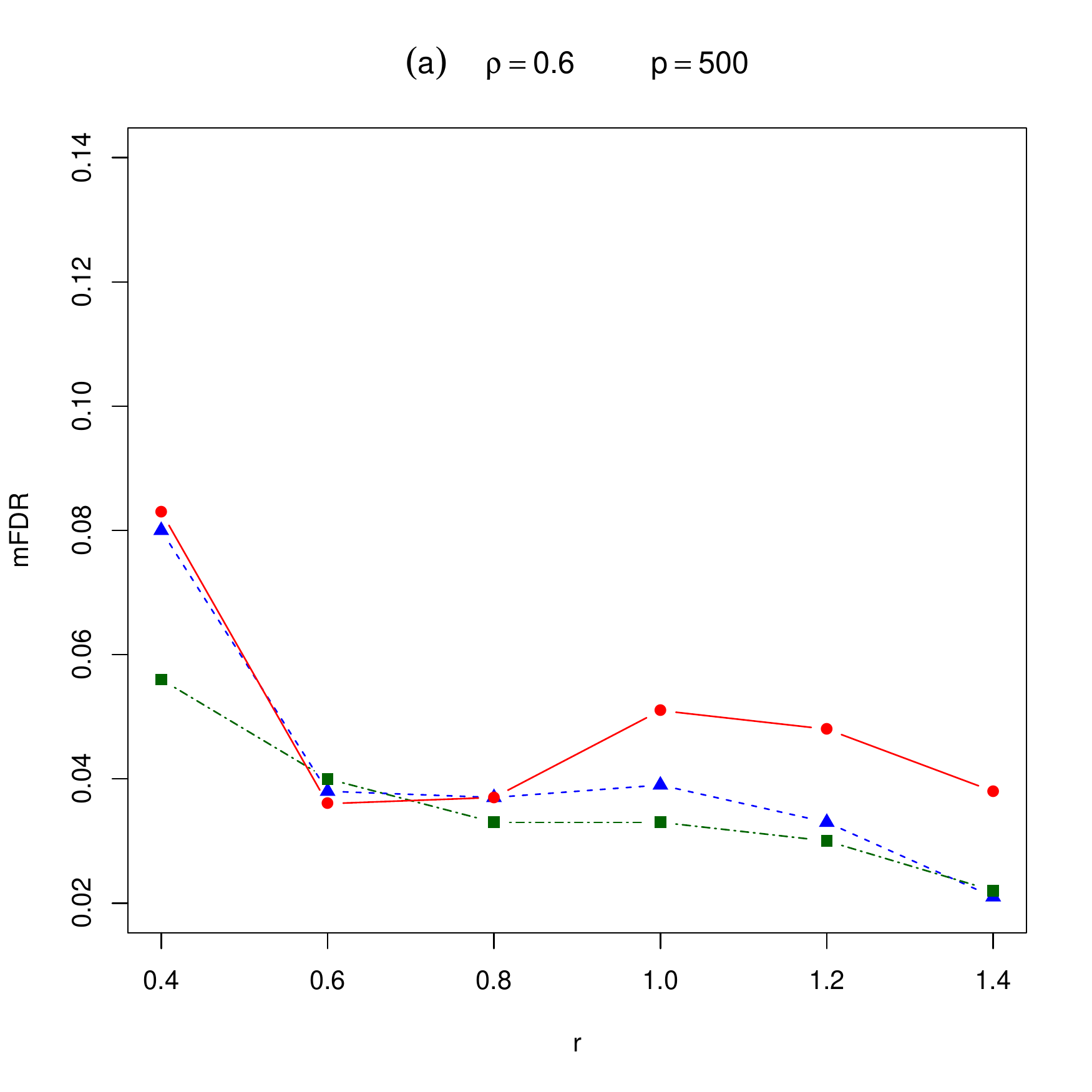}
\end{minipage}
\hspace{0.25cm}
\begin{minipage}[b]{0.32\linewidth}
\centering
\includegraphics[width=\textwidth]{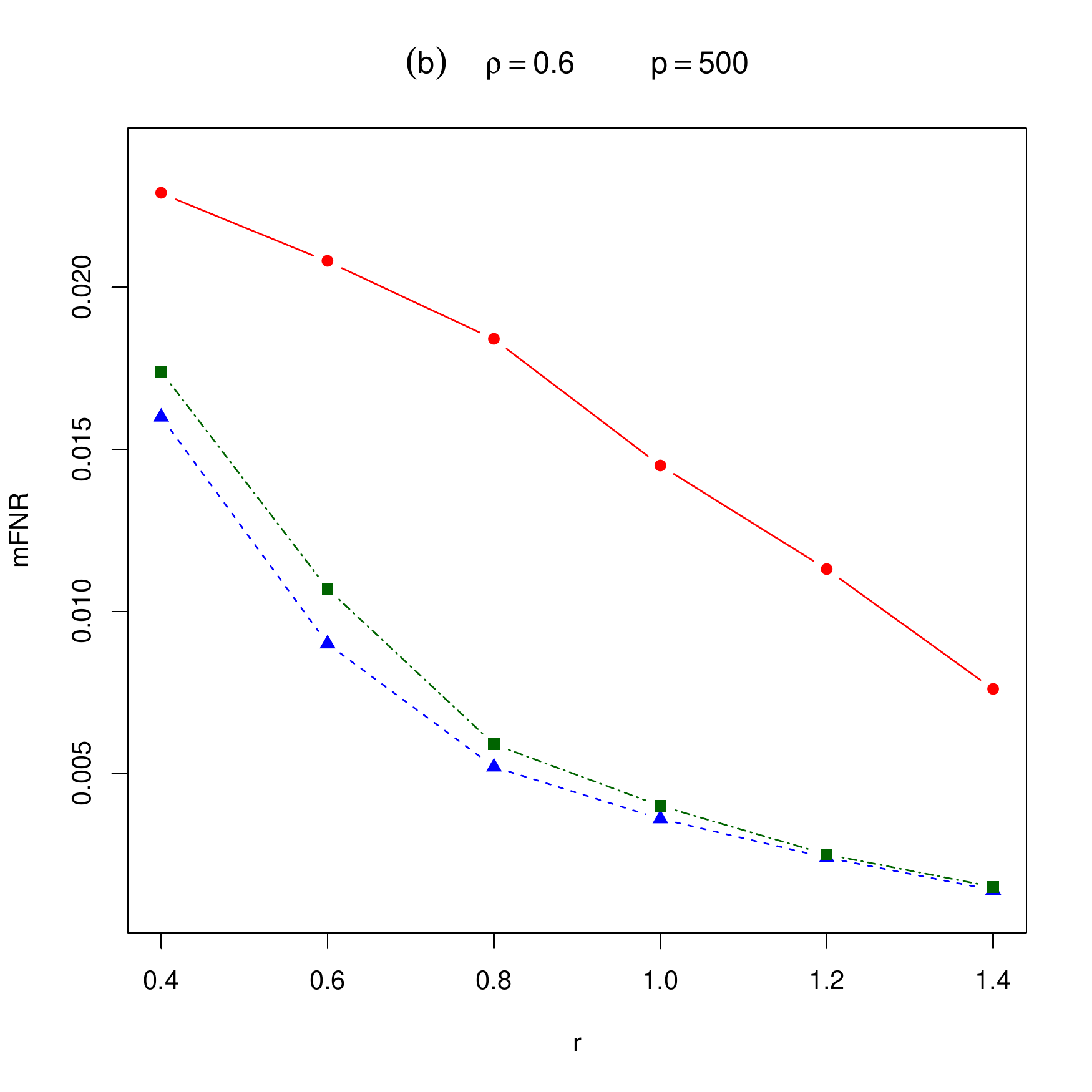}
\end{minipage}
\hspace{0.25cm}
\begin{minipage}[b]{0.32\linewidth}
\centering
\includegraphics[width=\textwidth]{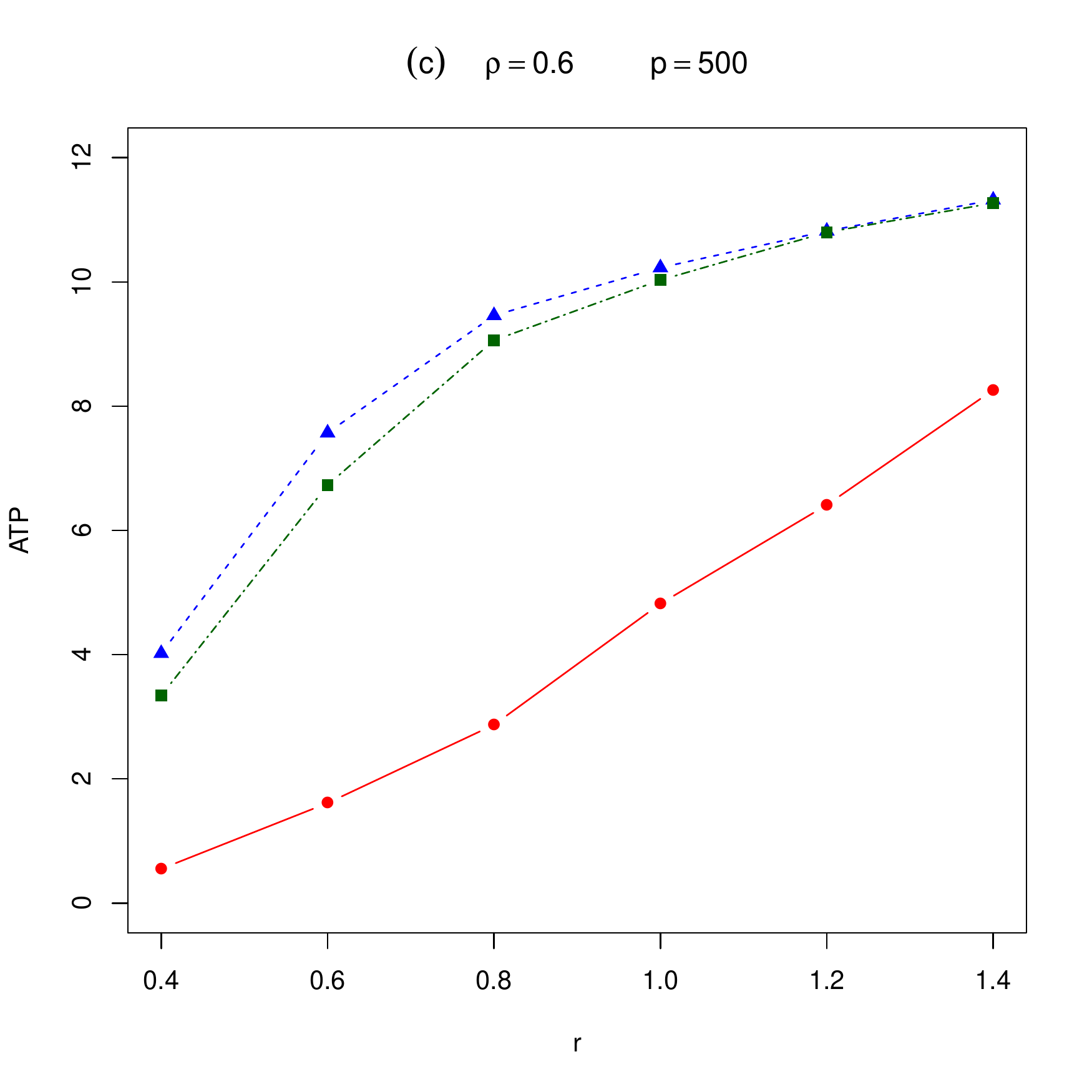}
\end{minipage}
\caption{The mFDR, mFNR and ATP yielded by $\mbox{DATE}_{\Omega}$, $\mbox{DATE}_{\hat{\Omega}}$ and the BH procedure under model (a). The dimension $p=500$, sample sizes $n_1=60$ and $n_2=60$ and $\beta=0.6$.}
\label{comp1}
\end{sidewaysfigure}

\begin{sidewaysfigure}
\begin{minipage}[b]{0.32\linewidth}
\centering
\includegraphics[width=\textwidth]{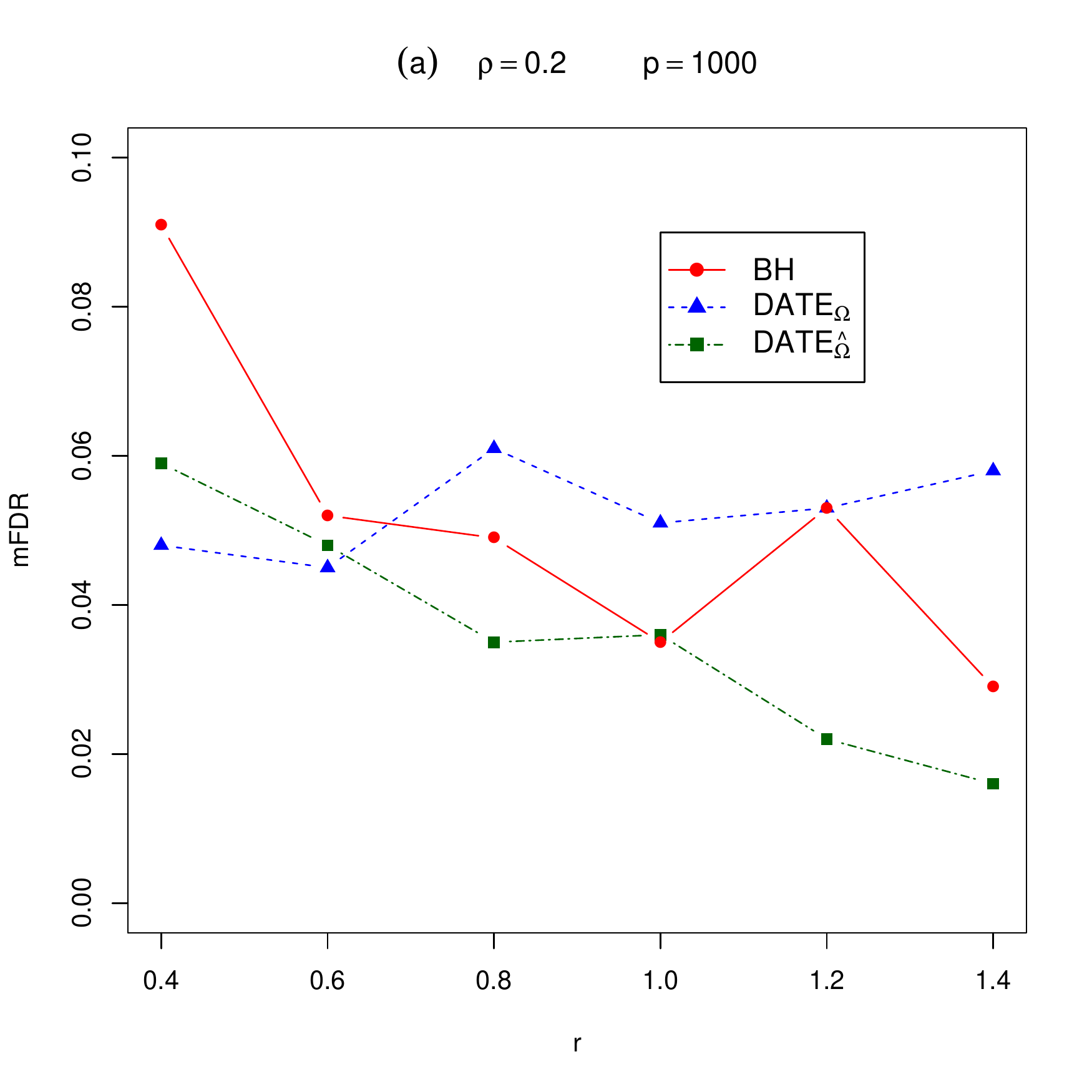}
\end{minipage}
\hspace{0.25cm}
\begin{minipage}[b]{0.32\linewidth}
\centering
\includegraphics[width=\textwidth]{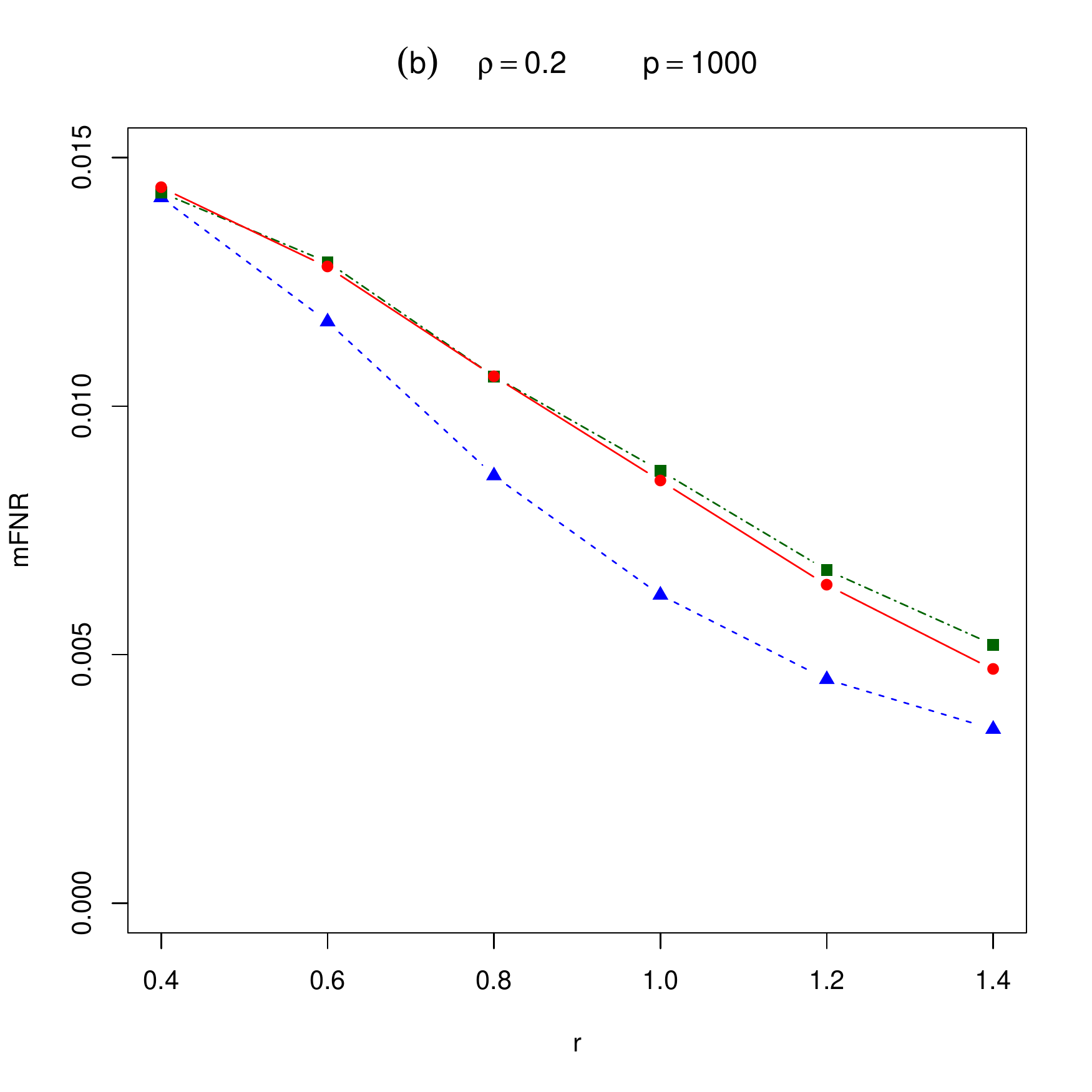}
\end{minipage}
\hspace{0.25cm}
\begin{minipage}[b]{0.32\linewidth}
\centering
\includegraphics[width=\textwidth]{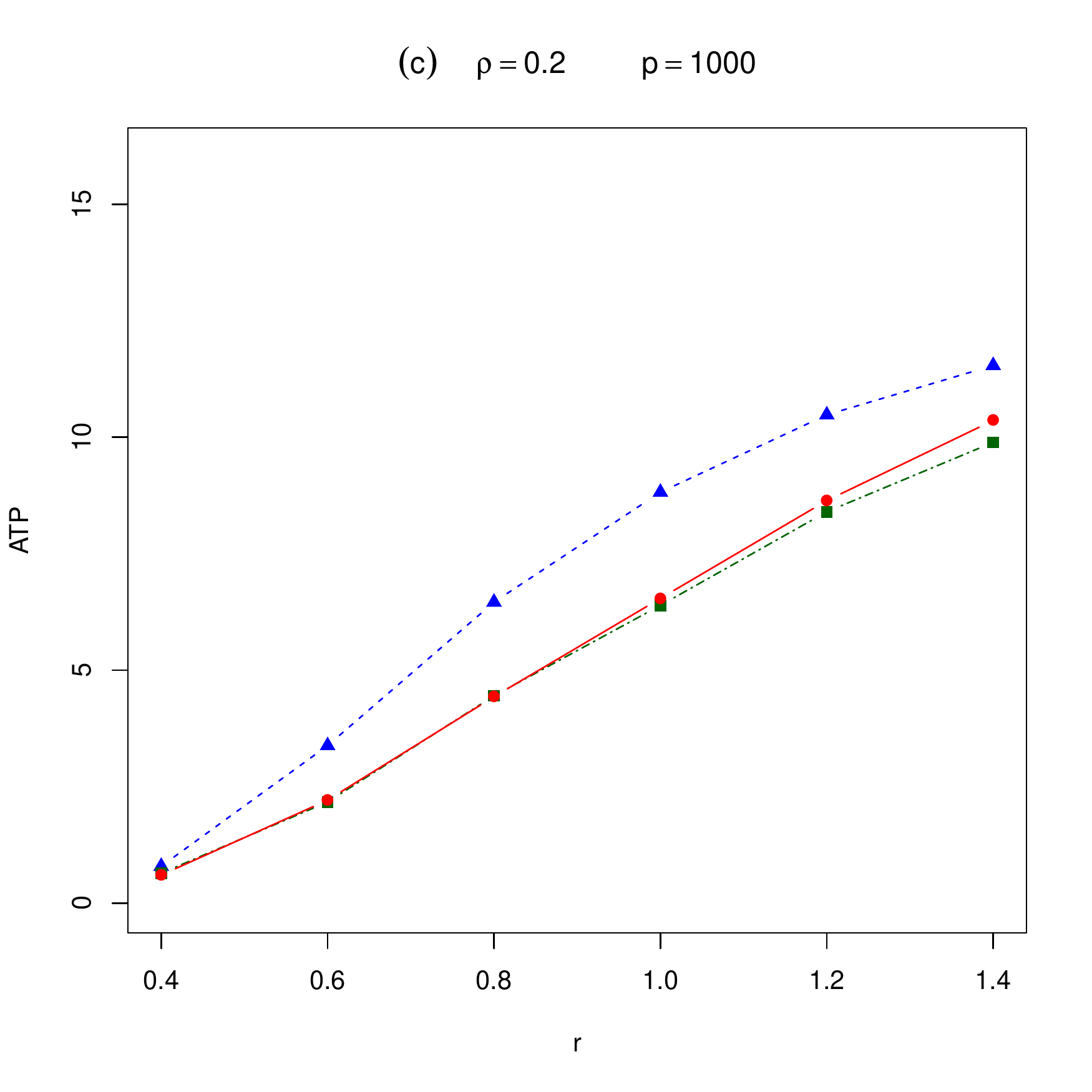}
\end{minipage}
\begin{minipage}[b]{0.32\linewidth}
\centering
\includegraphics[width=\textwidth]{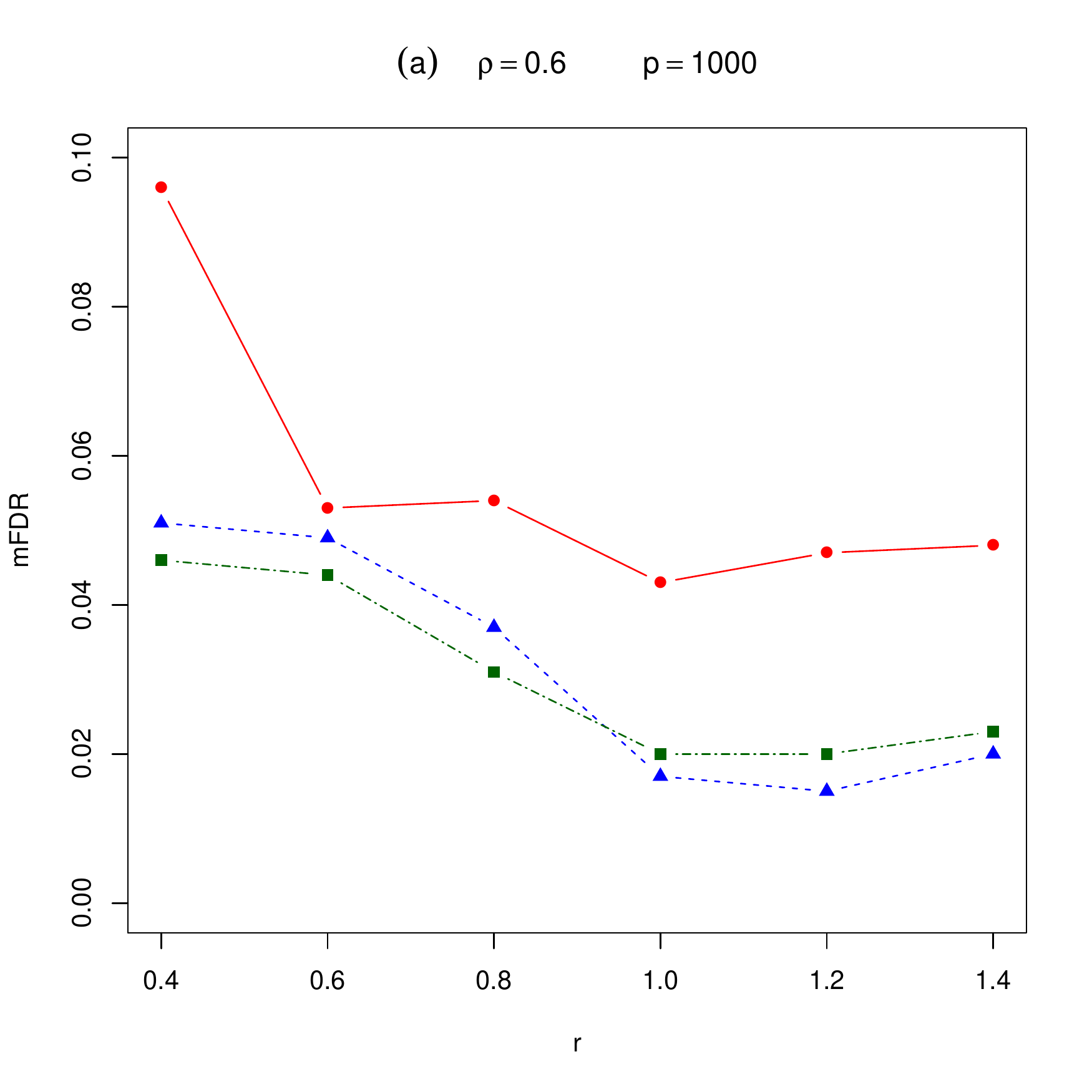}
\end{minipage}
\hspace{0.25cm}
\begin{minipage}[b]{0.32\linewidth}
\centering
\includegraphics[width=\textwidth]{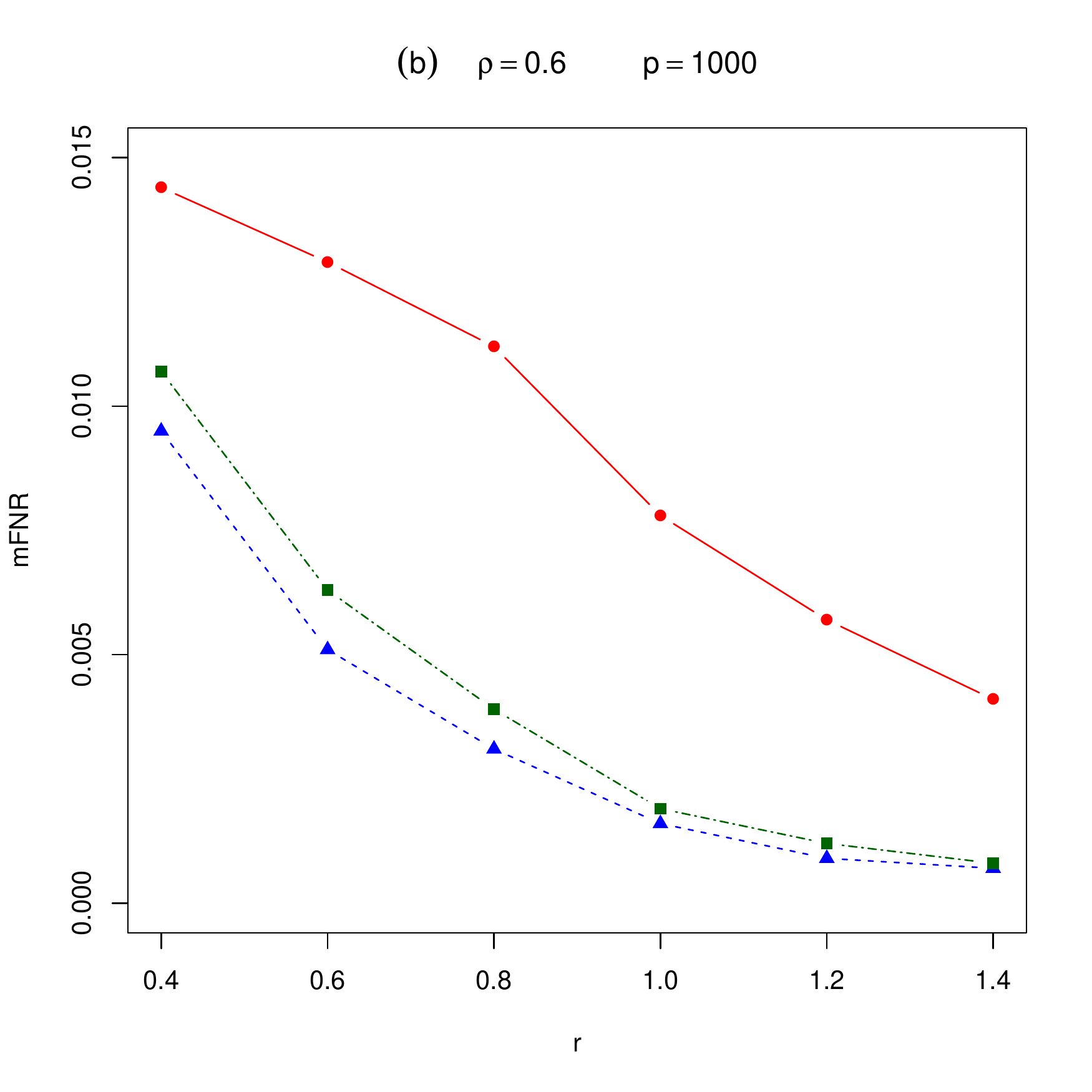}
\end{minipage}
\hspace{0.25cm}
\begin{minipage}[b]{0.32\linewidth}
\centering
\includegraphics[width=\textwidth]{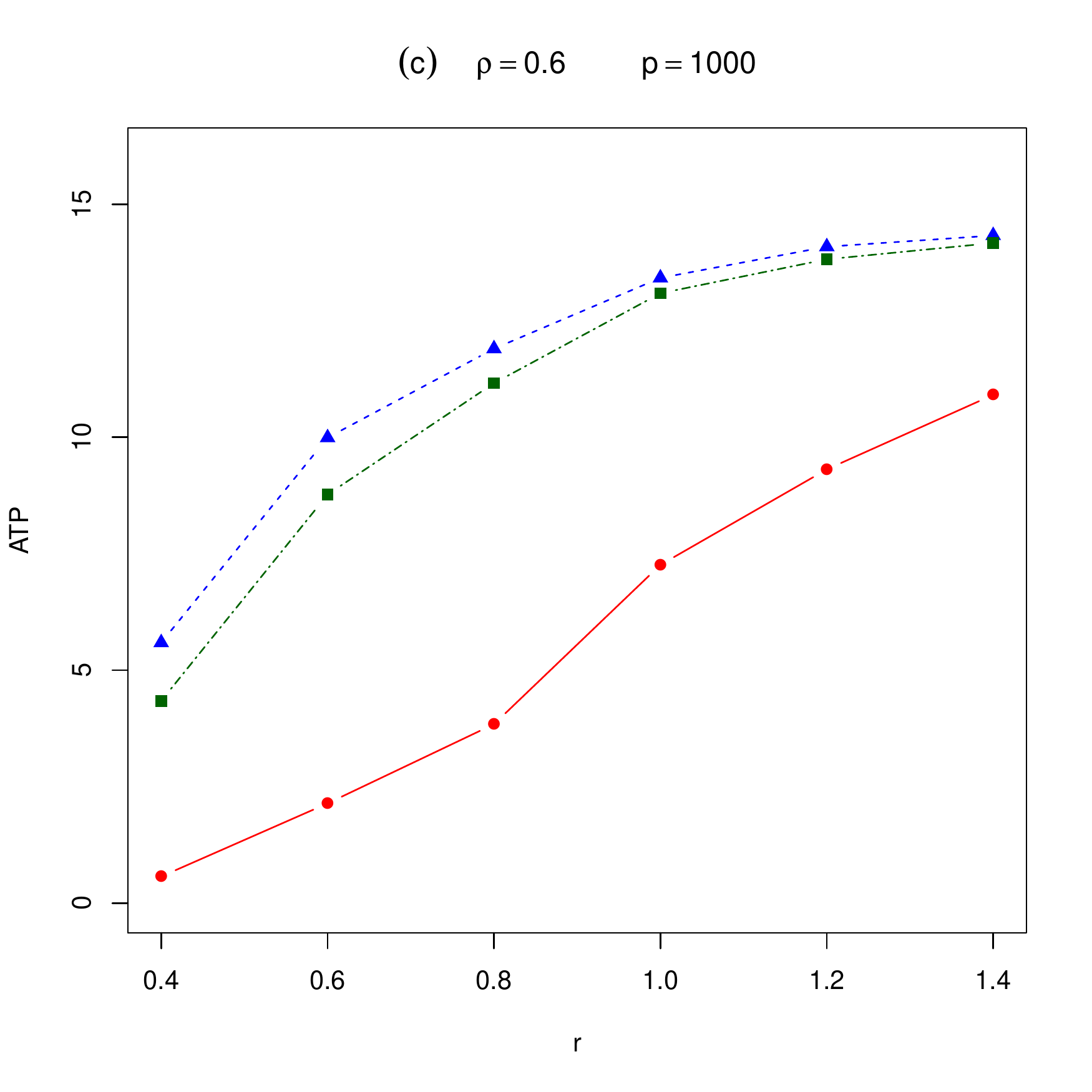}
\end{minipage}
\caption{The mFDR, mFNR and ATP yielded by $\mbox{DATE}_{\Omega}$, $\mbox{DATE}_{\hat{\Omega}}$ and the BH procedure under model (a). The dimension $p=1000$, sample sizes $n_1=60$ and $n_2=60$ and $\beta=0.6$.}
\label{comp2}
\end{sidewaysfigure}

The performance of each signal recovery procedure was evaluated by mFDR, mFNR and the average number of true positives ATP based on 100 replications. The nominal FDR level was set at $\alpha= 0.05$. 
Figure \ref{comp1} displays the performance of three procedures with different values of signal strength $r$ and data dependence $\rho$ under model (a) when $p=500$. In the first row of the Figure, data were weakly dependent and all three procedures had the mFDR controlled below the nominal level $0.05$ except $r=0.4$. The distortion of the mFDR at $r=0.4$ is due to the fact that the signals fall into the region of no recovery since $r< \beta/\bar{\omega}$ with $\bar{\omega}=1.08$ when $\rho=0.2$. With the dependence increased from $\rho=0.2$ to $0.6$, the inflation in mFDR was mitigated since $r>\beta/\underline{\omega}$ with $\underline{\omega}=1.56$ when $\rho=0.6$. Although the $\mbox{DATE}_{\Omega}$, $\mbox{DATE}_{\hat{\Omega}}$ and BH performed similarly in terms of the mFNR and ATP with  weakly dependent $\rho=0.2$, both $\mbox{DATE}_{\Omega}$ and $\mbox{DATE}_{\hat{\Omega}}$ had more ATP which is close to the number of true signals $[500^{0.4}]=12$ for strong signal strength $r$, and suffered less mFNR than the BH with moderate dependent $\rho=0.6$, which confirms that the data dependence is utilized by the proposed procedures for signal identification.  When dimension $p$ was increased from $500$ to $1000$, Figure \ref{comp2} demonstrates the results similar to Figure \ref{comp1}. Specially with strong signal strength $r$, the recovery of signals by both $\mbox{DATE}_{\Omega}$ and $\mbox{DATE}_{\hat{\Omega}}$ was close to the number of true signals $[1000^{0.4}]=16$.      

\begin{sidewaysfigure}
\begin{minipage}[b]{0.32\linewidth}
\centering
\includegraphics[width=\textwidth]{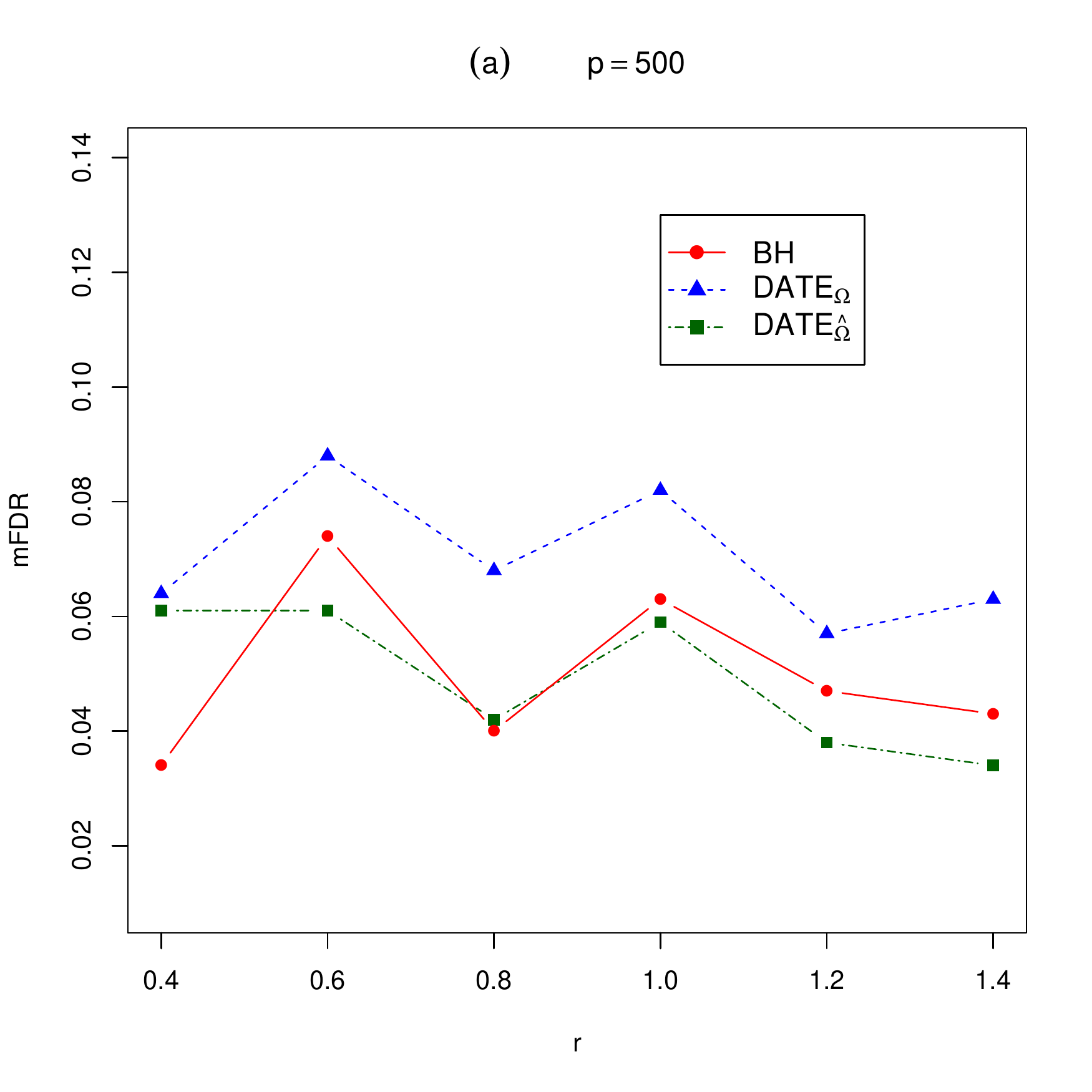}
\end{minipage}
\hspace{0.25cm}
\begin{minipage}[b]{0.32\linewidth}
\centering
\includegraphics[width=\textwidth]{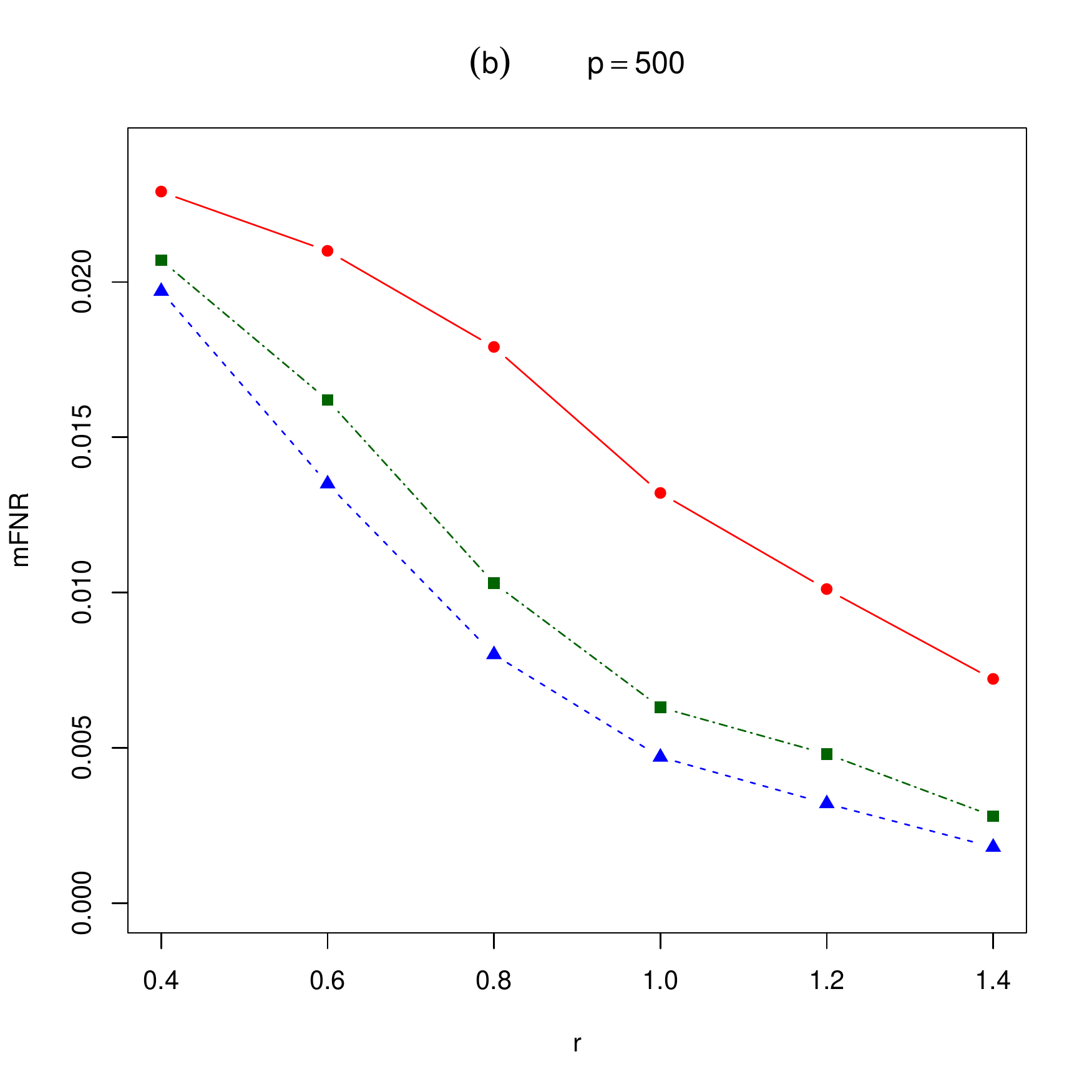}
\end{minipage}
\hspace{0.25cm}
\begin{minipage}[b]{0.32\linewidth}
\centering
\includegraphics[width=\textwidth]{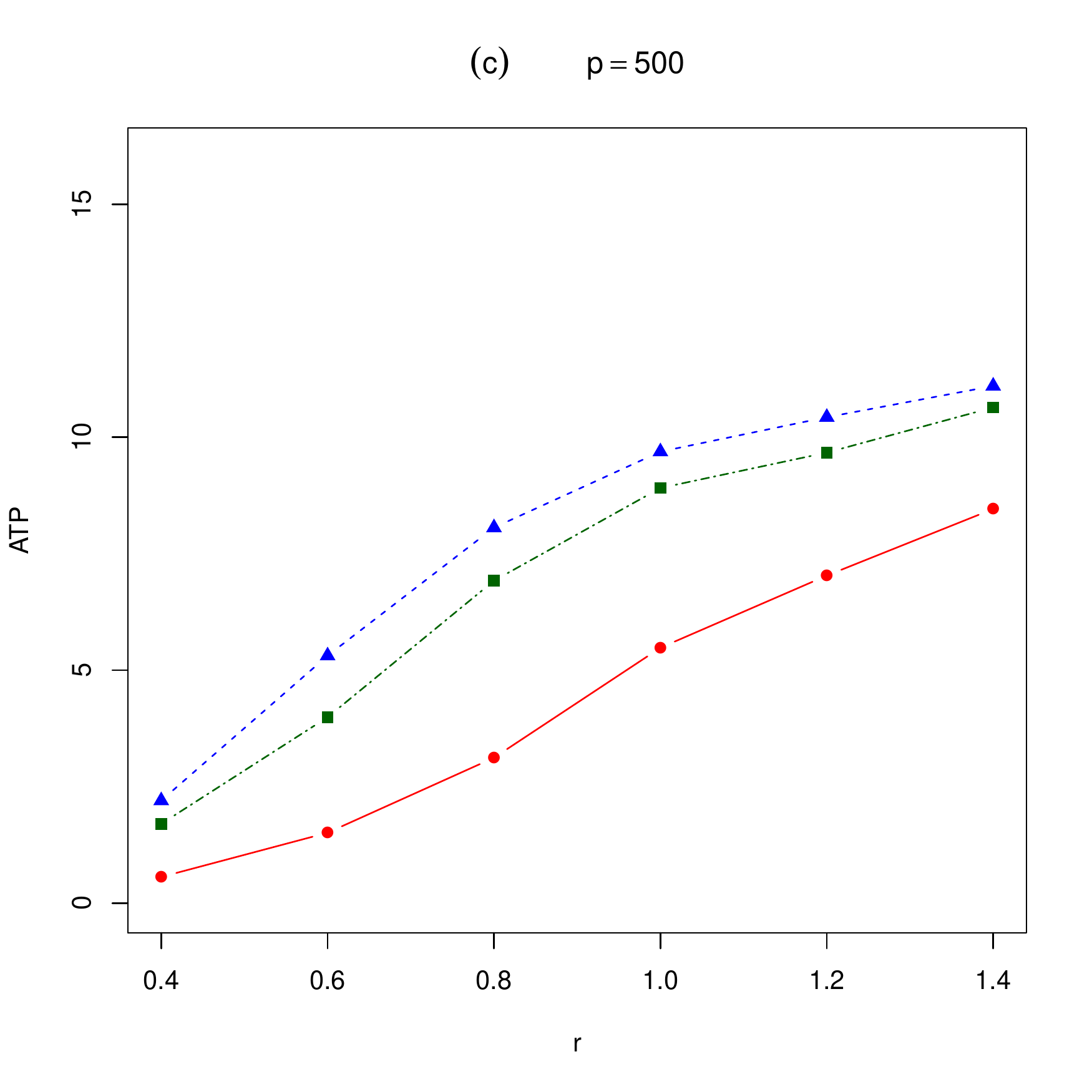}
\end{minipage}
\begin{minipage}[b]{0.32\linewidth}
\centering
\includegraphics[width=\textwidth]{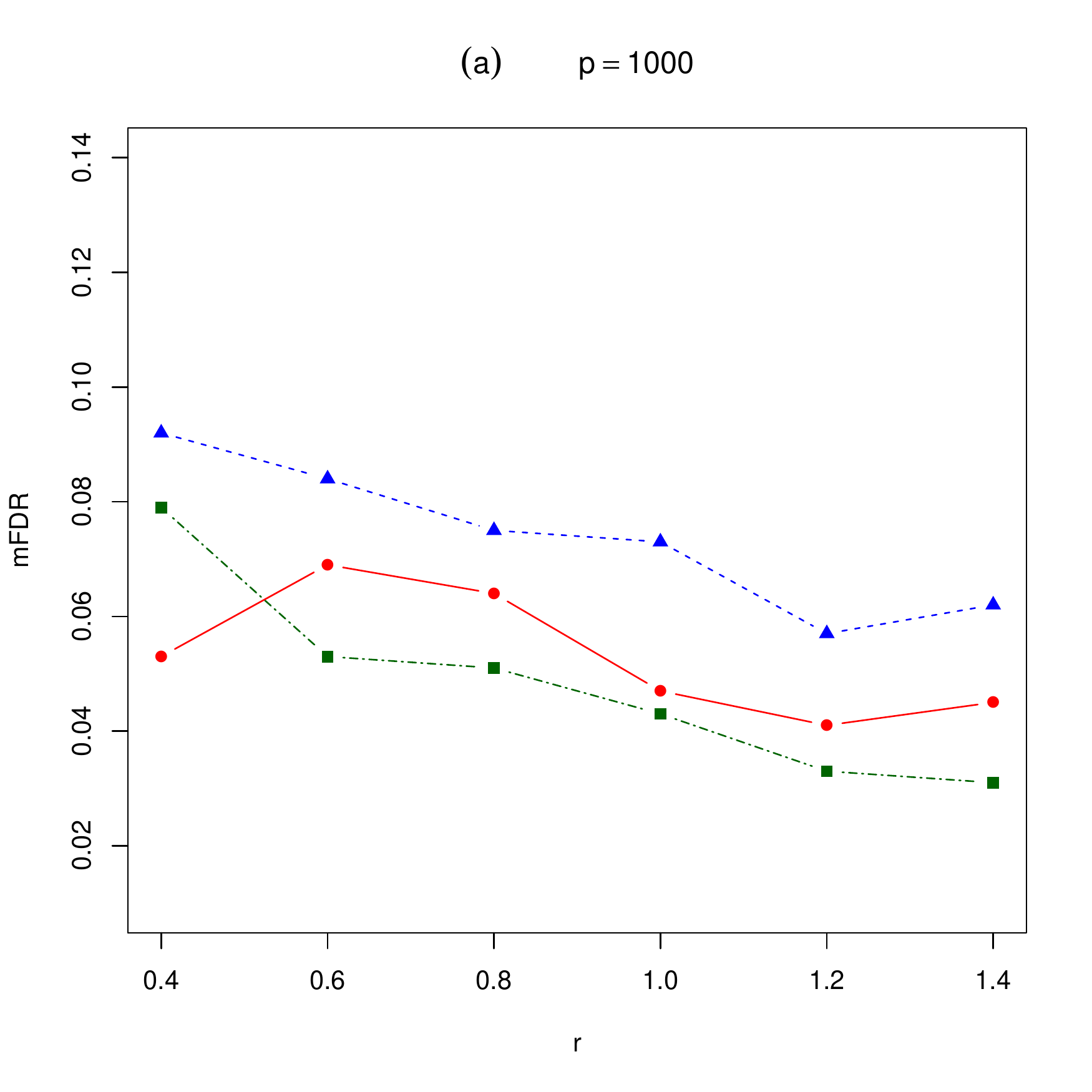}
\end{minipage}
\hspace{0.25cm}
\begin{minipage}[b]{0.32\linewidth}
\centering
\includegraphics[width=\textwidth]{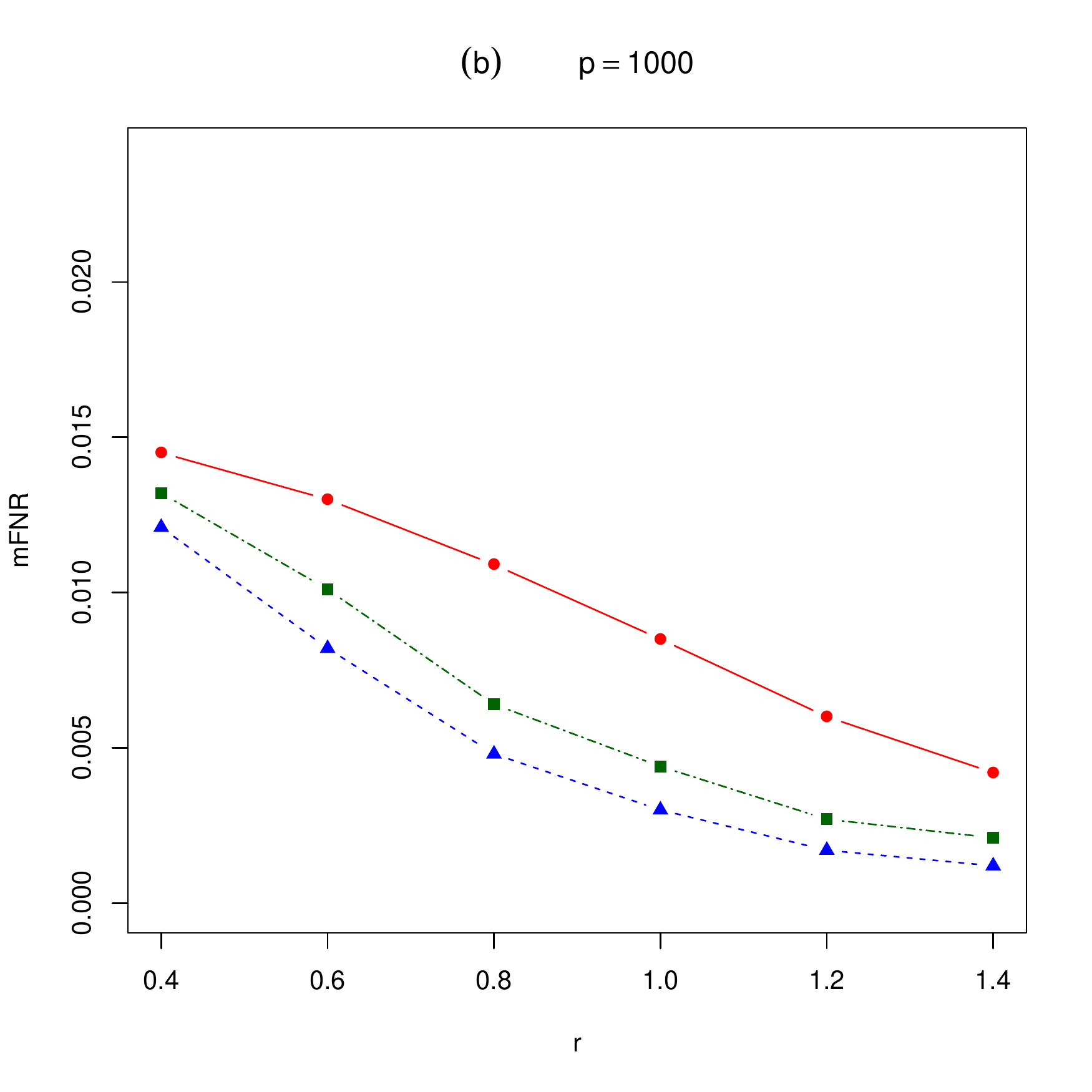}
\end{minipage}
\hspace{0.25cm}
\begin{minipage}[b]{0.32\linewidth}
\centering
\includegraphics[width=\textwidth]{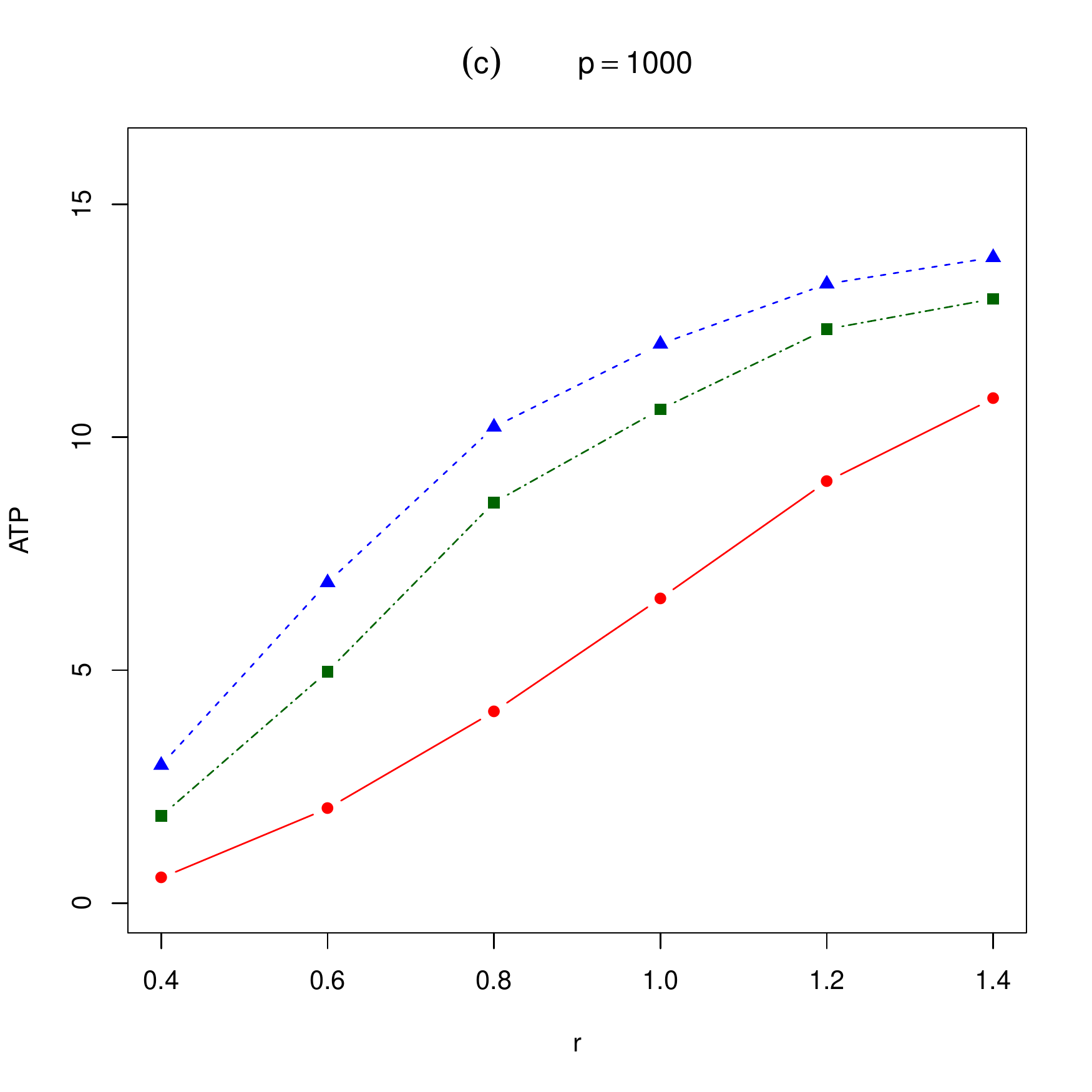}
\end{minipage}
\caption{The mFDR, mFNR and ATP yielded by $\mbox{DATE}_{\Omega}$, $\mbox{DATE}_{\hat{\Omega}}$ and the BH procedure under model (b). The sample sizes $n_1=60$ and $n_2=60$ and $\beta=0.6$.}
\label{comp3}
\end{sidewaysfigure}

\begin{sidewaysfigure}
\begin{minipage}[b]{0.32\linewidth}
\centering
\includegraphics[width=\textwidth]{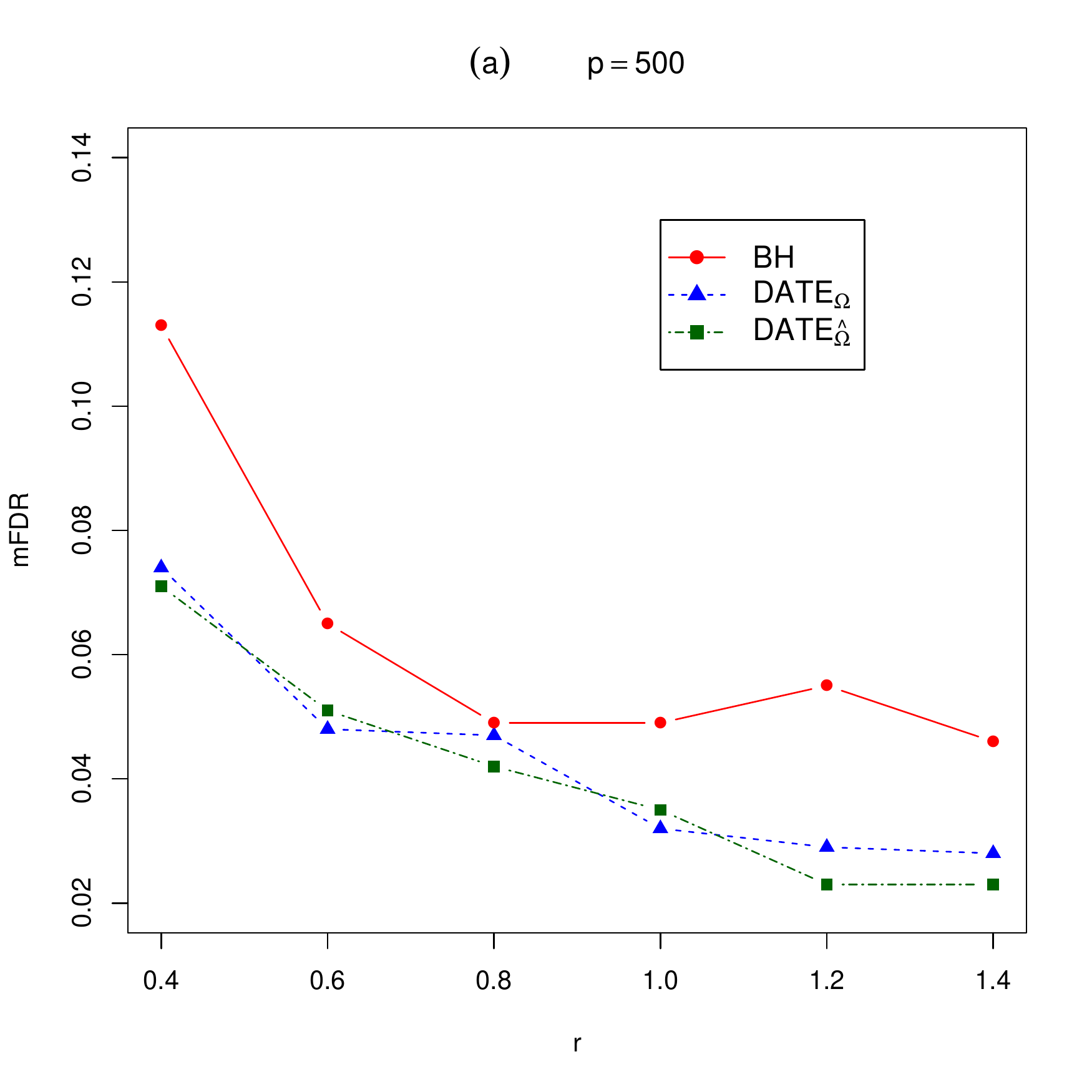}
\end{minipage}
\hspace{0.25cm}
\begin{minipage}[b]{0.32\linewidth}
\centering
\includegraphics[width=\textwidth]{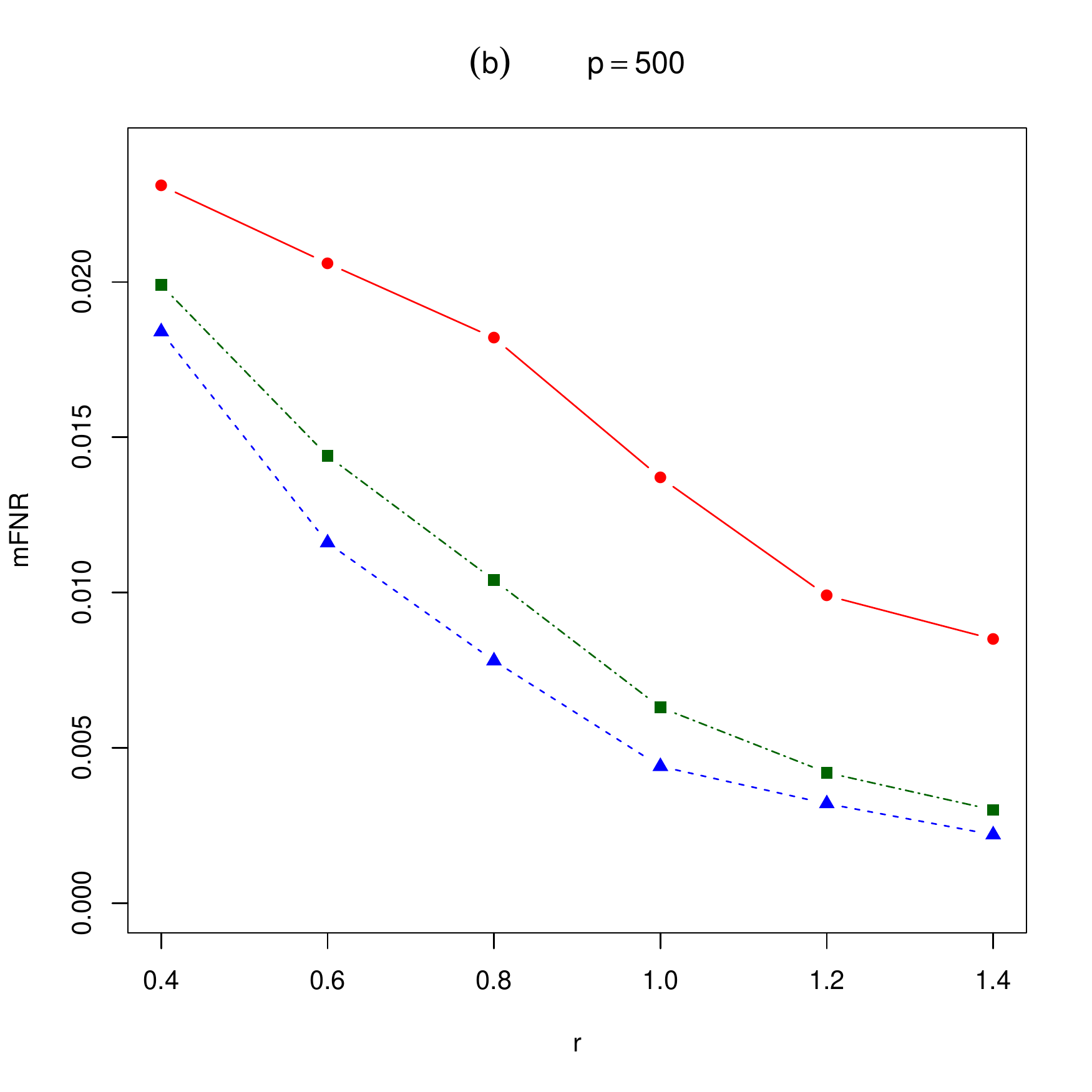}
\end{minipage}
\hspace{0.25cm}
\begin{minipage}[b]{0.32\linewidth}
\centering
\includegraphics[width=\textwidth]{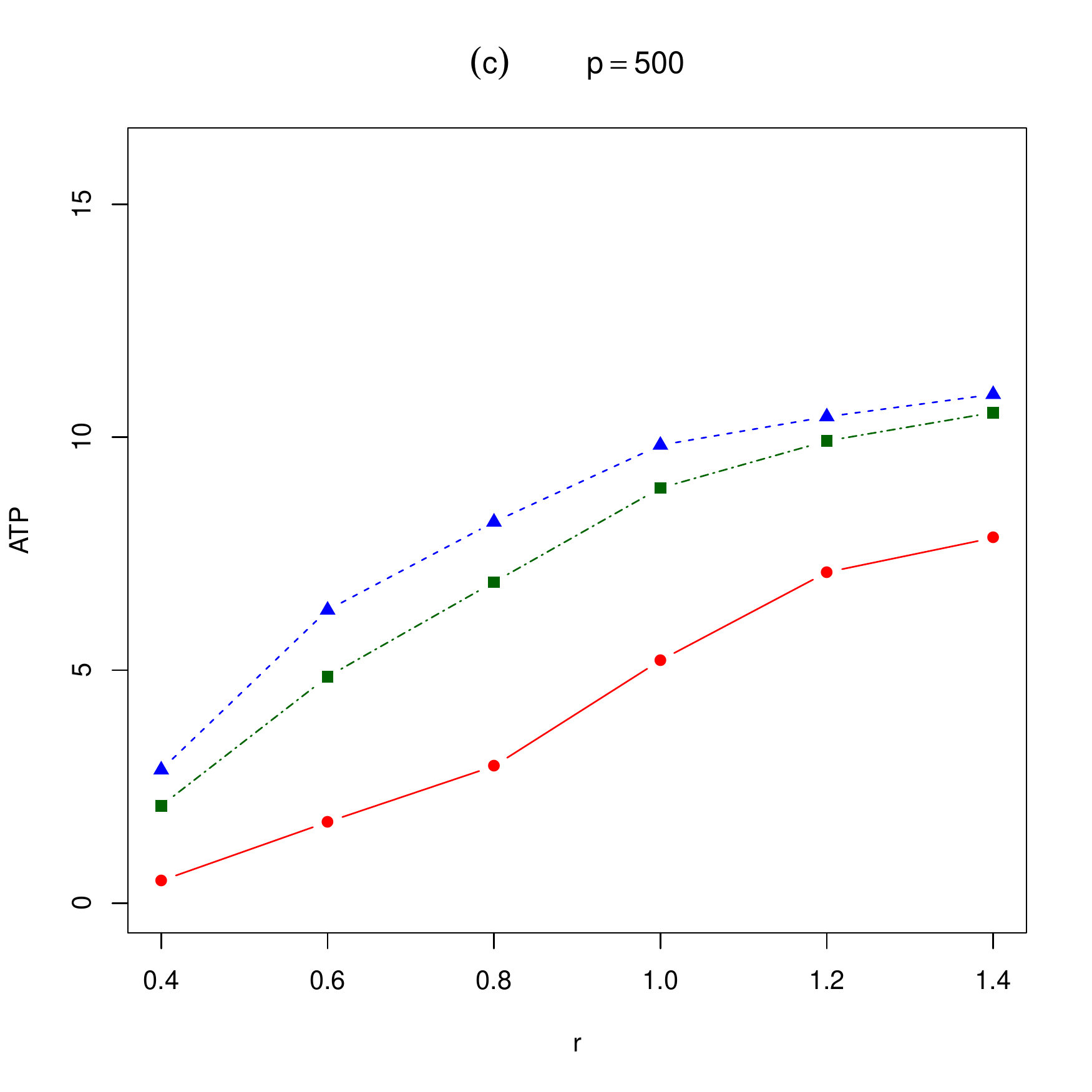}
\end{minipage}
\begin{minipage}[b]{0.32\linewidth}
\centering
\includegraphics[width=\textwidth]{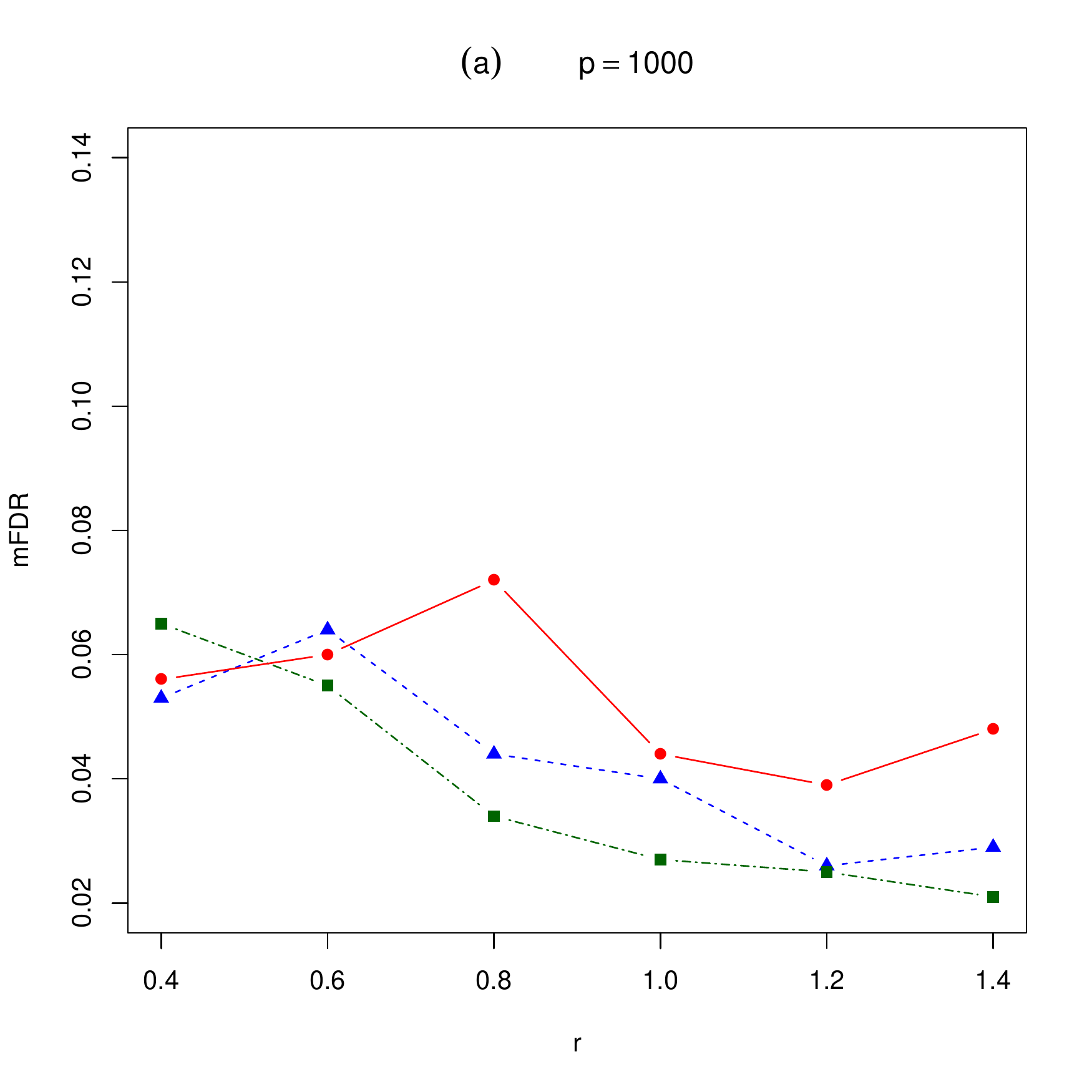}
\end{minipage}
\hspace{0.25cm}
\begin{minipage}[b]{0.32\linewidth}
\centering
\includegraphics[width=\textwidth]{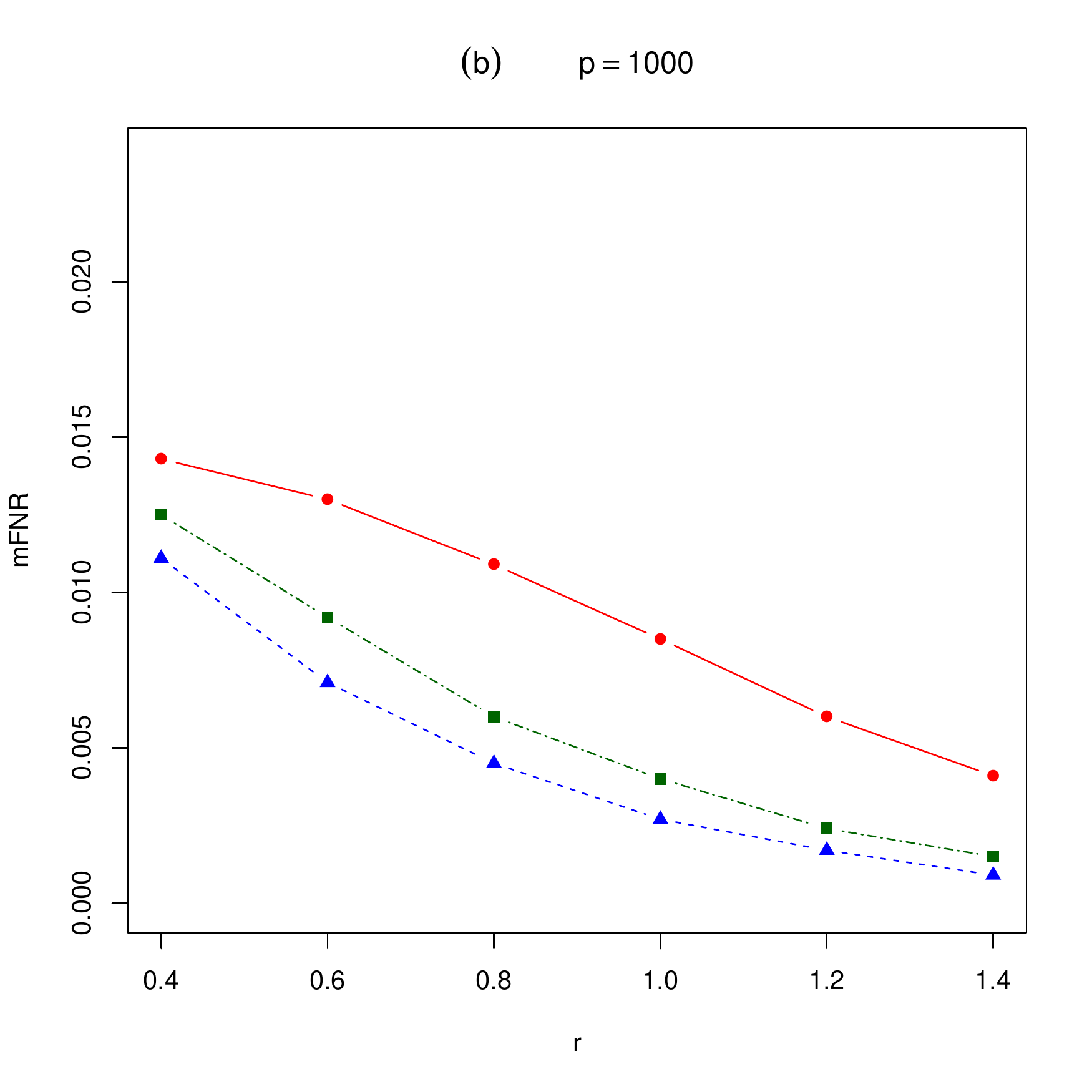}
\end{minipage}
\hspace{0.25cm}
\begin{minipage}[b]{0.32\linewidth}
\centering
\includegraphics[width=\textwidth]{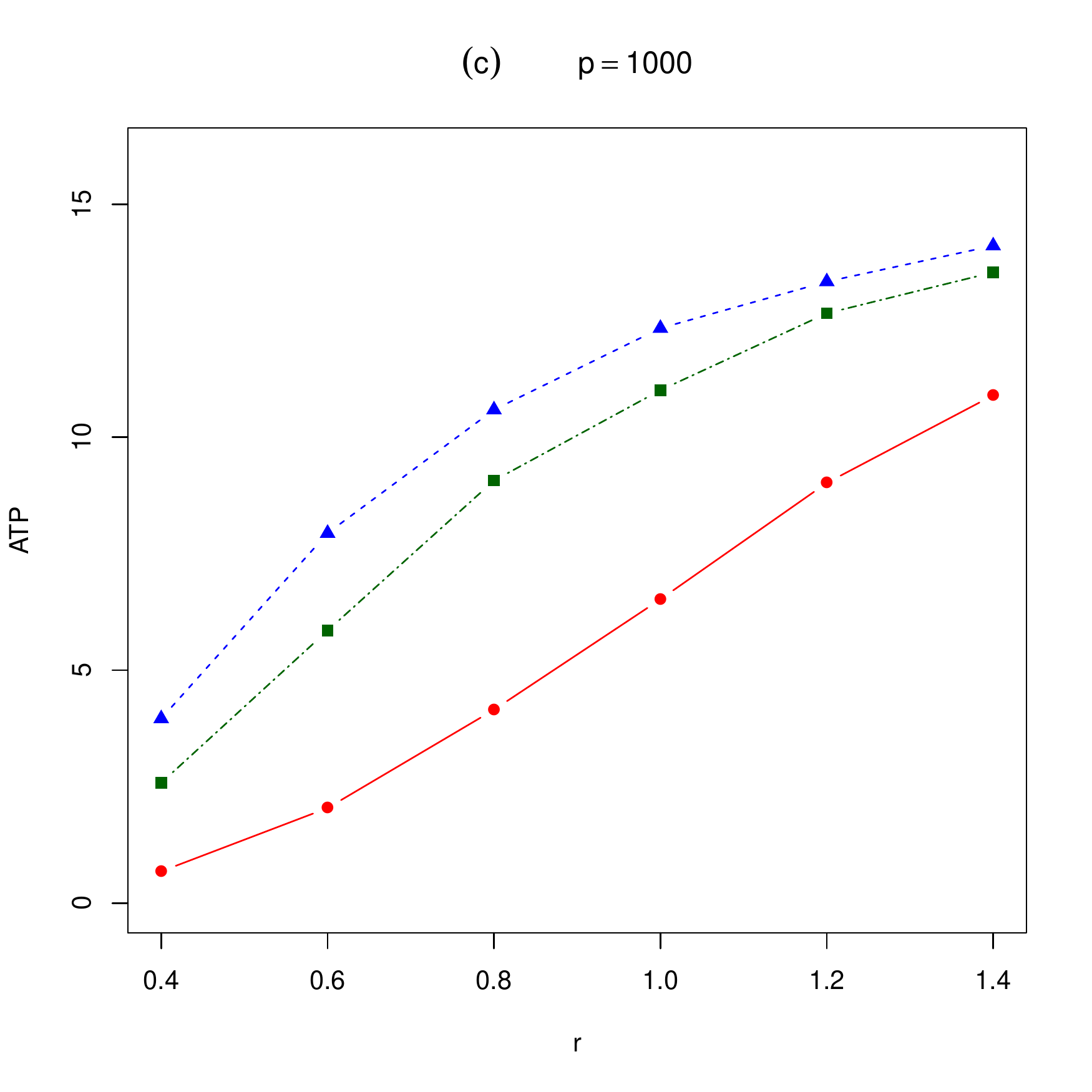}
\end{minipage}
\caption{The mFDR, mFNR and ATP yielded by $\mbox{DATE}_{\Omega}$, $\mbox{DATE}_{\hat{\Omega}}$ and the BH procedure under model (c). The sample sizes $n_1=60$ and $n_2=60$ and $\beta=0.6$.}
\label{comp4}
\end{sidewaysfigure}

\begin{sidewaysfigure}
\begin{minipage}[b]{0.32\linewidth}
\centering
\includegraphics[width=\textwidth]{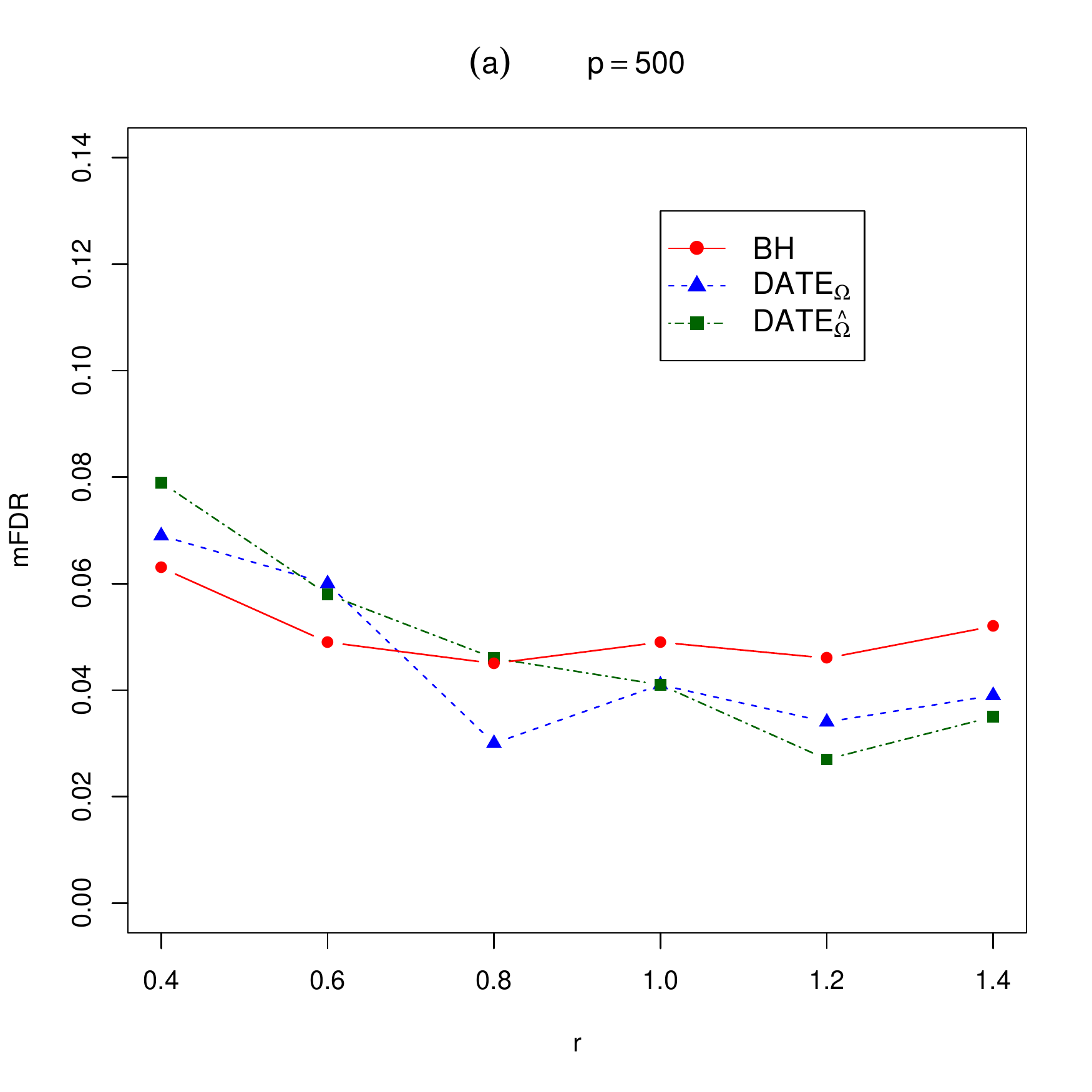}
\end{minipage}
\hspace{0.25cm}
\begin{minipage}[b]{0.32\linewidth}
\centering
\includegraphics[width=\textwidth]{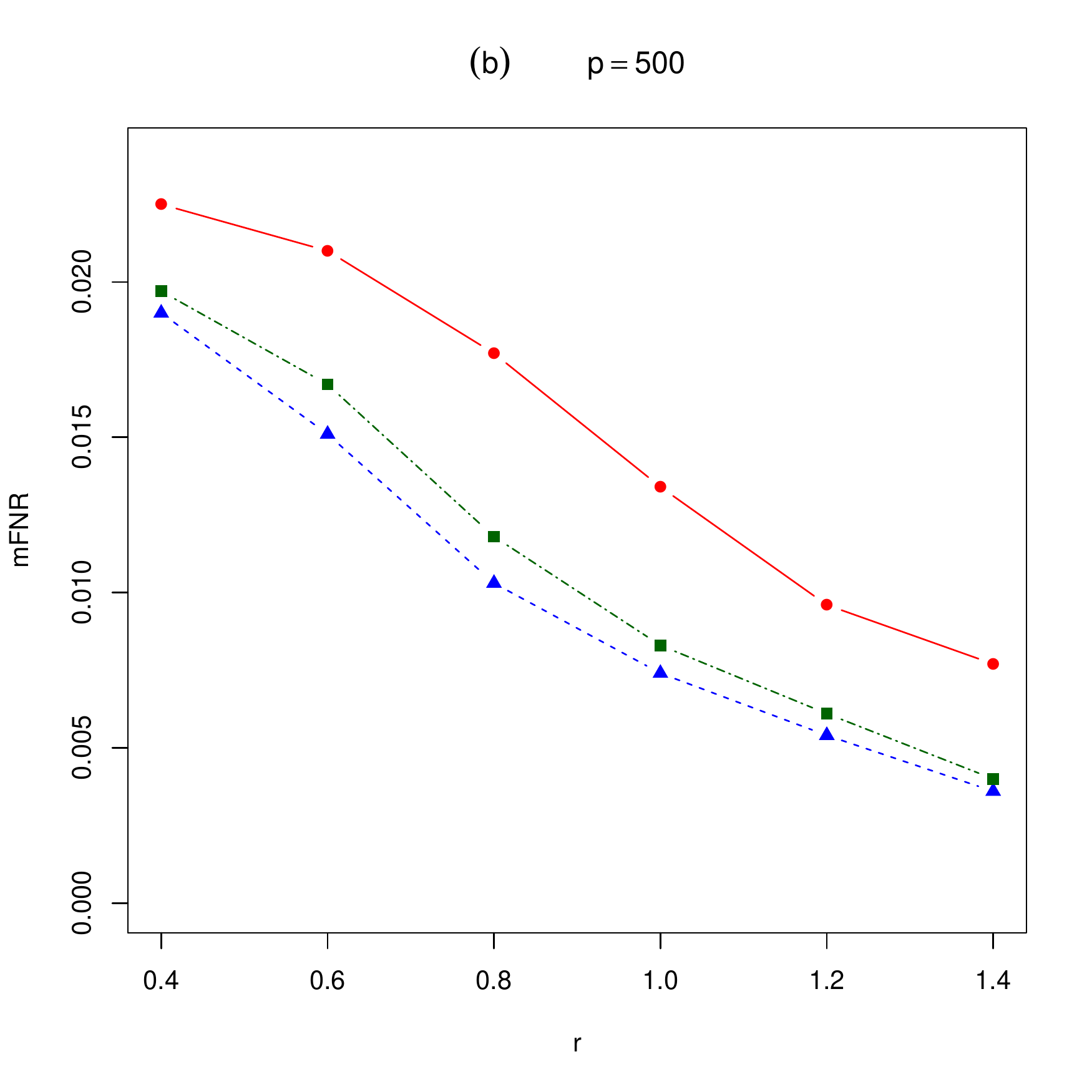}
\end{minipage}
\hspace{0.25cm}
\begin{minipage}[b]{0.32\linewidth}
\centering
\includegraphics[width=\textwidth]{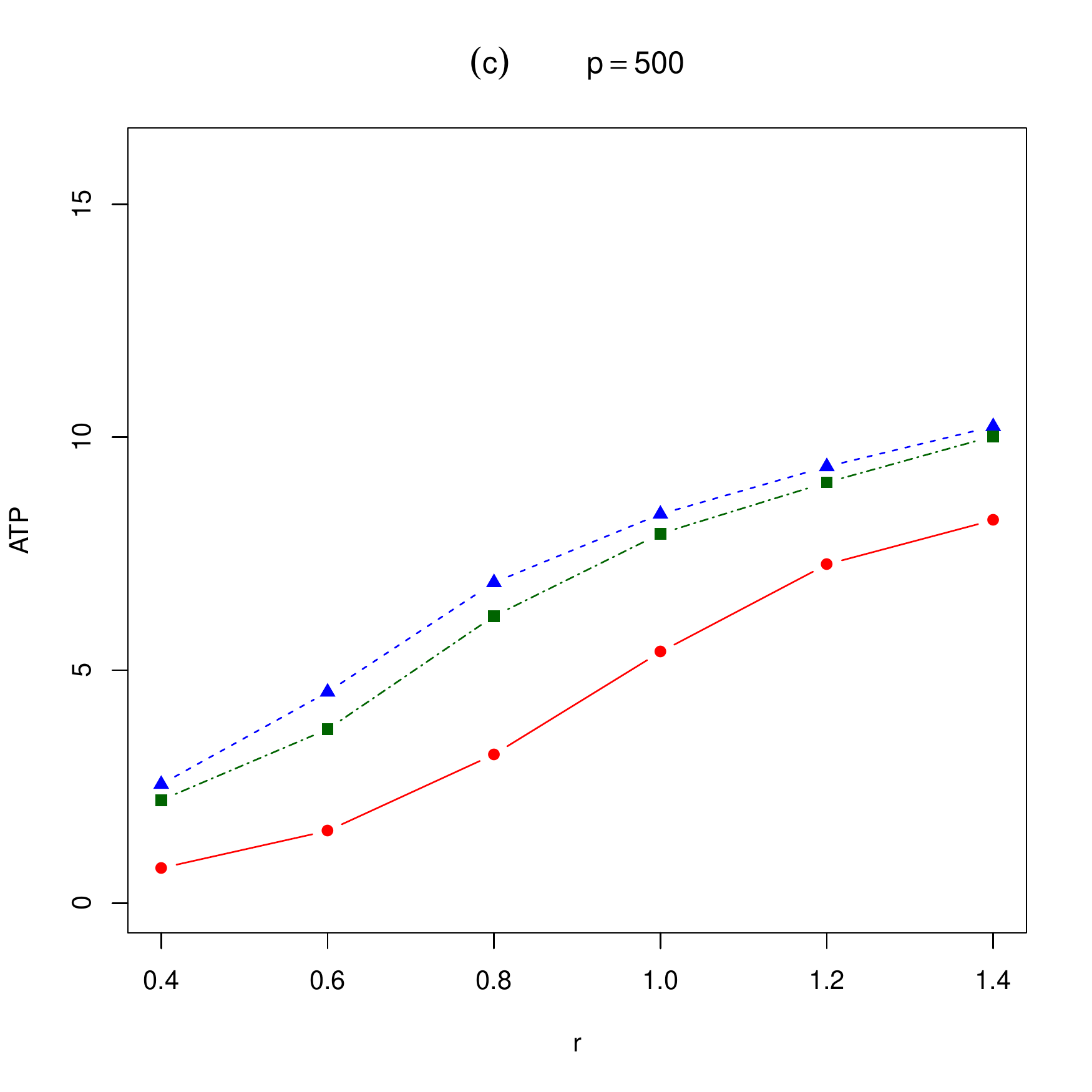}
\end{minipage}
\begin{minipage}[b]{0.32\linewidth}
\centering
\includegraphics[width=\textwidth]{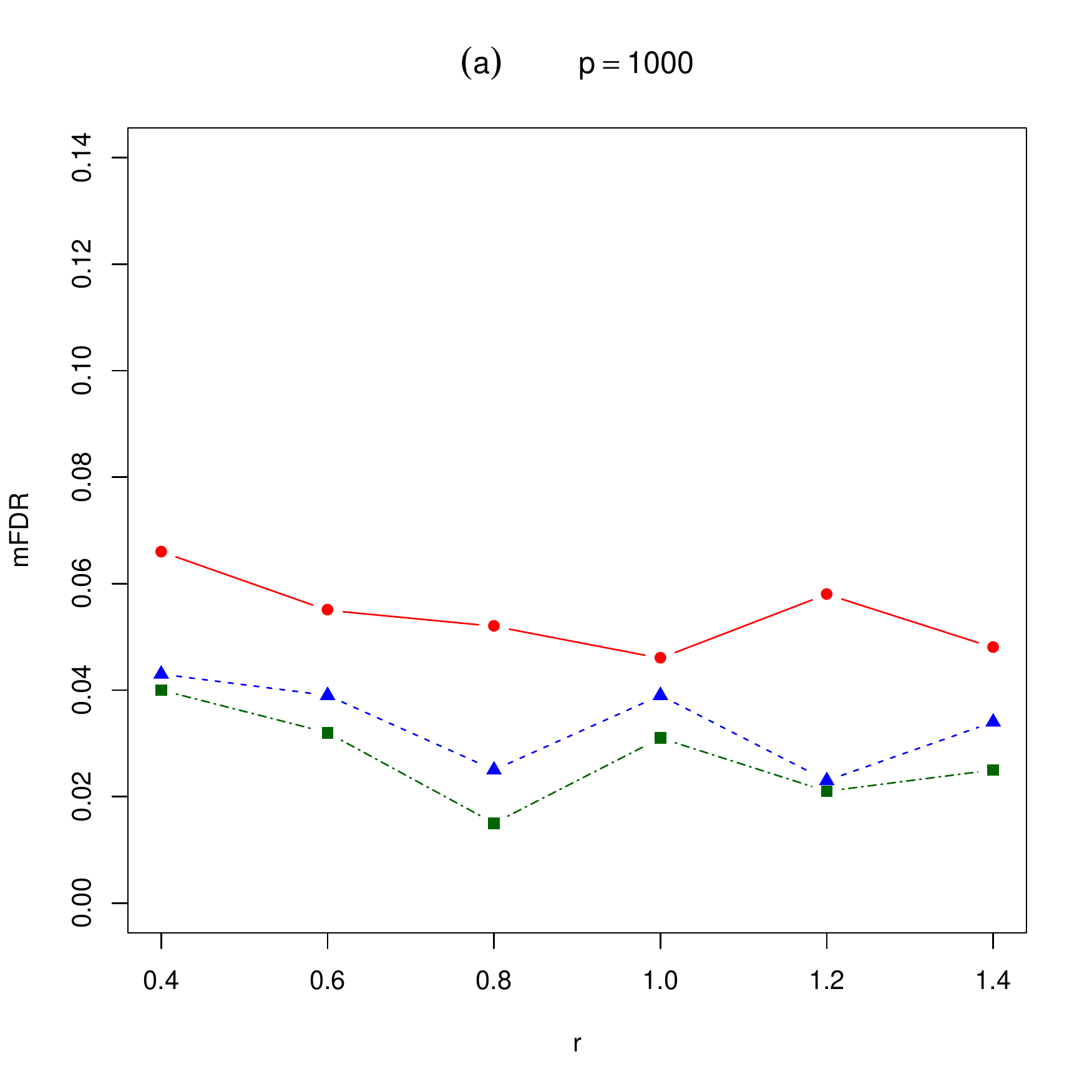}
\end{minipage}
\hspace{0.25cm}
\begin{minipage}[b]{0.32\linewidth}
\centering
\includegraphics[width=\textwidth]{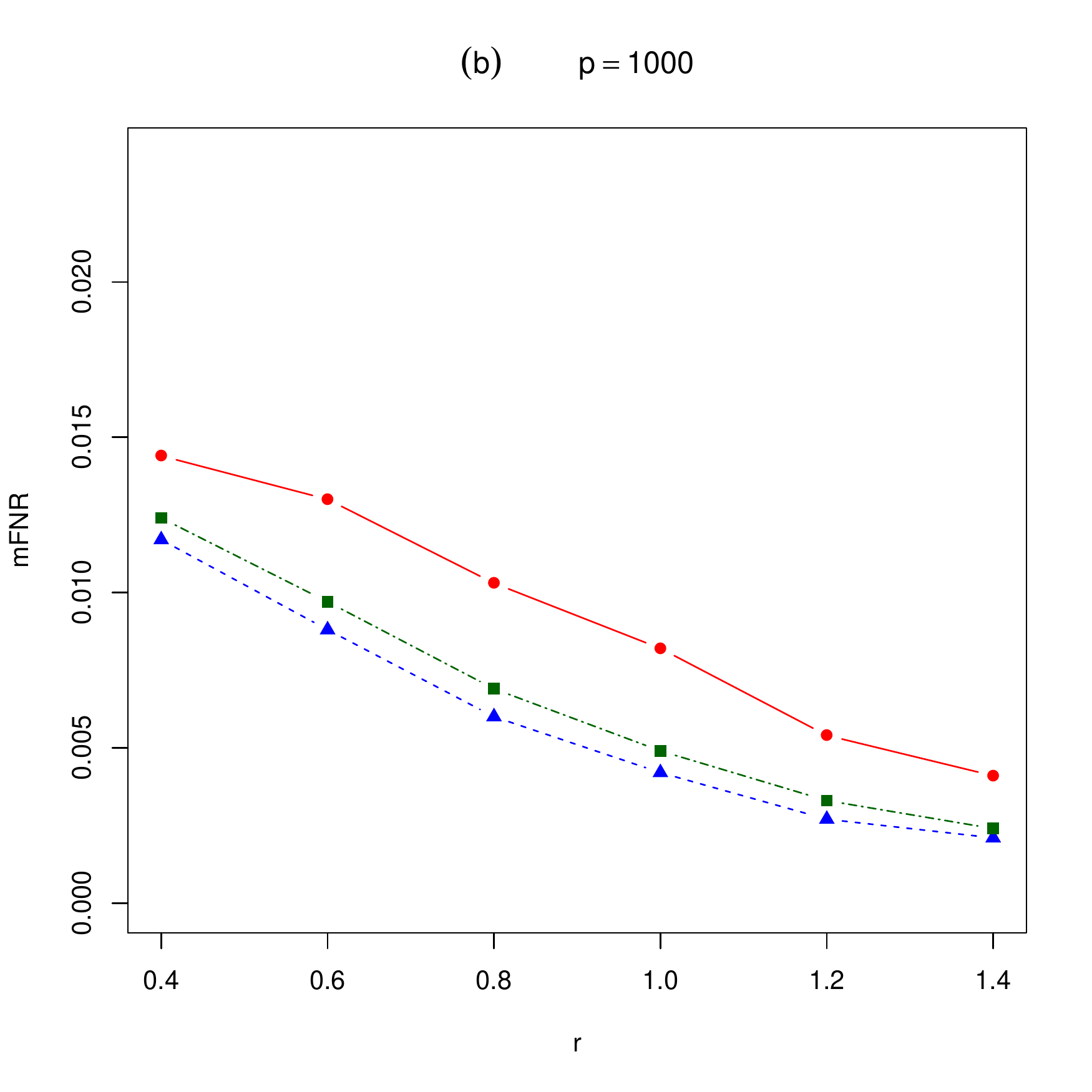}
\end{minipage}
\hspace{0.25cm}
\begin{minipage}[b]{0.32\linewidth}
\centering
\includegraphics[width=\textwidth]{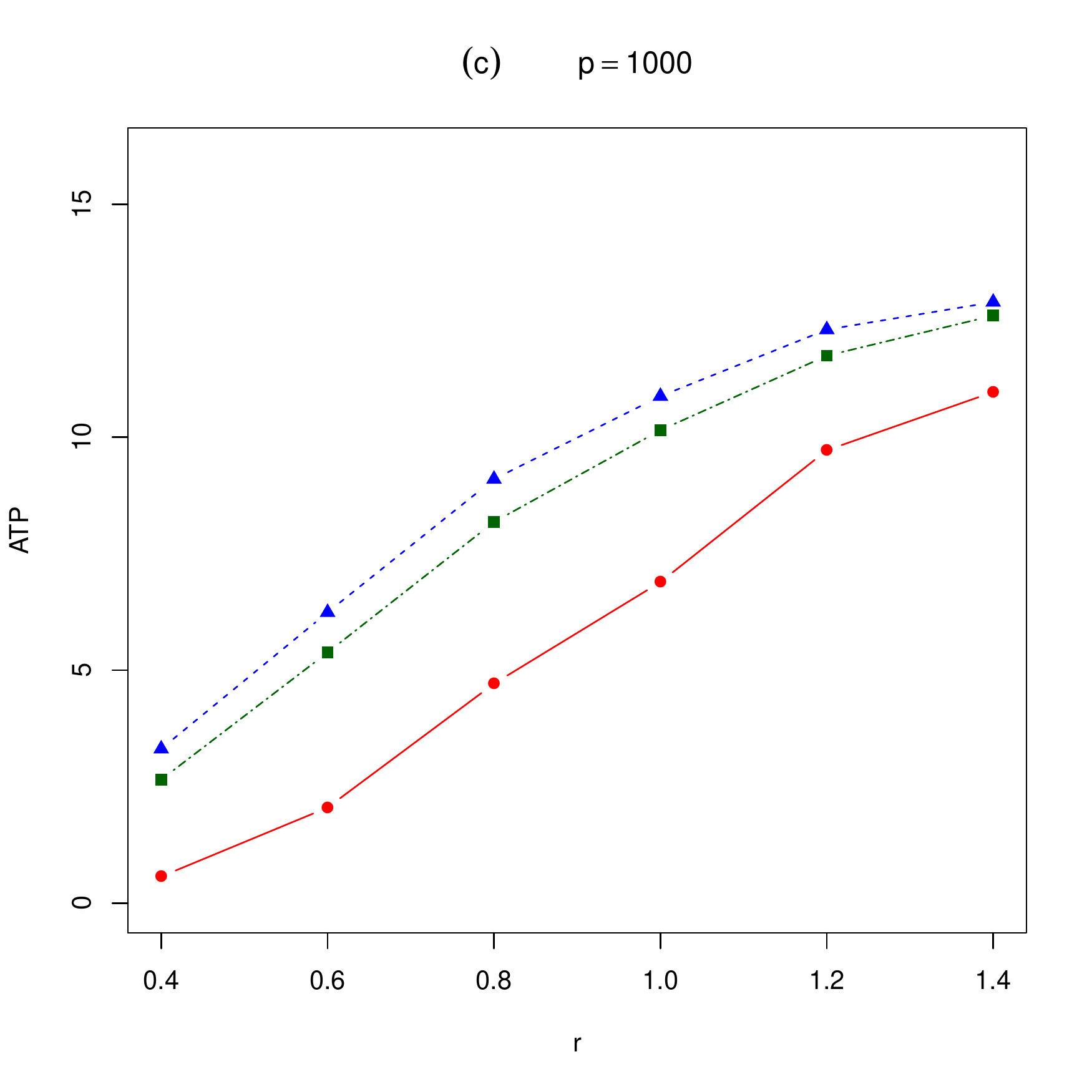}
\end{minipage}
\caption{The mFDR, mFNR and ATP yielded by $\mbox{DATE}_{\Omega}$, $\mbox{DATE}_{\hat{\Omega}}$ and the BH procedure under model (d). The sample sizes $n_1=100$ and $n_2=100$ and $\beta=0.6$.}
\label{comp5}
\end{sidewaysfigure}

The performance of three procedures with various dependent structures defined in models (b)-(d) were also displayed in Figures \ref{comp3}-\ref{comp5}. Again, both $\mbox{DATE}_{\Omega}$ and $\mbox{DATE}_{\hat{\Omega}}$ performed better than the BH in terms of mFNR and ATP even though all the procedures had the mFDR controlled at the nominal level $0.05$. 


$\mbox{DATE}_{\Omega}$ depends on the level of threshold $s$ and  $\mbox{DATE}_{\hat{\Omega}}$ depends on both $s$ and $q$, which are required to be chosen from intervals $(0, \beta)$ and $(\beta, \underline{\omega}r)$ respectively. Table 1 displays the performance of both $\mbox{DATE}_{\Omega}$ and $\mbox{DATE}_{\hat{\Omega}}$ in terms of mFDR and mFNR with different values of  $s$ and $q$ under model (a) where $\beta=0.6$, $\rho=0.6$ and $r=0.8$.  As we can see, the proposed procedure is  insensitive to the choice of $s$ and $q$ as long as they are chosen properly from the intervals.

\setcounter{section}{6} \setcounter{equation}{0}
\section*{\large 6. Empirical Study}

We applied the proposed DATE procedure to a human breast cancer dataset which  
is available at http://www.ncbi.nlm.nih.gov. The data were analyzed by
Richardson et al. (2006) to provide insight into
the molecular pathogenesis of Sporadic basal-like cancers (BLC) that is a
distinct class of human breast cancers. As discussed by Richardson et al. (2006), BLC specimens display X
chromosome abnormalities in the sense that most of the BLC cases
lack markers of a normal inactive X chromosome, which are rare in
non-BLC specimens. So our interest on this data set is to display these X
chromosome abnormalities by identifying the differentially expressed genes between the BLC and non-BLC. For this purpose, we formed two samples by taking 18 sporadic BLC specimens and 20 non-BLC specimens from the original data, 
and each sample contains $1438$ genes obtained from chromosome X.   

To apply the DATE procedure, 
we first estimated ${\Omega}$ in (\ref{transform_matrix}) where $\Sigma_1\ne \Sigma_2$ in general. To facilitate a simpler estimation,  
we changed the two-sample problem into an one-sample problem by defining
\[  
Y_i= X_{1i}-\sqrt{\frac{n_1}{n_2}} X_{2i}+\frac{1}{\sqrt{n_1n_2}}\sum_{j=1}^{n_1}X_{2j}-\frac{1}{n_2}\sum_{l=1}^{n_2}X_{2l} \qquad i=1,\cdots, n_1, 
\]
where we assume $n_1\le n_2$. It can be shown that $Y_i \stackrel{i.i.d.}{\sim} \mbox{N}(\delta, \Sigma_w)$ where $\Sigma_w=\Sigma_1+\frac{n_1}{n_2}\Sigma_2$ under the model (\ref{model}). Note that $\Omega=\frac{n_1+n_2}{n_2}\Sigma_w^{-1}$. To estimate   
$\Omega$, we only need to estimate $\Sigma_w^{-1}$ based on $Y_i$ for $i=1,\cdots,n_1$. 
The available packages for this purpose include {\it glasso, Covpath} and {\it CLIME}, which are coded based on different estimation approaches discussed in Section 3. To implement a fast algorithm, we adopted the {\it glasso} 
which chooses the non-negative definite matrix $\hat{\Omega}_{{Glasso}}$ to maximize a $L_1$-regularized log-likelihood:
\[
\mbox{log}\mbox{det}(\Sigma^{-1})-\mbox{tr}(S\Sigma^{-1})-\rho ||\Sigma^{-1}||_1,
\]
where $S$ is the sample covariance matrix and $\rho$ is a tuning parameter controlling the $L_1$ shrinkage. To select the regularization parameter $\rho$, 
we considered the package {\it huge} developed by Zhao, Liu, Roeder, Lafferty and Wasserman (2012) where three methods are provided: the stability approach for regularization selection, rotation information criterion and a likelihood-based extended Bayesian information criterion. Except the DATE procedure, we also considered the classical BH procedure integrated with two-sample t test as a comparison. 

In order to identify the differentially expressed genes, the FDR was chosen to be controlled at $\alpha=0.001, 0.005$ and $0.01$. Table 2 summarizes the number of differentially expressed genes identified by the BH only and the DATE only, and both procedures. By carefully investigating the genes identified by both procedures, we found that the XIST (X inactive specific transcript) gene was discovered. This gene is in charge of an early developmental process in females and provides dosage equivalence between males and females. The XIST difference is thought as one of the characteristics for the BLC according to Richardson et al. (2006). Moreover, the authors argue that there exists the overexpression of a small subset of genes on chromosome X for BLC. In Table 3, we list additional $17$ genes that are identified by the DATE but missed by the BH with the FDR controlled at $\alpha=0.001$.  The association of these genes with the BLC may deserve some further biological investigation.


\setcounter{section}{7} \setcounter{equation}{0}
\section*{\large 7. Discussion}

Signal identification is different from its closely related problem of signal detection. Whereas the detection focuses purely on the presence of signals, the signal identification is designated for locating the signals. The advantage of dependence for signal detection was exploited by Hall and Jin (2010) who showed that  the detection boundary can be lowered by incorporating the data correlation. However, it is unclear that the similar advantage can be offered by data dependence for signal identification. The current paper attempts to answer this question. Our analysis shows that both full and partial signal identification boundaries for dependent data are lower than those without dependence. 
Our result, combined with the findings in Hall and Jin (2010), shows that data dependence is advantageous in both signal detection and signal identification.

When data dependence is present, it becomes challenging to find a procedure which minimizes the FNR while controlling the FDR at a pre-specified level $\alpha <1$. When both signals and precision matrix are sparse, the proposed DATE procedure takes advantage of dependence through the transformation to enhance the signal strength and is shown to have the faster convergence rate in mFNR than other procedures without take data dependence into account. The current work is related with that of Ji and Jin (2012) and of Ji and Zhao (2014), where the authors considered the variable selection and multiple testing in the high dimensional regression problem. 
In our paper, the precision matrix for data transformation is assumed to be sparse. 
More research is needed to develop an optimal procedure for signal identification under general dependence structure.

\clearpage
\setcounter{equation}{0}
\def\theequation{A.\arabic{equation}}
\def\thesection{A}

\section*{\large Appendix: Technical Details.}
\medskip
\subsection*{\large A.1. Lemmas 1 and 2}

\textbf{Lemma 1.} For any positive definite matrix $A_{p, p}=(a_{ij})_{p \times p}$ and its inverse $B_{p, p}=(b_{ij})_{p \times p}$, the following inequality holds
\[
a_{ii}\cdot b_{ii} \ge 1 \quad i=1, \cdots, p.
\]

Proof. We first show that $a_{pp}\cdot b_{pp} \ge 1$. To this end, we write
\[
A_{p,p} =
 \begin{pmatrix}
  A_{p-1, p-1} & a_{p-1, 1}  \\
  a^{\prime}_{p-1 ,1} & a_{pp}
   \end{pmatrix}.
\]

Then using the result from matrix inversion in block form, we have
\be
b_{pp}=(a_{pp}-a^{\prime}_{p-1 ,1}A_{p-1,p-1}^{\prime}a_{p-1,1})^{-1}, \label{le5}
\ee
which implies that $a_{pp}\cdot b_{pp} \ge 1$ since $a^{\prime}_{p-1 ,1}A_{p-1,p-1}^{\prime}a_{p-1,1} \ge 0$.

For any $i$, we can switch $a_{ii}$ from its original position to the position $(p,p)$ using the permutation matrix $P_{p,p}$. Accordingly, $b_{ii}$ is moved from its original location to $(p,p)$ by the same matrix $P_{p,p}$. By the fact that the permutation matrix is also the orthogonal matrix, we have
\[
P_{p,p} A_{p,p} P_{p,p} P_{p,p} B_{p,p} P_{p,p}=I_{p,p}.
\]
Therefore, from (\ref{le5}), we have $a_{ii}\cdot b_{ii} \ge 1$ for any $i$. This completes the proof of Lemma 1.

\bigskip
For any $k \in \{1,\cdots,p\}$, we let
\[
A_k(\Omega)=\{l: 1\le l \le p, |\omega_{kl}|\ge L_p^{-1} \},
\]
and $B_k$ be the event that $\{\delta_l=0 \,\, \mbox{for all}\,\, l \ne k \,\, \mbox{and} \,\, l\in A_k\}$. If $\Omega$ is unknown, it can be estimated by $\hat{\Omega}$ (Cai, Liu and Luo (2011), which, with probability $1-O(p^{-\tau})$ where $\tau$ is a positive constant, satisfies
\[
||\hat{\Omega}-\Omega||_{L_1}=O_p\biggl\{(\frac{\log p}{n})^{\frac{1-\zeta}{2}} \biggr\}.
\]
Then, let $D_p$ be the event $\{||\hat{\Omega}-\Omega||_{L_1} \le (\frac{\log p}{n})^{\frac{1-\zeta}{2}}\}$.

\bigskip
\textbf{Lemma 2.} Assume conditions (C2)-(C4). 
Over the event $\{\delta_k \ne 0 \} \cap B_k \cap D_k$, 
\[
\hat{\delta}_{\Omega,k}= \omega_{kk}\delta_k+o(n^{-1/2}). 
\]

Proof:  We first consider that $\Omega$ is known.  By condition (C3), the number of elements in set $A_k(\Omega)$ satisfies that $|A_k(\Omega)|\le M L_p$. Since $\beta>1/2$, condition (C2) leads to
\begin{eqnarray}
\mbox{P}(\delta_k \ne 0, B_k^c)\le \sum_{l \in A_k, l\ne k} \mbox{P}(\delta_k \ne 0, \delta_l \ne 0)\le ML_p p^{-2\beta}=o(p^{-1}). \label{2-1}
\end{eqnarray}

Note that $\delta_{\Omega,k}=\sum_{l \in A_k}\omega_{kl} \delta_l +\sum_{l \in A_k^c} \omega_{kl}\delta_l$. Over the event $\{\delta_k\ne 0\} \cap B_k$, $\sum_{l \in A_k}\omega_{kl} \delta_l=\omega_{kk}\delta_k$. Moreover, for $l \in A_k^c$, $|\omega_{kl}|^{\zeta-1}/L_p^{1-\zeta}>1$. Therefore, using condition (C3) again, for some constant $c$, we have
\[
|\delta_{\Omega,k} -\omega_{kk}\delta_k |\le \max \limits_{l} |\delta_l|\,L_p^{\zeta-1} \sum_{l \in A_k^c} |\omega_{kl} |^{\zeta} \le  \max \limits_{l} |\delta_l|\,L_p^{\zeta-1} c.
\]
Since $|\delta_l| \sim \sqrt{2r\mbox{log}p/n}$ and $\zeta <1$, we can choose a large enough slowly varying function $L_p$ such that $\max \limits_{l} |\delta_l|\,L_p^{\zeta-1}=o(n^{-1/2})$. Thus,  
we have $\delta_{\Omega,k}= \omega_{kk}\delta_k+o(n^{-1/2})$. 

Next, we consider that $\Omega$ is unknown. If $\tau >1$, $\mbox{P}(D_p^c)=o(p^{-1})$ by the definition of the event of $D_p$. Note that $\hat{\delta}_{\Omega,k}=\delta_{\Omega,k}+\{(\hat{\Omega}-\Omega)\delta\}_k$. Then over the event $D_p$ and by condition (C4),
\[
\{(\hat{\Omega}-\Omega)\delta\}_k\le \max \limits_{l}|\delta_l| \cdot ||\hat{\Omega}-\Omega||_{L_1}\le (\frac{\mbox{log}p}{n})^{1-\frac{\zeta}{2}}=o(n^{-1/2}).
\]
Then, over the event $\{\delta_k \ne 0\} \cap B_k \cap D_k$, $\hat{\delta}_{\Omega,k}= \omega_{kk}\delta_k+o(n^{-1/2})$.
This completes the proof of Lemma 2.

\medskip
\subsection*{\large A.2. Proof of Lemma 3}

Recall that in Lemma 2, $B_k$ is the event that $\{\delta_l=0 \,\, \mbox{for all}\,\, l \ne k \,\, \mbox{and} \,\, l\in A_k\}$.
Since $\underline{\omega}r< (\sqrt{1-\Lambda}+\sqrt{1-\beta})^2$, it can be shown that $\{(\underline{\omega}r-\beta+\Lambda)^2+4\underline{\omega}r\beta\}/(4\underline{\omega}r)<1$. Also using (\ref{2-1}),  we know that it is sufficient to prove Lemma 3 over the event $B_k$. Without loss of generality, we assume $\sqrt{n} \delta_k=\sqrt{2r \mbox{log}p}$. The result for negative signals can be derived similarly. 
Note that $s < (\underline{\omega}r+\beta-\Lambda)^2/(4\underline{\omega}r)=\{(\underline{\omega}r+\beta-\Lambda)^2/(2\underline{\omega}r)^2\} (\underline{\omega}r) < \underline{\omega}r$ since $\underline{\omega}r > \beta-\Lambda$. Then, over the event $B_k$, 
\begin{eqnarray}
&\quad&\mbox{P}(T_k < 2s \mbox{log}p, \delta_k \ne 0)\nonumber\\
&\le &\mbox{P}(\delta_k \ne 0) \mbox{P}(T_k < 2s \mbox{log}p|\delta_k \ne 0 \cap B_k)\nonumber\\
&=&p^{-\beta}\biggl\{\mbox{P}\biggl(\frac{\sqrt{n}(\bar{Z}_1^{(k)}-\bar{Z}_2^{(k)})}{\sqrt{\omega_{kk}}}-\frac{\sqrt{n}\delta_{\Omega_k}}{\sqrt{\omega_{kk}}}<\sqrt{2s \mbox{log}p}-\frac{\sqrt{n}\delta_{\Omega_k}}{\sqrt{\omega_{kk}}}|\delta_k \ne 0 \cap B_k\biggr)\nonumber\\
&-&\mbox{P}\biggl(\frac{\sqrt{n}(\bar{Z}_1^{(k)}-\bar{Z}_2^{(k)})}{\sqrt{\omega_{kk}}}-\frac{\sqrt{n}\delta_{\Omega_k}}{\sqrt{\omega_{kk}}}<-\sqrt{2s \mbox{log}p}-\frac{\sqrt{n}\delta_{\Omega_k}}{\sqrt{\omega_{kk}}}|\delta_k \ne 0 \cap B_k\biggr)\biggr\}\nonumber\\
&\le&p^{-\beta}L_pp^{-(\sqrt{\omega_{kk} r}-\sqrt{s})^2}\{1+o(1)\}\nonumber\\
&\le&p^{-\beta}L_pp^{-(\sqrt{\underline{\omega} r}-\sqrt{s})^2}\{1+o(1)\}.\nonumber
 \end{eqnarray}
Since $s < (\underline{\omega}r+\beta-\Lambda)^2/(4\underline{\omega}r)$, we have 
\begin{eqnarray}
\sum_{k=1}^p \mbox{P}(T_k < 2s \mbox{log}p, \delta_k \ne 0) \le L_p p^{1-\{(\underline{\omega}r-\beta+\Lambda)^2+4\underline{\omega}r\beta\}/(4\underline{\omega}r)}.\nonumber
\end{eqnarray} 

Next we consider that $\Omega$ is unknown. Let $D_p$ be the event
\[
\biggl\{\max \limits_{1\le k \le p} |\sum_l(\hat{\Omega}_{kl}-\Omega_{kl})(\bar{X}_1^{(l)}-\bar{X}_2^{(l)})|\le (\frac{\mbox{log}p}{n})^{1-\frac{\alpha}{2}}, \max \limits_{1\le k \le p} |\hat{\omega}_{kk}-\omega_{kk}| \le (\frac{\mbox{log}p}{n})^{\frac{1-\alpha}{2}} \biggr\}.
\]

Note that
\[
|\hat{T}_k^{\frac{1}{2}}|=\biggl| \biggl\{\frac{\sqrt{n}\sum_l \Omega_{kl}(\bar{X}_1^{(l)}-\bar{X}_2^{(l)})}{\sqrt{\omega_{kk}}}+\frac{\sqrt{n}\sum_l(\hat{\Omega}_{kl}-\Omega_{kl})(\bar{X}_1^{(l)}-\bar{X}_2^{(l)})}{\sqrt{\omega_{kk}}}  \biggr\}\frac{1}{1+\frac{\sqrt{\hat{\omega}_{kk}}-\sqrt{\omega_{kk}}}{\sqrt{\omega_{kk}}}}\biggr|,
\] 
which by condition (C4), leads to
\begin{eqnarray}
\mbox{P}(\hat{T}_k < 2s \mbox{log}p, \delta_k \ne 0)&\le&p^{-\beta} \mbox{P}(\hat{T}_k < 2s \mbox{log}p |\delta_k \ne 0 \cap B_k\cap D_p)+\mbox{P}(\delta_k\ne 0 \cap B_k^c)+\mbox{P}(D_p^c)\nonumber\\
&\le& p^{-\beta}\mbox{P}(T_k < 2s \mbox{log}p| \delta_k \ne 0)+\mbox{P}(\delta_k\ne 0 \cap B_k^c)+\mbox{P}(D_p^c)\nonumber\\
&\le&L_p p^{\{(\underline{\omega}r-\beta+\Lambda)^2+4\underline{\omega}r\beta\}/(4\underline{\omega}r)}+\mbox{P}(\delta_k\ne 0 \cap B_k^c)+\mbox{P}(D_p^c).\nonumber
\end{eqnarray}


Since $\mbox{P}(\delta_k\ne 0 \cap B_k^c)=o(p^{-1})$ by (\ref{2-1}), $\mbox{P}(D_p^c)=o(p^{-1})$ and $\{(\underline{\omega}r-\beta+\Lambda)^2+4\underline{\omega}r\beta\}/(4\underline{\omega}r)<1$, 
\[
\mbox{P}(\hat{T}_k < 2s \mbox{log}p, \delta_k \ne 0)\le L_p p^{-\{(\underline{\omega}r-\beta+\Lambda)^2+4\underline{\omega}r\beta\}/(4\underline{\omega}r)}\{1+o(1)\}.
\] 
This completes the proof of Lemma 3.

\medskip
\subsection*{\large A.3. Proof of Lemma 4}

Similar to Lemma 2.2 of Ji and Jin (2012), it can be shown that with probability $1-o(p^{-1})$, each row of the regularized $\Omega^*$ defined by (\ref{reg_omega}) has no more than $L_p$ nonzero components and also $|| \Omega-\Omega^*||_{L_1}\le C(\mbox{log}p)^{-(1-\alpha)}$. 

Note that $\{(\underline{\omega}r-\beta+\Lambda)^2+4\underline{\omega}r\beta\}/(4\underline{\omega}r)<1$. And a connected graph with size $l+1$ for $l \ge 1$ always contains a connected graph with size $l$. 
Then with respect to $(V_0, \Omega^*)$, it is sufficient to show that there exists an integer $m$ such that
\[
\mbox{P}\{\mathcal{U}(s) \,\, \mbox{contains a connected graph with size} \,\, m \} \le o(p^{-1}).
\]

Recall that $D_p$ is the event
\[
\biggl\{\max \limits_{1\le k \le p} |\sum_l(\hat{\Omega}_{kl}-\Omega_{kl})(\bar{X}_1^{(l)}-\bar{X}_2^{(l)})|\le (\frac{\mbox{log}p}{n})^{1-\frac{\alpha}{2}}, \max \limits_{1\le k \le p} |\hat{\omega}_{kk}-\omega_{kk}| \le (\frac{\mbox{log}p}{n})^{\frac{1-\alpha}{2}} \biggr\}.
\]
And $\mbox{P}(D_p^c)= o(p^{-1})$. Therefore, it is enough to show that
\[
\mbox{P}\{\mathcal{U}(s) \, \mbox{contains a connected graph with size} \, m, D_p \} \le o(p^{-1}).
\]

Since there are logarithmically large number of nonzero elements in each row or column of $\Omega^*$, there are at most $pL_p^m$ connected graphs with size $m$ by the result from Frieze and Molloy (1999). As a result, by the union bound, it is sufficient to show that for any connected graph of size $m$, say $\mathcal{I}=\{k_1, \cdots, k_m \}$, 
\[
\mbox{P}\{\mathcal{I} \subset \mathcal{U}(s), D_p \} \le o(p^{-2}).
\] 
Let $\hat{T}= \{\hat{T}_1, \cdots, \hat{T}_p\}$ and $\mathbf{1}_p=\{1,\cdots, 1\}$. Then we need to show that 
\[
\mbox{P}\{\hat{T}^{\mathcal{I}}\ge 2s\mbox{log}p \mathbf{1}_p^{\mathcal{I}} , D_p \} \le o(p^{-2}).
\]
Since for any $1\le k \le p$, 
\[
\hat{T}_k=\biggl\{\frac{\sqrt{n}\sum_l \Omega_{kl}(\bar{X}_1^{(l)}-\bar{X}_2^{(l)})}{\omega_{kk}^{1/2}}+\frac{\sqrt{n}\sum_l(\hat{\Omega}_{kl}-\Omega_{kl})(\bar{X}_1^{(l)}-\bar{X}_2^{(l)})}{\omega_{kk}^{1/2}}  \biggr\}^2\biggl(\frac{1}{1+\frac{\hat{\omega}_{kk}-\omega_{kk}}{\omega_{kk}}}\biggr),
\]  
we only need to show that
\[
\mbox{P}\{|(T^{\mathcal{I}})^{1/2}|\ge (2s\mbox{log}p)^{1/2} \mathbf{1}_p^{\mathcal{I}} , D_p \} \le o(p^{-2}).
\] 
Let $\tilde{\delta}_{\Omega}^{\mathcal{I}}=(\delta_{\Omega, k_1}/\omega_{k_1 k_1}^{1/2}, \cdots, \delta_{\Omega, k_m}/\omega_{k_m k_m}^{1/2} )$ and $\mathcal{E}=(T^{\mathcal{I}})^{1/2}-\sqrt{n}\tilde{\delta}_{\Omega}^{\mathcal{I}}$ . Then it can be shown that 
\[
\mathcal{E} \sim \mbox{N}(0, \bar{\Omega}^{\mathcal{I},\mathcal{I}}),
\]   
where $\bar{\Omega}_{ij}=\omega_{ij}/(\omega_{ii}\omega_{jj})^{1/2}$ for $i, j \in \mathcal{I}$. 
By Cauchy-Schwartz inequality, 
\be
||\mathcal{E}||^2 \ge \frac{1}{2} ||(T^{\mathcal{I}})^{1/2}||^2-||\sqrt{n}\tilde{\delta}_{\Omega}^{\mathcal{I}}||^2.\label{ine1}
\ee
Since the largest eigenvalue of $\bar{\Omega}^{\mathcal{I},\mathcal{I}}$ is not greater than that of $\bar{\Omega}$. The latter has the largest eigenvalue no greater than $C_0/ \underline{\omega} \le C_0$ since $\underline{\omega}\ge 1$. Therefore, 
\be
\mathcal{E}^{\prime}(\bar{\Omega}^{\mathcal{I},\mathcal{I}})^{-1}\mathcal{E} \ge \frac{1}{C_0} ||\mathcal{E}||^2.\label{ine2}  
\ee
Moreover, by the construction of the thresholding, 
\be
||(T^{\mathcal{I}})^{1/2}||^2 \ge 2ms \mbox{log}p. \label{ine3}
\ee
Combining (\ref{ine1}), (\ref{ine2}) and (\ref{ine3}), we have
\[
\mathcal{E}^{\prime}(\bar{\Omega}^{\mathcal{I},\mathcal{I}})^{-1}\mathcal{E} \ge \frac{1}{C_0}\{ms \mbox{log}p-||\sqrt{n}\tilde{\delta}_{\Omega}^{\mathcal{I}}||^2 \}.
\]

Using Lemma A.3 of Ji and Jin (2012), we can show that
\[
\mbox{P}(||\sqrt{n}\tilde{\delta}_{\Omega}^{\mathcal{I}}||^2 \ge (C_0 k + cm (\mbox{log}p)^{-2(1-\alpha)})(2s\mbox{log}p), D_p) \le L_p p^{-\beta k},
\]
where $k$ is chosen to satisfy $(C_0 k + cm (\mbox{log}p)^{-2(1-\alpha)})(2s\mbox{log}p) \le  \frac{1}{2} ms \mbox{log}p$. Denote $A_p$ to be the event
\[
\{||\sqrt{n}\tilde{\delta}_{\Omega}^{\mathcal{I}}||^2 \ge (C_0 k + cm (\mbox{log}p)^{-2(1-\alpha)})(2s\mbox{log}p) \}.
\]
Then, we have $\mbox{P}(A_p \cap D_p) \le L_p p^{-\beta k}$. Then, for sufficiently large $m$,
\begin{eqnarray}
\mbox{P}\{\mathcal{I} \subset \mathcal{U}(s), D_p \} &\le& \mbox{P} \{\mathcal{E}^{\prime}(\bar{\Omega}^{\mathcal{I},\mathcal{I}})^{-1}\mathcal{E} \ge \frac{1}{2C_0}(ms \mbox{log}p) \}+\mbox{P}(A_p \cap D_p)  \nonumber\\
&\le & L_p (p^{-\frac{1}{4C_0}ms}+p^{-\beta k})\nonumber\\
&=& o(p^{-2}).\nonumber  
\end{eqnarray}
This completes the proof of Lemma 4.

\medskip
\subsection*{\large A.4. Proof of Theorem 1}

To make the discussion earlier, we change the two-sample problem into an one-sample problem. Without loss of generality, we assume $n_1 \le n_2$ and define
\be  
Y_i= X_{1i}-\sqrt{\frac{n_1}{n_2}} X_{2i}+\frac{1}{\sqrt{n_1n_2}}\sum_{j=1}^{n_1}X_{2j}-\frac{1}{n_2}\sum_{l=1}^{n_2}X_{2l} \qquad i=1,\cdots, n_1. \label{var_t}
\ee
It can be shown that $Y_i \stackrel{i.i.d.}{\sim} \mbox{N}(\delta, \Sigma_1+\frac{n_1}{n_2}\Sigma_2)$ for $i=1,\cdots, n_1$ under the model (\ref{model}).

Note that the loss function for $j$th dimension is $L(\theta_j, \hat{\theta}_j)=\theta_j (1-\hat{\theta}_j)+p^{-\Lambda}(1-\theta_j)\hat{\theta}_j$ where $\theta_j=0$ if $\delta_j=0$ and $\theta_j=1$ otherwise, and $\hat{\delta}_j$ is the decision rule with value equal to either 0 or 1. 
Clearly, $L(\theta, a)=\sum_j L(\theta_j, \hat{\theta}_j)$. The following derivation for $j$th dimension can be also extended to other dimensions. Therefore, without making any confusion, we drop the subscript $j$. Let $\tilde{\delta}= \delta-\alpha e_j$ where $e_j$ is a $p \times 1$ vector with $j$th element equal to 1. Let $h(\mathcal{Y}; \tilde{\delta}, \alpha)$ be the joint density of $(Y_1, \cdots, Y_{n_1})$ where $Y_i$ is defined in (\ref{var_t}):
\begin{eqnarray}
h(\mathcal{Y}; \tilde{\delta}, \alpha)&=& (2\pi)^{-n_1p/2}|\tilde{\Sigma}|^{-n_1/2}\mbox{exp}(-\frac{1}{2}\sum_{i=1}^{n_1}\{(Y_i-\tilde{\delta})^{\prime}\tilde{\Sigma}^{-1}(Y_i-\tilde{\delta}) \})\nonumber\\
&\quad&\mbox{exp}\{\alpha e_j^{\prime}\tilde{\Sigma}^{-1}\sum_{i=1}^{n_1}(Y_i-\tilde{\delta})-n_1/2\alpha^2 \tilde{\omega}_{jj}\}\nonumber\\
&=&h(\mathcal{Y}; \tilde{\delta}, 0)\mbox{exp}\{\alpha e_j^{\prime}\tilde{\Sigma}^{-1}\sum_{i=1}^{n_1}(Y_i-\tilde{\delta})-n_1/2\alpha^2 \tilde{\omega}_{jj}\},\nonumber
\end{eqnarray} 
where $\tilde{\Sigma}=(\Sigma_1+n_1/n_2\Sigma_2)$ and $\tilde{\omega}_{jj}$ is the $j$th diagonal element of $\tilde{\Sigma}^{-1}$. If we let
\begin{eqnarray}
f_0(\mathcal{Y})= \int h(\mathcal{Y}; \tilde{\delta}, 0) dF(\tilde{\delta}) \quad \mbox{and} \quad f_1(\mathcal{Y})= \int h(\mathcal{Y}; \tilde{\delta}, \alpha) d \pi_p(\alpha)dF(\tilde{\delta}),\label{A1.1}
\end{eqnarray}
where $F(\tilde{\delta})$ is the joint CDF of $\tilde{\delta}$ and $\pi_p(\alpha)$ is the CDF of $\alpha$ defined in condition (C2). Then
the following Bayesian decision rule minimizes the risk function for $j$th dimension: 
\[
\hat{\theta}_j= \mbox{I}\biggl\{\frac{(1-\epsilon_p)f_0(\mathcal{Y})}{\epsilon_pf_1(\mathcal{Y})} \le p^{\Lambda}\biggr\},
\] 
where $\epsilon_p =p^{-\beta}$. The corresponding risk function is
\begin{eqnarray}
H_j&=&\mbox{E}\{L(\theta_j, \hat{\theta}_j) \}\nonumber\\
  &=& \mbox{E} (\mbox{E}\{L(\theta_j, \hat{\theta}_j) \}| \mathcal{Y} )\nonumber\\
  &=& \mbox{E} \biggl\{\frac{\epsilon_pf_1(\mathcal{Y})}{(1-\epsilon_p)f_0(\mathcal{Y})+\epsilon_pf_1(\mathcal{Y})}(1-\hat{\theta}_j) +p^{-\Lambda}\frac{(1-\epsilon_p)f_0(\mathcal{Y})}{(1-\epsilon_p)f_0(\mathcal{Y})+\epsilon_pf_1(\mathcal{Y})} \hat{\theta}_j\biggr\}\nonumber\\
  &=& \int_{A^c} \epsilon_pf_1(\mathcal{Y})d\mathcal{Y} + p^{-\Lambda} \int_{A}(1-\epsilon_p)f_0(\mathcal{Y})d\mathcal{Y}\nonumber\\
  &=&\epsilon_p - \int_{A}|p^{-\Lambda}(1-\epsilon_p)f_0(\mathcal{Y})-\epsilon_pf_1(\mathcal{Y})|d\mathcal{Y}, \nonumber   
\end{eqnarray}
where the set $A=\{\mathcal{Y}: \hat{\theta}_j=1 \}$, and from line two to line three, we have used the fact that
\[
\mbox{E}(\theta_j|\mathcal{Y})=\mbox{P}(\theta_j=1|\mathcal{Y})=\frac{\mbox{P}(\mathcal{Y}| \theta_j=1)\mbox{P}(\theta_j=1)}{\mbox{P}(\mathcal{Y}| \theta_j=0)\mbox{P}(\theta_j=0)+\mbox{P}(\mathcal{Y}| \theta_j=1)\mbox{P}(\theta_j=1)}.
\]

Similarly, 
\begin{eqnarray}
H_j &=& \int_{A^c} \epsilon_pf_1(\mathcal{Y})d\mathcal{Y} + p^{-\Lambda} \int_{A}(1-\epsilon_p)f_0(\mathcal{Y})d\mathcal{Y}\nonumber\\
  &=&p^{-\Lambda}(1-\epsilon_p) - \int_{A^c}|p^{-\Lambda}(1-\epsilon_p)f_0(\mathcal{Y})-\epsilon_pf_1(\mathcal{Y})|d\mathcal{Y}. \nonumber   
\end{eqnarray}
Then the following result can be derived: 
\begin{eqnarray}
H_j= \frac{1}{2}\biggl\{p^{-\Lambda} (1-\epsilon_p)+\epsilon_p- \int_{A} |p^{-\Lambda}(1-\epsilon_p)f_0-\epsilon_p f_1 |d\mathcal{Y}  \biggr\},\label{A1-2}
\end{eqnarray}
where, by Fubini's Theorem,
\begin{eqnarray}
&\quad&\int_{A} |p^{-\Lambda}(1-\epsilon_p)f_0-\epsilon_p f_1 |d\mathcal{Y}\nonumber\\
&=& \int \biggl|\int \{p^{-\Lambda}(1-\epsilon_p)h(\mathcal{Y}; \tilde{\delta}, 0)-\epsilon_p h(\mathcal{Y}; \tilde{\delta}, \alpha) \}d\pi _p(\alpha)dF(\tilde{\delta}) \biggr| d \mathcal{Y}\nonumber\\
&\le& \int H(\tilde{\delta}, \alpha)d\pi _p(\alpha)dF(\tilde{\delta}), \nonumber 
\end{eqnarray}
where $H(\tilde{\delta}, \alpha)=\int |p^{-\Lambda}(1-\epsilon_p)h(\mathcal{Y}; \tilde{\delta}, 0)-\epsilon_p h(\mathcal{Y}; \tilde{\delta}, \alpha)| d \mathcal{Y}$. It can be shown that $H(\tilde{\delta}, \alpha)=H(\tilde{\delta}, -\alpha)$ and $H(\tilde{\delta}, \alpha)$ is an increasing function of $\alpha>0$. Hence, for $\alpha \in [-\tau_p, 0) \cup (0, \tau_p]$ where $\tau_p=\sqrt{2r \mbox{log}p/n}$, we have $H(\tilde{\delta}, \alpha) \le H(\tilde{\delta}, \tau_p)$. As a result, 
\begin{eqnarray}
\int_{A} |p^{-\Lambda}(1-\epsilon_p)f_0-\epsilon_p f_1 |d\mathcal{Y}
&\le& \int H(\tilde{\delta}, \tau_p)dF(\tilde{\delta}), \label{A1-3}
\end{eqnarray}
where, if we let $D_p= \{\mathcal{Y}: \epsilon_p \mbox{exp}\{\tau_p e_j^{\prime}\tilde{\Sigma}^{-1}\sum_{i=1}^{n_1}(Y_i-\tilde{\delta})-n_1/2\tau_p^2 \tilde{\omega}_{jj}\} > p^{-\Lambda}(1-\epsilon_p) \}$, then 
\begin{eqnarray}
H(\tilde{\delta}, \tau_p)&=&-\int_{D_p} \{p^{-\Lambda}(1-\epsilon_p)h(\mathcal{Y}; \tilde{\delta}, 0)-\epsilon_p h(\mathcal{Y}; \tilde{\delta}, \tau_p)\} d \mathcal{Y}\nonumber\\
&+&\int_{D_p^c} \{p^{-\Lambda}(1-\epsilon_p)h(\mathcal{Y}; \tilde{\delta}, 0)-\epsilon_p h(\mathcal{Y}; \tilde{\delta}, \tau_p)\} d \mathcal{Y}.\nonumber
\end{eqnarray} 
This, together with the fact that
\begin{eqnarray}
p^{-\Lambda}(1-\epsilon_p)+\epsilon_p&=&\int_{D_p} \{p^{-\Lambda}(1-\epsilon_p)h(\mathcal{Y}; \tilde{\delta}, 0)+\epsilon_p h(\mathcal{Y}; \tilde{\delta}, \tau_p)\} d \mathcal{Y}\nonumber\\
&+&\int_{D_p^c} \{p^{-\Lambda}(1-\epsilon_p)h(\mathcal{Y}; \tilde{\delta}, 0)+\epsilon_p h(\mathcal{Y}; \tilde{\delta}, \tau_p)\} d \mathcal{Y},\nonumber
\end{eqnarray}
leads to 
\begin{eqnarray}
H(\tilde{\delta}, \tau_p)=p^{-\Lambda}(1-\epsilon_p)+\epsilon_p-2\biggl\{p^{-\Lambda}(1-\epsilon_p)\int_{D_p}h(\mathcal{Y}; \tilde{\delta}, 0)d \mathcal{Y}+\epsilon_p\int_{D_p^c}h(\mathcal{Y}; \tilde{\delta}, \tau_p) d \mathcal{Y}\biggr\}.\nonumber
\end{eqnarray}
Define $W_j(\tilde{\delta})=e_j^{\prime}\tilde{\Sigma}^{-1}\sum_{i=1}^{n_1}(Y_i-\tilde{\delta})$. Then under $H_{0j}$, $W_j(\tilde{\delta}) \sim \mbox{N}(0, n_1\tilde{\omega}_{jj})$ since $\alpha=0$, and under $H_{1j}$, $W_j(\tilde{\delta}) \sim \mbox{N}(n_1\tau_p \tilde{\omega}_{jj}, n_1\tilde{\omega}_{jj})$. Then,
\begin{eqnarray}
H(\tilde{\delta}, \tau_p)=p^{-\Lambda}(1-\epsilon_p)+\epsilon_p-2\biggl\{p^{-\Lambda}(1-\epsilon_p)\bar{\Phi}(\frac{\lambda_p}{\sqrt{n_1\tilde{\omega}_{jj}}})+\epsilon_p\Phi(\frac{\lambda_p-n_1\tau_p\tilde{\omega}_{jj}}{\sqrt{n_1 \tilde{\omega}_{jj}}})\biggr\},\nonumber
\end{eqnarray}
where $\lambda_p=1/\tau_p \{\mbox{log}p^{-\Lambda}+\mbox{log}(\frac{1-\epsilon_p}{\epsilon_p})+\frac{n_1}{2} \tau_p^2 \tilde{\omega}_{jj}\}$. Then combining (\ref{A1-2}) and (\ref{A1-3}), we have  
\begin{eqnarray}
H_j \ge p^{-\Lambda}(1-\epsilon_p)\bar{\Phi}(\frac{\lambda_p}{\sqrt{n_1\tilde{\omega}_{jj}}})+\epsilon_p\Phi(\frac{\lambda_p-n_1\tau_p\tilde{\omega}_{jj}}{\sqrt{n_1 \tilde{\omega}_{jj}}}). \label{A1-4}
\end{eqnarray}

Using $\epsilon_p=p^{-\beta}$, $\tau_p^2= 2r \mbox{log}p/n$ and $n_1\tilde{\omega}_{jj}=n\omega_{jj}$ where ${\omega}_{jj}$ is the $j$th diagonal element of $\Omega$, we have 
\[
\frac{\lambda_p}{\sqrt{n_1\tilde{\omega}_{jj}}}=(\frac{\beta-\Lambda}{\sqrt{2r\omega_{jj}}}+\frac{\sqrt{r{\omega}_{jj}}}{\sqrt{2}})\sqrt{\mbox{log}p},   
\]
and
\[
\frac{\lambda_p-n_1\tau_p\tilde{\omega}_{jj}}{\sqrt{n_1 \tilde{\omega}_{jj}}}=(\frac{\beta-\Lambda-r{\omega}_{jj}}{\sqrt{2r{\omega}_{jj}}})\sqrt{\mbox{log}p}.
\]
First, if $\beta-r {\omega}_{jj}< \Lambda < \beta+r {\omega}_{jj}$, then (\ref{A1-4}) becomes  
\begin{eqnarray}
H_j&\ge&p^{-\Lambda} L_p p^{-\frac{(r{\omega}_{jj}+\beta-\Lambda)^2}{4r{\omega}_{jj}}}+ p^{-\beta} L_pp^{-\frac{(r{\omega}_{jj}-\beta+\Lambda)^2}{4r{\omega}_{jj}}}\nonumber\\
&=&p^{-\beta} L_pp^{-\frac{(r{\omega}_{jj}-\beta+\Lambda)^2}{4r{\omega}_{jj}}}.\label{A1-5}
\end{eqnarray}
Next, we consider $\Lambda< \beta-r{\omega}_{jj}$, then (\ref{A1-4}) becomes
\begin{eqnarray}
H_j&\ge&p^{-\Lambda} L_p p^{-\frac{(r{\omega}_{jj}+\beta-\Lambda)^2}{4r{\omega}_{jj}}}+ p^{-\beta}\nonumber\\
&=&p^{-\beta}\{1+o(1)\}.\label{A1-6}
\end{eqnarray}
Last, if $\Lambda>\beta+r {\omega}_{jj}$, then (\ref{A1-4}) becomes
\begin{eqnarray}
H_j&\ge&p^{-\Lambda}+p^{-\beta} L_pp^{-\frac{(r{\omega}_{jj}-\beta+\Lambda)^2}{4r{\omega}_{jj}}}\nonumber\\
&=&p^{-\Lambda}\{1+o(1)\}. \label{A1-7}
\end{eqnarray}

Recall that $H=\sum_{j=1}^p H_j$. Using the fact that $\underline{\omega}\le {\omega}_{jj}\le \bar{\omega}$ and $(r{\omega}_{jj}+\beta-\Lambda)^2/(4r{\omega}_{jj})$ is an increasing function of ${\omega}_{jj}$, Theorem 1 can be derived based on the results given in (\ref{A1-5}), (\ref{A1-6}) and (\ref{A1-7}).

\medskip
\subsection*{\large A.5. Proof of Theorem 2}

Recall that in the proof of Theorem 1, we have defined the loss function $L(\theta, \hat{\theta})=\sum_{i=1}^p \{\theta_i(1-\hat{\theta}_i)+p^{-\Lambda}(1-\theta_i)\hat{\theta}_i\}$. For any decision rule $\hat{\theta}_i$, the marginal false discovery rate
\begin{eqnarray}
\mbox{mFDR}=\frac{\mbox{E}\{\sum_i(1-\theta_i)\hat{\theta}_i\}}{\mbox{E}(\sum_i \hat{\theta}_i)}=1-\frac{\mbox{E}(\sum_i \theta_i\hat{\theta}_i)}{\mbox{E}(\sum_i \hat{\theta}_i)}. \label{A_2.1}
\end{eqnarray}
Since $\mbox{E}(\sum_i \theta_i\hat{\theta}_i)\le \mbox{min}\{p^{1-\beta}, \mbox{E}(\sum_i \hat{\theta}_i) \}$, $\mbox{mFDR}=1+o(1)$ if $p^{1-\beta}=o\{\mbox{E}(\sum_i \hat{\theta}_i)\}$. Hence, if mFDR is controlled at a level $\alpha<1$, we must have either $p^{1-\beta} \sim {\mbox{E}(\sum_i \hat{\theta}_i)}$ or $\mbox{E}(\sum_i \hat{\theta}_i)=o(p^{1-\beta})$. For the latter,
\[
\mbox{mFNR}=\frac{\mbox{E}\{\sum_i\theta_i(1-\hat{\theta}_i)\}}{\mbox{E}\{\sum_i (1-\hat{\theta}_i)\}}=\frac{p^{1-\beta}\{1+o(1)\}}{p\{1+o(1)\}}=p^{-\beta}\{1+o(1)\}.
\]
Next, we consider the mFNR under the constraint $p^{1-\beta}\sim {\mbox{E}(\sum_i \hat{\theta}_i)}$. Toward this end, we first note that if $\mbox{mFDR}\le \alpha <1$, the following result can be derived from (\ref{A_2.1}):
\be
p^{-\beta} \alpha \sum_{i=1}^p \mbox{E}(\hat{\theta}_i| \theta_i=1)\ge (1-\alpha)\sum_{i=1}^p\mbox{E}(\hat{\theta}_i| \theta_i=0). \label{A_2.2}
\ee 
Recall that in the proof of Theorem 1, the optimal decision rule 
\[
\hat{\theta}_i= \mbox{I}\biggl\{\frac{(1-p^{\beta})f_0(\mathcal{Y})}{p^{-\beta}f_1(\mathcal{Y})} \le p^{\Lambda}\biggr\},
\] 
where $f_0(\mathcal{Y})$ and $f_1(\mathcal{Y})$ are defined in (\ref{A1.1}). For simplicity, we choose point mass for $\alpha=\sqrt{2r \mbox{log}p/n}$ and $\tilde{\delta}$ in (\ref{A1.1}). Then the decision rule can be simplified as
\[
\hat{\theta}_i= \mbox{I}\biggl\{e_i^{\prime}\tilde{\Sigma}^{-1}\sum_{i=1}^{n_1}(Y_i-\tilde{\delta}) \ge \frac{(\beta-\Lambda)\mbox{log}p}{\sqrt{2r \mbox{log}p/n}}+\frac{\omega_{ii}r \mbox{log}p}{\sqrt{2r \mbox{log}p/n}}\biggr\}.
\] 
Since under $H_{0i}$, $e_i^{\prime}\tilde{\Sigma}^{-1}\sum_{i=1}^{n_1}(Y_i-\tilde{\delta}) \sim \mbox{N}(0, n{\omega}_{ii})$, and under $H_{1i}$, $e_i^{\prime}\tilde{\Sigma}^{-1}\sum_{i=1}^{n_1}(Y_i-\tilde{\delta}) \sim \mbox{N}(\sqrt{2rn\mbox{log}p} {\omega}_{ii}, n{\omega}_{ii})$, we have
\be 
\mbox{E}(\hat{\theta}_i| \theta_i=0)=\bar{\Phi}\biggl\{(\frac{\beta-\Lambda}{\sqrt{2\omega_{ii}r}}+\sqrt{\frac{\omega_{ii}r}{2}})\sqrt{\mbox{log}p} \biggr\}, \label{e1}
\ee
and
\be 
\mbox{E}(\hat{\theta}_i| \theta_i=1)=\bar{\Phi}\biggl\{(\frac{\beta-\Lambda}{\sqrt{2\omega_{ii}r}}-\sqrt{\frac{\omega_{ii}r}{2}})\sqrt{\mbox{log}p} \biggr\}, \label{e2}
\ee
Since $\beta-\underline{\omega}r<\Lambda< \beta+\underline{\omega}r$, 
\[
\mbox{E}(\hat{\theta}_i| \theta_i=0)=L_pp^{-\frac{(\beta-\Lambda+\omega_{ii}r)^2}{4\omega_{ii}r}},
\]
and 
\[
\mbox{E}(\hat{\theta}_i| \theta_i=1)=1-L_pp^{-\frac{(\omega_{ii}r-\beta+\Lambda)^2}{4\omega_{ii}r}}.
\]

First note that in both (\ref{e1}) and (\ref{e2}), if we choose $\Lambda$ to be $\Lambda_{ii}={\omega_{ii}}r+\beta-2\sqrt{{\omega_{ii}}r\beta\bigg(1-\frac{g(\alpha,p)}{\beta}\biggr)}$ where $g(\alpha,p)={\mbox{log}\{\frac{\alpha}{(1-\alpha)}\sqrt{4\pi \beta \mbox{log}p}\}}{\mbox{log}^{-1}p}$, then the ``=" holds in (\ref{A_2.2}). To have a universal $\Lambda$ which does not depend on index $i$, we can choose $\Lambda=\underline{\omega}r+\beta-2\sqrt{\underline{\omega}r\beta\bigg(1-\frac{g(\alpha,p)}{\beta}\biggr)}$ such that the right hand is no greater than the left hand of (\ref{A_2.2}). Equivalently, this implies that  $\mbox{mFDR}\le \alpha <1$.

Given $\Lambda$, 
\begin{eqnarray}
\mbox{mFNR}&=&\frac{\mbox{E}\{\sum_i\theta_i(1-\hat{\theta}_i)\}}{\mbox{E}\{\sum_i (1-\hat{\theta}_i)\}}
=\frac{\sum_i L_pp^{-\beta-\frac{({\omega_{ii}}r-\beta+\Lambda)^2}{4{\omega_{ii}}r}}}{p\{1+o(1)\}}\nonumber\\
&\ge& \frac{L_pp^{1-\beta-\frac{(\bar{\omega}r-\beta+\Lambda)^2}{4\bar{\omega}r}}}{p\{1+o(1)\}}\nonumber\\
&\ge& L_p p^{-\beta-\bigl\{\sqrt{\bar{\omega}r}-\sqrt{\beta-g(\alpha, p)}\bigr\}^2}.\nonumber
\end{eqnarray}
This completes the proof of Theorem 2.

\medskip
\subsection*{\large A.6. Proof of Theorem 3}

Note that the loss function $L\{\theta_j, \mbox{sgn}(\hat{\delta}_j)\}=\theta_j \{1-\mbox{sgn}(\hat{\delta}_j)\}+p^{-\Lambda}(1-\theta_j)\mbox{sgn}(\hat{\delta}_j)$ where $\theta_j=0$ if $\delta_j=0$ and $\theta_j = 1$ otherwise, and $\hat{\delta}_j$ is estimated to be one of three values from $\{-\delta^{data}, 0, \delta^{date}\}$ by the DATE procedure. Since after the thresholding step, all the coordinates are assigned into either $ \mathcal{U}(s)$ or $\mathcal{U}^c(s)$, the corresponding risk is 
\[
H(\Lambda)=\sum_{j=1}^p \mbox{E} (L\{\theta_j,\mbox{sgn}(\hat{\delta}_j)\})= I + II,
\]
where $I$ is the risk in the thresholding step and $II$ is the risk in the excising step, i.e.,
\[
I=\sum_{j=1}^p  \mbox{E} (L\{\theta_j,\mbox{sgn}(\hat{\delta}_j)\}\mbox{I}\{ j \notin \mathcal{U}(s)\}), \quad II=\sum_{j=1}^p  \mbox{E} (L\{\theta_j,\mbox{sgn}(\hat{\delta}_j)\}\mbox{I}\{ j \in \mathcal{U}(s)\}).
\]

For $I$, we know that if $j \notin \mathcal{U}(s)$, the estimated signal $\hat{\delta}_j =0$ based on the DATE procedure. By Lemma 3,   
\[
I = \sum_{j=1}^p \mbox{P}(\hat{T}_j < 2s \mbox{log}p, \delta_j\ne 0) \le L_p p^{1-\{(\underline{\omega}r-\beta+\Lambda)^2+4\underline{\omega}r\beta\}/(4\underline{\omega}r)},
\]
which is not greater than the upper bound of $H(\Lambda)$ except a slowly varying function. Hence, we only need to show that $II \le L_p p^{1-\{(\underline{\omega}r-\beta+\Lambda)^2+4\underline{\omega}r\beta\}/(4\underline{\omega}r)}$.

Let event $A_p=\{\mathcal{U}(s)$ are split into disconnected clusters of size no more than $K$ with respect to $(V_0, \Omega^{*}) \}$. By Lemma 4, $\mbox{P}(A_p^c)\le  L_p p^{-\{(\underline{\omega}r-\beta+\Lambda)^2+4\underline{\omega}r\beta\}/(4\underline{\omega}r)}$. Therefore, it is sufficient to show that for all $1\le j \le p$,
\[ 
 \mbox{E}\biggl(L\{\theta_j, \mbox{sgn}(\hat{\delta}_j)\}\mbox{I}\{( j \in \mathcal{U}(s))\cap A_p\}\biggr) \le L_p p^{-\{(\underline{\omega}r-\beta+\Lambda)^2+4\underline{\omega}r\beta\}/(4\underline{\omega}r)}.
\]
By Lemma 4, we know that over the event $\{j\in \mathcal{U}(s) \} \cap A_p$, there exists a unique component $\mathcal{I}_0=\{i_i, \cdots, i_m \}$ with size $m \le K$ satisfying $ j \in \mathcal{I}_0$. Therefore, it is sufficient to show that for any fixed connected subgroup $\mathcal{I}_0$ that contains $j$,
\be
 \mbox{E}\biggl(L\{\theta_j, \mbox{sgn}(\hat{\delta}_j)\}\mbox{I}\{( j \in \mathcal{I}_0)\cap A_p\}\biggr) \le L_p p^{-\{(\underline{\omega}r-\beta+\Lambda)^2+4\underline{\omega}r\beta\}/(4\underline{\omega}r)}.\label{ee1}
\ee

Recall that $L\{\theta_j, \mbox{sgn}(\hat{\delta}_j)\}$ consists of the expected false positive and false negative. Accordingly if we define two events 
\[
R_1=\{\theta_j=0, \mbox{sgn}(\hat{\delta}_j)\ne 0, j \in \mathcal{I}_0, A_p\}, 
\]
and
\[
R_2=\{\theta_j = 1, \mbox{sgn}(\hat{\delta}_j)=0, j \in \mathcal{I}_0, A_p\},
\]
then to show (\ref{ee1}), we only need to show that 
\begin{eqnarray}
p^{-\Lambda}\mbox{P}(R_1) \le L_p p^{-\{(\underline{\omega}r-\beta+\Lambda)^2+4\underline{\omega}r\beta\}/(4\underline{\omega}r)}, \quad \mbox{P}(R_2) \le L_p p^{-\{(\underline{\omega}r-\beta+\Lambda)^2+4\underline{\omega}r\beta\}/(4\underline{\omega}r)}. \label{ee2}
\end{eqnarray}

Within the component $\mathcal{I}_0$, by conducting the DATE procedure, it is possible that some signals are wrongly identified as noise and some noise can be identified as signals. For convenience, we let $B_{nn}$ be the number of true negatives, $B_{ns}$ be the number of false positives, $B_{sn}$ be the number of false negatives, and $B_{ss}$ be the number of true positives. Then, the total number of signals in $\mathcal{I}_0$ is $B_{sn}+B_{ss}$. 
Let the event $M_p=\{\mbox{sgn}(\hat{\delta}(\mathcal{I}_0)\ne \mbox{sgn}(\delta^{\mathcal{I}_0}), j \in \mathcal{I}_0, A_p\}$. Since $j \in \mathcal{I}_0$, the event $R_1$ is contained in $M_p$ and the event $R_2$ is contained in the event $M_p \cap \{B_{sn}+B_{ss}\ge 1\}$. 
Therefore, to show (\ref{ee2}), we only need to show 
\begin{eqnarray}
p^{-\Lambda}\mbox{P}(M_p) &\le& L_p p^{-\{(\underline{\omega}r-\beta+\Lambda)^2+4\underline{\omega}r\beta\}/(4\underline{\omega}r)}, \nonumber\\
\mbox{P}(M_p \cap \{B_{sn}+B_{ss}\ge 1\}) &\le& L_p p^{-\{(\underline{\omega}r-\beta+\Lambda)^2+4\underline{\omega}r\beta\}/(4\underline{\omega}r)}. \label{ee3}
\end{eqnarray}

Let the event $D_p=\{||\hat{\Omega}-\Omega||_{L_1} \le (\mbox{log}p/n)^{(1-\alpha)/2} \}$. Since $\mbox{P}(D_p^c)=o(p^{-1})$, it is sufficient to show (\ref{ee3}) over the event $D_p$. Moreover, define the event $B_p(\mathcal{I}_0)$ through its complement: $B_p^c (\mathcal{I}_0)=\{$ there exist indices $i \notin \mathcal{I}_0$ and $j \in \mathcal{I}_0$ such that $\delta_i \neq 0$, $\Omega^*(i, j) \neq 0 \}$. 
Similar to Ji and Jin (2012), we can show that
\[
\mbox{P}(j \in \mathcal{I}_0, B_p^c \cap A_p) \le L_p p^{-\{(\underline{\omega}r-\beta+\Lambda)^2+4\underline{\omega}r\beta\}/(4\underline{\omega}r)}.
\]
Therefore, in order to show (\ref{ee3}), it is sufficient to show that 
\begin{eqnarray}
p^{-\Lambda}\mbox{P}(M_p \cap B_p \cap D_p) &\le& L_p p^{-\{(\underline{\omega}r-\beta+\Lambda)^2+4\underline{\omega}r\beta\}/(4\underline{\omega}r)}, \nonumber\\
\mbox{P}(M_p \cap B_p \cap D_p \cap \{B_{sn}+B_{ss}\ge 1\}) &\le& L_p p^{-\{(\underline{\omega}r-\beta+\Lambda)^2+4\underline{\omega}r\beta\}/(4\underline{\omega}r)}. \label{ee4}
\end{eqnarray}

Most importantly, by Lemma A.4 of Ji and Jin (2012), over the event $\{( j \in \mathcal{I}_0) \cap A_p \cap B_p\}$, 
\[
||(\Omega \delta)^{\mathcal{I}_0}-\Omega^{\mathcal{I}_0 \mathcal{I}_0} \delta^{\mathcal{I}_0}||_{\infty}=o(\sqrt{\mbox{log}p/n}),  
\]
which implies that $(\Omega \delta)^{\mathcal{I}_0} \approx \Omega^{\mathcal{I}_0 \mathcal{I}_0} \delta^{\mathcal{I}_0}$. This enables us to find $\hat{\delta}(\mathcal{I}_0)$ over the event $M_p \cap B_p \cap D_p$, each components of which has the value taken from $\{-\delta^{date}, 0, \delta^{date}\}$ to minimize 
\be
n\{(\bar{{Z}}_1-\bar{{Z}}_2)^{\mathcal{I}_0}-{A}\delta \}^{\prime} {A}^{-1}\{(\bar{{Z}}_1-\bar{{Z}}_2)^{\mathcal{I}_0}-{A}\delta \}+ (\lambda^{date})^2 ||\delta||_0, \label{A_mle} 
\ee
where ${A}={\Omega}^{\mathcal{I}_0 \mathcal{I}_0}$. 

If the event $\{M_p \cap B_p \cap D_p\}$ happens, then by (\ref{A_mle}), 
\begin{eqnarray}
&\quad &n\{(\bar{{Z}}_1-\bar{{Z}}_2)^{\mathcal{I}_0}-\Omega^{\mathcal{I}_0 \mathcal{I}_0}{\delta}(\mathcal{I}_0) \}^{\prime} (\Omega^{\mathcal{I}_0 \mathcal{I}_0})^{-1}\{(\bar{{Z}}_1-\bar{{Z}}_2)^{\mathcal{I}_0}-\Omega^{\mathcal{I}_0 \mathcal{I}_0}\hat{\delta}(\mathcal{I}_0) \}+ (\lambda^{date})^2 ||\hat{\delta}(\mathcal{I}_0)||_0\nonumber\\
&\le&n\{(\bar{{Z}}_1-\bar{{Z}}_2)^{\mathcal{I}_0}-\Omega^{\mathcal{I}_0 \mathcal{I}_0}\delta^*({\mathcal{I}_0}) \}^{\prime} (\Omega^{\mathcal{I}_0 \mathcal{I}_0})^{-1}\{(\bar{{Z}}_1-\bar{{Z}}_2)^{\mathcal{I}_0}-\Omega^{\mathcal{I}_0 \mathcal{I}_0}\delta^*({\mathcal{I}_0}) \}+ (\lambda^{date})^2 ||\delta^*({\mathcal{I}_0})||_0,\nonumber
\end{eqnarray}
where $\delta^*({\mathcal{I}_0})$ is defined to be a vector on ${\mathcal{I}_0}$ each component of which corresponds to the true signals or noise in the sense that for $l \in \mathcal{I}_0$, $\delta^*_l=0$ if $\delta_l=0$ and $\delta^*_l=\delta^{date}\,\mbox{sgn}(\delta)$ if $\delta_l \ne 0$. If we let $d=||\delta^*({\mathcal{I}_0})||_0-||\hat{\delta}(\mathcal{I}_0)||_0=B_{sn}-B_{ns}$, it follows that
\be
n\{\delta^*({\mathcal{I}_0})-\hat{\delta}(\mathcal{I}_0)\}^{\prime}(\bar{{Z}}_1-\bar{{Z}}_2)^{\mathcal{I}_0}\le \frac{1}{2}\biggl((\lambda^{date})^2 d+n\{\delta^*({\mathcal{I}_0})\}^{\prime}\Omega^{\mathcal{I}_0 \mathcal{I}_0}\delta^*({\mathcal{I}_0})-\{\hat{\delta}(\mathcal{I}_0)\}^{\prime}\Omega^{\mathcal{I}_0 \mathcal{I}_0}\hat{\delta}(\mathcal{I}_0) \biggr).\label{A_diff}
\ee
Since $\sqrt{n}\{(\bar{Z}_1-\bar{Z}_2)^{\mathcal{I}_0}\}=\sqrt{n}\,\Omega^{\mathcal{I}_0 \mathcal{I}_0}\delta^{\mathcal{I}_0}+z$ 
where $z\sim \mbox{N}(0, \Omega^{\mathcal{I}_0 \mathcal{I}_0})$. Then, (\ref{A_diff}) can be written as
\be
\frac{\Delta_1^{\prime} z}{\sqrt{\Delta_1^{\prime} \Omega^{\mathcal{I}_0 \mathcal{I}_0}\Delta_1}}\le -\frac{\sqrt{2r \mbox{log}p}}{2\sqrt{\Delta_1^{\prime} \Omega^{\mathcal{I}_0 \mathcal{I}_0}\Delta_1}}\biggl\{-d\, \frac{\beta-\Lambda}{r}+2\Delta_1^{\prime} \Omega^{\mathcal{I}_0 \mathcal{I}_0}\Delta_2+\Delta_1^{\prime} \Omega^{\mathcal{I}_0 \mathcal{I}_0}\Delta_1 \biggr\},\label{A_diff2}
\ee
where 
\[
\Delta_1=\frac{\sqrt{n} \{(\delta^*)^{\mathcal{I}_0}-\hat{\delta}(\mathcal{I}_0)\}}{\sqrt{2r\mbox{log}p}}, \quad \Delta_2=\frac{\sqrt{n} \{\delta^{\mathcal{I}_0}-(\delta^*)^{\mathcal{I}_0}\}}{\sqrt{2r\mbox{log}p}}.
\]


In (\ref{A_diff2}), both $\delta^{\mathcal{I}_0}$ and $z$ are random. Given $\delta^{\mathcal{I}_0}$, 
\[
\frac{\Delta_1^{\prime}z}{\sqrt{\Delta_1^{\prime} \Omega^{\mathcal{I}_0 \mathcal{I}_0}\Delta_1}} \sim \mbox{N}(0,1).
\]

Note that if the event $\{M_p \cap B_p \cap D_p\}$ happens, then the inequality (\ref{A_diff2}) holds. Therefore,
\begin{eqnarray}
&\quad&\mbox{P}(M_p \cap B_p \cap D_p)\nonumber\\
&\le& \mbox{P} \biggl(\frac{\Delta_1^{\prime} z}{\sqrt{\Delta_1^{\prime} \Omega^{\mathcal{I}_0 \mathcal{I}_0}\Delta_1}}\le -\frac{\sqrt{2r \mbox{log}p}}{2\sqrt{\Delta_1^{\prime} \Omega^{\mathcal{I}_0 \mathcal{I}_0}\Delta_1}}\biggl\{-d\, \frac{\beta-\Lambda}{r}+2\Delta_1^{\prime} \Omega^{\mathcal{I}_0 \mathcal{I}_0}\Delta_2+\Delta_1^{\prime} \Omega^{\mathcal{I}_0 \mathcal{I}_0}\Delta_1 \biggr\}\biggr).\nonumber\\\label{A_2}
\end{eqnarray}

Next, we evaluate the right hand side of the inequality (\ref{A_2}) for different values of $B_{sn}+B_{ss}$. To this end, we first notice that the right hand side is bounded by $p^{-\beta(B_{sn}+B_{ss})}$, which is the probability of having $B_{sn}+B_{ss}$ signals in $\mathcal{I}_0$. Therefore, if $B_{sn}+B_{ss} \ge \{(\underline{\omega}r-\beta+\Lambda)^2+4\underline{\omega}r\beta\}/(4\underline{\omega}r \beta)$, from (\ref{A_2}), it is easy to show that (\ref{ee4}) is true. 
Hence, we only need to consider the case where $B_{sn}+B_{ss} < \{(\underline{\omega}r-\beta+\Lambda)^2+4\underline{\omega}r\beta\}/(4\underline{\omega}r \beta)$. Note that the value of $B_{nn}$ does not affect the inequality in (\ref{A_2}). Therefore, we assume $B_{nn}=0$. Also similar to Lemma A.6 of Ji and Jin (2012), it can be shown that $\Delta_1^{\prime} \Omega^{\mathcal{I}_0 \mathcal{I}_0}\Delta_1 \ge \underline{\omega}$. Moreover, since the support of any signal $|\delta_k|$ is $[\sqrt{2r \mbox{log}p/n},(1+\eta)\sqrt{2r \mbox{log}p/n}]$, $\Delta_2 \ge 0$ if $\mbox{sgn}(\delta_k)=1$ and $\Delta_2 < 0$ otherwise. 

\begin{itemize}
\item $B_{sn}+B_{ss}=0$;

For this case, we have $d=B_{sn}-B_{ns}=-B_{ns}\le -1$. Using the fact that $\Delta_1^{\prime} \Omega^{\mathcal{I}_0 \mathcal{I}_0}\Delta_1 \ge \underline{\omega}$ and $\Delta_2=0$, we have
\[
\frac{-d\, \frac{\beta-\Lambda}{r}+2\Delta_1^{\prime} \Omega^{\mathcal{I}_0 \mathcal{I}_0}\Delta_2+\Delta_1^{\prime} \Omega^{\mathcal{I}_0 \mathcal{I}_0}\Delta_1
}{2\sqrt{\Delta_1^{\prime} \Omega^{\mathcal{I}_0 \mathcal{I}_0}\Delta_1}} \ge \frac{\frac{\beta-\Lambda}{r}+\Delta_1^{\prime} \Omega^{\mathcal{I}_0 \mathcal{I}_0}\Delta_1}{2\sqrt{\Delta_1^{\prime} \Omega^{\mathcal{I}_0 \mathcal{I}_0}\Delta_1}}\ge \frac{\frac{\beta-\Lambda}{r}+\underline{\omega}}{2\sqrt{\underline{\omega}}}.
\]
Then, from (\ref{A_2}), by using the fact that $\bar{\Phi}(a)\approx \phi(a)/a$ if $a \to \infty$, we have
\[
\mbox{P}(M_p \cap B_p \cap D_p) \le \bar{\Phi}(-\frac{\frac{\beta-\Lambda}{r}+\underline{\omega}}{2\sqrt{\underline{\omega}}}\sqrt{2r\mbox{log}p}) \le L_p p^{-\frac{(\underline{\omega}r+\beta-\Lambda)^2}{4\underline{\omega}r}}.
\]
Then, for this case, 
\be
\mbox{P}(R_1) \le L_p p^{-\frac{(\underline{\omega}r+\beta-\Lambda)^2}{4\underline{\omega}r}}.\label{R1-1}
\ee

\item $B_{sn}+B_{ss} = 1$ but $B_{ns}=0$;

For this case, since $\mbox{sgn}(\hat{\delta}(\mathcal{I}_0)\ne \mbox{sgn}(\delta^{\mathcal{I}_0})$, we must have $B_{sn}\ne 0$. Otherwise, both $B_{sn}= 0$ and $B_{ns}=0$ leads to $\mbox{sgn}(\hat{\delta}(\mathcal{I}_0)= \mbox{sgn}(\delta^{\mathcal{I}_0})$. As a result, $B_{sn}=1$ and $B_{ss}=0$. It follows that $d=B_{sn}-B_{ns}=1$, $\Delta_1^{\prime} \Omega^{\mathcal{I}_0 \mathcal{I}_0}\Delta_1\ge \underline{\omega}$, and $\Delta_1^{\prime} \Omega^{\mathcal{I}_0 \mathcal{I}_0}\Delta_2 \ge 0$. Then,
\[
\frac{-d\, \frac{\beta-\Lambda}{r}+2\Delta_1^{\prime} \Omega^{\mathcal{I}_0 \mathcal{I}_0}\Delta_2+\Delta_1^{\prime} \Omega^{\mathcal{I}_0 \mathcal{I}_0}\Delta_1
}{2\sqrt{\Delta_1^{\prime} \Omega^{\mathcal{I}_0 \mathcal{I}_0}\Delta_1}} \ge \frac{-\frac{\beta-\Lambda}{r}+\Delta_1^{\prime} \Omega^{\mathcal{I}_0 \mathcal{I}_0}\Delta_1}{2\sqrt{\Delta_1^{\prime} \Omega^{\mathcal{I}_0 \mathcal{I}_0}\Delta_1}}\ge \frac{-\frac{\beta-\Lambda}{r}+\underline{\omega}}{2\sqrt{\underline{\omega}}}.
\]
This, together with $B_{sn}+B_{ss}=1$, shows that (\ref{A_2}) satisfies
\[
\mbox{P}(M_p \cap B_p \cap D_p) \le L_p p^{-\beta}p^{-\frac{(\underline{\omega}r-\beta+\Lambda)^2}{4\underline{\omega}r}},
\]
which implies that 
\be
\mbox{P}(R_1) \le L_p p^{-\beta}p^{-\frac{(\underline{\omega}r-\beta+\Lambda)^2}{4\underline{\omega}r}},  \mbox{P}(R_2) \le L_p p^{-\beta}p^{-\frac{(\underline{\omega}r-\beta+\Lambda)^2}{4\underline{\omega}r}}. \label{R1-2}
\ee

In the following, we consider $1\le B_{sn}+B_{ss} \le {\{(\underline{\omega}r-\beta+\Lambda)^2+4\underline{\omega}r\beta\}/(4\underline{\omega}r \beta)}$, $B_{nn}=0$, and when $B_{ns}=0$, $B_{sn}+B_{ss}\ge 2$. To this end, we apply the Cauchy-Schwartz to get
\be
|\Delta_1^{\prime} \Omega^{\mathcal{I}_0 \mathcal{I}_0}\Delta_2| \le \sqrt{\Delta_1^{\prime} \Omega^{\mathcal{I}_0 \mathcal{I}_0}\Delta_1}\sqrt{\Delta_2^{\prime} \Omega^{\mathcal{I}_0 \mathcal{I}_0}\Delta_2}. \label{ee6}
\ee  
Using the spectral decomposition, $\Delta_2^{\prime} \Omega^{\mathcal{I}_0 \mathcal{I}_0}\Delta_2 \le C_0 ||\Delta_2||_2^2$ where $C_0$ is defined in condition (C2). Since the support of signal is $[\sqrt{2s\mbox{log}p/n}, (1+\eta)\sqrt{2s\mbox{log}p/n}]$, and $\Delta_2$ has $(B_{ss}+B_{sn})$ nonzero signals, 
\[
\Delta_2^{\prime} \Omega^{\mathcal{I}_0 \mathcal{I}_0}\Delta_2 \le C_0 (B_{ss}+B_{sn}) \eta^2.
\]
Moreover, with assumption $B_{sn}+B_{ss}\le \{(\underline{\omega}r-\beta+\Lambda)^2+4\underline{\omega}r\beta\}/(4\underline{\omega}r \beta)$, (\ref{ee6}) can be written as 
\[
|\Delta_1^{\prime} \Omega^{\mathcal{I}_0 \mathcal{I}_0}\Delta_2| \le \sqrt{C}\cdot \sqrt{\Delta_1^{\prime} \Omega^{\mathcal{I}_0 \mathcal{I}_0}\Delta_1},
\] 
where $C=C_0 \eta^2 \{(\underline{\omega}r-\beta+\Lambda)^2+4\underline{\omega}r\beta\}/(4\underline{\omega}r \beta)$. Then it  follows that $\Delta_1^{\prime} \Omega^{\mathcal{I}_0 \mathcal{I}_0}\Delta_2 \ge -\sqrt{C}\cdot \sqrt{\Delta_1^{\prime} \Omega^{\mathcal{I}_0 \mathcal{I}_0}\Delta_1}$, which implies that
\[
\frac{-d\, \frac{\beta-\Lambda}{r}+2\Delta_1^{\prime} \Omega^{\mathcal{I}_0 \mathcal{I}_0}\Delta_2+\Delta_1^{\prime} \Omega^{\mathcal{I}_0 \mathcal{I}_0}\Delta_1
}{2\sqrt{\Delta_1^{\prime} \Omega^{\mathcal{I}_0 \mathcal{I}_0}\Delta_1}} \ge \frac{-d\frac{\beta-\Lambda}{r}+\Delta_1^{\prime} \Omega^{\mathcal{I}_0 \mathcal{I}_0}\Delta_1}{2\sqrt{\Delta_1^{\prime} \Omega^{\mathcal{I}_0 \mathcal{I}_0}\Delta_1}}-\sqrt{C}.
\]
Next we consider three different cases: $B_{ns}=B_{sn}\ge 1$; $B_{ns}>B_{sn}$; $B_{ns}<B_{sn}$. 

For the case $B_{ns}=B_{sn}\ge 1$, we have $d=B_{sn}-B_{ns}=0$. Then,
\[
\frac{-d\, \frac{\beta-\Lambda}{r}+2\Delta_1^{\prime} \Omega^{\mathcal{I}_0 \mathcal{I}_0}\Delta_2+\Delta_1^{\prime} \Omega^{\mathcal{I}_0 \mathcal{I}_0}\Delta_1
}{2\sqrt{\Delta_1^{\prime} \Omega^{\mathcal{I}_0 \mathcal{I}_0}\Delta_1}} \ge \frac{\sqrt{\Delta_1^{\prime} \Omega^{\mathcal{I}_0 \mathcal{I}_0}\Delta_1}}{2}-\sqrt{C}.
\]
Recall that $\Delta_1^{\prime} \Omega^{\mathcal{I}_0 \mathcal{I}_0}\Delta_1 \ge \underline{\omega}$ and $\eta \le \frac{\beta-\Lambda}{\sqrt{C_0}r} \frac{\sqrt{\beta r}}{\sqrt{(\underline{\omega}r-\beta+\Lambda)^2+4\underline{\omega}r\beta}}$. Then, $\sqrt{C} \le (\beta-\Lambda)/(2\sqrt{\underline{\omega}}r)$, and 
\[
\mbox{P}(M_p \cap B_p \cap D_p) \le L_p p^{-\beta}p^{-\frac{(\underline{\omega}r-\beta+\Lambda)^2}{4\underline{\omega}r}}.
\]

For the case $B_{ns}> B_{sn}$, $d=B_{sn}-B_{ns} \le -1$. Since $\eta \le \frac{\beta-\Lambda}{\sqrt{C_0}r} \frac{\sqrt{\beta r}}{\sqrt{(\underline{\omega}r-\beta+\Lambda)^2+4\underline{\omega}r\beta}}$, we have $\sqrt{C}\le (\beta-\Lambda)/(\sqrt{\underline{\omega}}r)$, and
\begin{eqnarray}
\frac{-d\, \frac{\beta-\Lambda}{r}+2\Delta_1^{\prime} \Omega^{\mathcal{I}_0 \mathcal{I}_0}\Delta_2+\Delta_1^{\prime} \Omega^{\mathcal{I}_0 \mathcal{I}_0}\Delta_1
}{2\sqrt{\Delta_1^{\prime} \Omega^{\mathcal{I}_0 \mathcal{I}_0}\Delta_1}} &\ge& \frac{\frac{\beta-\Lambda}{r}+\Delta_1^{\prime} \Omega^{\mathcal{I}_0 \mathcal{I}_0}\Delta_1}{2\sqrt{\Delta_1^{\prime} \Omega^{\mathcal{I}_0 \mathcal{I}_0}\Delta_1}}-\sqrt{C}\nonumber\\
&\ge& \frac{1}{2}(\frac{\beta-\Lambda}{\sqrt{\underline{\omega}}r}+\sqrt{\underline{\omega}})-\sqrt{C}\nonumber\\
&\ge& \frac{1}{2}(\sqrt{\underline{\omega}}-\frac{\beta-\Lambda}{\sqrt{\underline{\omega}}r}).\nonumber
\end{eqnarray}
Since $B_{sn}+B_{ss}\ge 1$, then 
\[
\mbox{P}(M_p \cap B_p \cap D_p) \le L_p p^{-\beta}p^{-\frac{(\underline{\omega}r-\beta+\Lambda)^2}{4\underline{\omega}r}}.
\]

For the case $B_{ns}<B_{sn}$, we have either $B_{ns}=0$ or $B_{ns}\ge 1$. If $B_{ns}=0$, $B_{sn}+B_{ss}\ge 2$ as we have required. If $B_{ns}\ge 1$, we also have $B_{sn}+B_{ss}\ge 2$ due to the fact that $B_{sn}>B_{ns}$. Since $\beta+(\underline{\omega}r-\beta+\Lambda)^2/(4\underline{\omega}r)=\Lambda+(\underline{\omega}r+\beta-\Lambda)^2/(4\underline{\omega}r)$ and $\beta-\Lambda < \underline{\omega}r$, then
\[
\beta+(\underline{\omega}r-\beta+\Lambda)^2/(4\underline{\omega}r) \le \Lambda+\underline{\omega}r. 
\]
Then using the fact that $\beta(B_{sn}+B_{ss}) \ge 2\beta$ and by assuming $\beta > 1/2$ and $\underline{\omega}r < (\sqrt{1-\Lambda}+\sqrt{1-\beta})^2$, we can derive the following inequality  
\[
\beta(B_{sn}+B_{ss}) \ge \beta+\frac{(\underline{\omega}r-\beta+\Lambda)^2}{4\underline{\omega}r}.
\]
Therefore, for three different cases: $B_{ns}=B_{sn}\ge 1$; $B_{ns}>B_{sn}$; $B_{ns}<B_{sn}$, we have
\[
\mbox{P}(M_p \cap B_p \cap D_p) \le L_p p^{-\beta}p^{-\frac{(\underline{\omega}r-\beta+\Lambda)^2}{4\underline{\omega}r}},
\]
which implies that 
\be
\mbox{P}(R_1) \le L_p p^{-\beta}p^{-\frac{(\underline{\omega}r-\beta+\Lambda)^2}{4\underline{\omega}r}}, \mbox{P}(R_2) \le L_p p^{-\beta}p^{-\frac{(\underline{\omega}r-\beta+\Lambda)^2}{4\underline{\omega}r}}. \label{R1-3}
\ee
\end{itemize}

In summary, from (\ref{R1-1}), (\ref{R1-2}) and (\ref{R1-3}), we know that if $B_{sn}+B_{ss}=0$, $\mbox{P}(R_1) \le L_p p^{-\frac{(\underline{\omega}r+\beta-\Lambda)^2}{4\underline{\omega}r}}$. And if $B_{sn}+B_{ss}\ge 1$, $\mbox{P}(R_1) \le L_p p^{-\beta}p^{-\frac{(\underline{\omega}r-\beta+\Lambda)^2}{4\underline{\omega}r}}$. Since
\[
\frac{(\underline{\omega}r+\beta-\Lambda)^2}{4\underline{\omega}r} \le \beta+\frac{(\underline{\omega}r-\beta+\Lambda)^2}{4\underline{\omega}r},
\]
we have $\mbox{P}(R_1) \le L_p p^{-\frac{(\underline{\omega}r+\beta-\Lambda)^2}{4\underline{\omega}r}}$. Similarly, we have $\mbox{P}(R_2) \le L_p p^{-\beta}p^{-\frac{(\underline{\omega}r-\beta+\Lambda)^2}{4\underline{\omega}r}}$. Then, 
\[
p^{-\Lambda}\mbox{P}(R_1)+\mbox{P}(R_2)\le L_p p^{-\beta}p^{-\frac{(\underline{\omega}r-\beta+\Lambda)^2}{4\underline{\omega}r}},
\]
which shows that (\ref{ee2}) is true. This completes the proof of Theorem 3.

\medskip
\subsection*{\large A.7. Proof of Theorem 4}

Let $\theta_i=0$ if $\delta_i=0$ and $\theta_i=1$ otherwise, and $\hat{\delta}_i$ is the corresponding estimate by the DATE procedure.
Recall that the marginal false discovery rate is defined as
\[
\mbox{mFDR}=\frac{\sum_i\mbox{P}(\theta_i=0, \mbox{sgn}(\hat{\delta}_i)\ne 0)}{\sum_i\mbox{P}(\theta_i=0, \mbox{sgn}(\hat{\delta}_i)\ne 0)+\sum_i\mbox{P}(\theta_i=1, \mbox{sgn}(\hat{\delta}_i) \ne 0)}.
\]

Similar to (\ref{A_2}) in the proof of Theorem 3, 
\be
\mbox{P}(\theta_i \ne \mbox{sgn}(\hat{\delta}_i))\le p^{-\beta(B_{sn}+B_{ss})}\bar{\Phi}\biggl(\frac{-d\, \frac{\beta-\Lambda-\frac{\Upsilon}{2{log}p}}{r}+2\Delta_1^{\prime} \Omega^{\mathcal{I}_0 \mathcal{I}_0}\Delta_2+\Delta_1^{\prime} \Omega^{\mathcal{I}_0 \mathcal{I}_0}\Delta_1
}{2\sqrt{\Delta_1^{\prime} \Omega^{\mathcal{I}_0 \mathcal{I}_0}\Delta_1}}\sqrt{2r\mbox{log}p} \biggr).\label{A7-1}
\ee

First, note that 
\begin{eqnarray}
\mbox{P}(\theta_i=1, \mbox{sgn}(\hat{\delta}_i)\ne 0)&=&\mbox{P}(\theta_i=1)-\mbox{P}(\theta_i=1, \mbox{sgn}(\hat{\delta}_i)=0)\nonumber\\
&=&p^{-\beta}-\mbox{P}(\theta_i=1, \mbox{sgn}(\hat{\delta}_i)=0).\nonumber
\end{eqnarray}
Following the similar derivations for Theorem 3, from (\ref{A7-1}), we can show that
\[
\mbox{P}(\theta_i=1, \mbox{sgn}(\hat{\delta}_i)=0)\le p^{-\beta}\bar{\Phi}\biggl(\frac{(-\beta+\Lambda+\underline{\omega}r)\mbox{log}p+\Upsilon/2
}{\sqrt{2\underline{\omega}r\mbox{log}p}} \biggr),
\]
where $\Upsilon$ is defined in Theorem 4. Then,
\[
\mbox{P}(\theta_i=1, \mbox{sgn}(\hat{\delta}_i)\ne 0)\ge p^{-\beta}\bar{\Phi}\biggl(\frac{(\beta-\Lambda-\underline{\omega}r)\mbox{log}p-\Upsilon/2
}{\sqrt{2\underline{\omega}r\mbox{log}p}} \biggr).
\]


Similarly,
\begin{eqnarray}
\mbox{P}(\theta_i=0, \mbox{sgn}(\hat{\delta}_i)\ne 0))&=&\mbox{P}(\theta_i=0, \mbox{sgn}(\hat{\delta}_i)=1))+\mbox{P}(\theta_i=0, \mbox{sgn}(\hat{\delta}_i)=-1))\nonumber\\
&\le&2(1-p^{-\beta}) \bar{\Phi}\biggl(\frac{(\beta-\Lambda+\underline{\omega}r)\mbox{log}p-\Upsilon/2
}{\sqrt{2\underline{\omega}r\mbox{log}p}} \biggr).\nonumber
\end{eqnarray}

To require mFDR to be controlled at $\alpha$, we need to find $\Upsilon$ such that
\begin{eqnarray}
\alpha \ge \frac{2(p-p^{1-\beta})\bar{\Phi}\biggl(\frac{(\beta-\Lambda+\underline{\omega}r)\mbox{log}p-\Upsilon/2
}{\sqrt{2\underline{\omega}r\mbox{log}p}} \biggr)}{2(p-p^{1-\beta}) \bar{\Phi}\biggl(\frac{(\beta-\Lambda+\underline{\omega}r)\mbox{log}p-\Upsilon/2
}{\sqrt{2\underline{\omega}r\mbox{log}p}} \biggr)+p^{1-\beta}\bar{\Phi}\biggl(\frac{(\beta-\Lambda-\underline{\omega}r)\mbox{log}p-\Upsilon/2
}{\sqrt{2\underline{\omega}r\mbox{log}p}} \biggr)}.\label{A7-2}
\end{eqnarray}

When $p \to \infty$, (\ref{A7-2}) can be solved asymptotically. If we assume $\Upsilon=o(\mbox{log}p)$, then by the fact that $\underline{\omega}r>\beta-\Lambda$, 
\[
\bar{\Phi}\biggl(\frac{(\beta-\Lambda-\underline{\omega}r)\mbox{log}p-\Upsilon/2
}{\sqrt{2\underline{\omega}r\mbox{log}p}} \biggr) \to 1.
\]
Then using the fact that $\bar{\Phi}(a) \approx \phi(a)/a$ for $a \to \infty$, we can solve (\ref{A7-2}) by choosing  
\[
\Upsilon=\frac{4\underline{\omega}r}{\underline{\omega}r+\beta-\Lambda}\biggl(\frac{1}{2}\mbox{log}\mbox{log}p+\mbox{log}\biggl\{\frac{\alpha\sqrt{\pi}(\underline{\omega}r+\beta-\Lambda)}{2\sqrt{\underline{\omega}r}(1-\alpha)} \biggr\} \biggr)
\]
such that 
\begin{eqnarray}
\mbox{mFDR} &\le& \frac{ \frac{\alpha}{1-\alpha} p^{-\frac{(\underline{\omega}r+\beta-\Lambda)^2}{4\underline{\omega}r}}}{ \frac{\alpha}{1-\alpha} p^{-\frac{(\underline{\omega}r+\beta-\Lambda)^2}{4\underline{\omega}r}}+p^{-\beta}\{1+o(1)\}}\nonumber\\
&\le& \alpha\{1+o(1)\}, \nonumber
\end{eqnarray}
where we have used the result that $\Lambda=(\sqrt{\underline{\omega}r}-\sqrt{\beta})^2$.

Similarly, the marginal false non-discovery rate
\begin{eqnarray}
\mbox{mFNR}&=&\frac{\sum_i\mbox{P}(\theta_i=1, \mbox{sgn}(\hat{\delta}_i)=0)}{\sum_i\mbox{P}(\theta_i=1, \mbox{sgn}(\hat{\delta}_i)=0)+\sum_i\mbox{P}(\theta_i=0, \mbox{sgn}(\hat{\delta}_i)=0)}\nonumber\\
&\le& L_p  p^{-\beta-(\sqrt{\underline{\omega}r}-\sqrt{\beta})^2}.\nonumber
\end{eqnarray}
This completes the proof of Theorem 4.

\medskip
\subsection*{\large A.8. Proof of Theorem 5}

Note that the optimal rate does not change if we add a slowly varying logarithm function to the tuning parameters in Theorem 4 by carefully reviewing its proof. Therefore, if we can show that there exists a constant $C$ such that
\begin{eqnarray}
\mbox{P}(|\hat{\beta}-\beta|>(\mbox{log}p)^{-C})=o(p^{-1}), \label{A8.1}
\end{eqnarray}
\begin{eqnarray}
\mbox{P}(|\hat{r}-r|>(\mbox{log}p)^{-C})=o(p^{-1}), \label{A8.2}
\end{eqnarray}
and
\begin{eqnarray}
\mbox{P}(|\hat{\underline{\omega}}-\underline{\omega}|>(\mbox{log}p)^{-C})=o(p^{-1}), \label{A8.3}
\end{eqnarray}
where $\hat{\beta}$, $\hat{r}$ and $\hat{\underline{\omega}}$ are defined in (\ref{tuning_e}), then Theorem 5 can be proved. 

First, let's prove (\ref{A8.3}). Note that with probability $1-O(p^{-\tau})$ and for some constant $C$,
\[
||\hat{\Omega}-\Omega||_{L_1}\le C\biggl\{(\frac{\mbox{log}p}{n})^{\frac{1-\zeta}{2}} \biggr\}.
\]
Under condition (C4), 
\[
(\frac{\mbox{log}p}{n})^{\frac{1-\zeta}{2}}=(\mbox{log}p)^{-\frac{(1-\theta)(1-\zeta)}{2\theta}} < (\mbox{log}p)^{-1/2}.
\]
If $\tau>1$, for large enough $p$,
\begin{eqnarray}
\mbox{P}(|\hat{\underline{\omega}}-\underline{\omega}|>(\mbox{log}p)^{-C})
&\le&\mbox{P}(\min \limits_{1\le k \le p} |\hat{{\omega}}_{kk}-\omega_{kk}|>(\mbox{log}p)^{-C})\nonumber\\ 
 &\le& \mbox{P}(||\hat{\Omega}-\Omega||_{L_1}>(\mbox{log}p)^{-C})=o(p^{-1}).
\end{eqnarray}

Next, let's prove (\ref{A8.1}) or equivalently, we need to show that 
\begin{eqnarray}
\mbox{P}\biggl(|\frac{1}{p^{1-\beta}}\sum_{k=1}^p\mbox{I}(\hat{T}_k > 2q\mbox{log}p)-1 | > L_p p^{-C}\biggr)=o(p^{-1}). \nonumber
\end{eqnarray}

To this end, we first notice that
\[
\sum_{k=1}^p\mbox{I}(\hat{T}_k > 2q\mbox{log}p)=\sum_{k=1}^p\mbox{I}(\hat{T}_k > 2q\mbox{log}p)\mbox{I}(\delta_k=0)+\sum_{k=1}^p\mbox{I}(\hat{T}_k > 2q\mbox{log}p)\mbox{I}(\delta_k\ne 0).
\]
Then, 
\begin{eqnarray}
&\quad&\mbox{P}\biggl(|\frac{1}{p^{1-\beta}}\sum_{k=1}^p\mbox{I}(\hat{T}_k > 2q\mbox{log}p)-1 | > L_p p^{-C}\biggr)\nonumber\\
 &\le& \mbox{P}\biggl(|\frac{1}{p^{1-\beta}}\sum_{k=1}^p\mbox{I}(\hat{T}_k > 2q\mbox{log}p)\mbox{I}(\delta_k=0)| > L_p p^{-C}\biggr)\nonumber\\
&+&\mbox{P}\biggl(|\frac{1}{p^{1-\beta}}\sum_{k=1}^p\mbox{I}(\hat{T}_k > 2q\mbox{log}p)\mbox{I}(\delta_k\ne 0)-1| > L_p p^{-C}\biggr).\nonumber
\end{eqnarray}

Using Chebyshev's inequality and the result in Lemma A.7 of Ji and Jin (2012), we have
\begin{eqnarray}
\mbox{P}\biggl(|\frac{1}{p^{1-\beta}}\sum_{k=1}^p\mbox{I}(\hat{T}_k > 2q\mbox{log}p)\mbox{I}(\delta_k=0)| > L_p p^{-C}\biggr) &\le& \frac{\mbox{E}(\{\sum_{k=1}^p \mbox{I}(\hat{T}_k > 2q\mbox{log}p) \mbox{I}(\delta_k=0)\}^m)}{p^{m-m\beta}L_p^mp^{-mC}}\nonumber\\
&\le&\frac{L_pp^m p^{-mq}}{p^{m-m\beta-mC}}\nonumber\\
&=&L_p p^{-m(q-\beta+C)}.\nonumber
\end{eqnarray}
Then we can choose $m$ large enough to have $p^{-m(q-\beta+C)}=o(p^{-1})$ since $q>\beta$. Based on similar derivations, (\ref{A8.2}) can be shown accordingly. This completes the proof of Theorem 5.  

\section*{Reference}
\parindent 0.2in
\textsc{Benjamini, Y. and Hochberg, Y.} (1995). Controlling the false discovery rate: A practical and powerful approach to multiple testing. \textit{Journal of the Royal Statistical Society: Series B}, \textbf{57} 289-300.

\parindent -0.2in
\leftskip 0.2in

\textsc{Benjamini, Y. and Yekutieli, D.} (2001). The control of the false discovery rate in multiple testing under dependency. \textit{The Annals of Statistics}, \textbf{29} 1165-1188.

\textsc{Bickel, P. and Levina, E.} (2008a). Regularized estimation of large covariance matrices. \textit{The Annals of Statistics}, \textbf{36}, 199-227.

\textsc{Bickel, P. and Levina, E.} (2008b). Covariance regularization by threholding. \textit{The Annals of Statistics}, \textbf{36}, 2577-2604.

\textsc{Cai, T. , Liu, W. and Xia, Y.} (2014). Two-sample test of high dimensional means under dependence. \textit{Journal of the Royal Statistical Society: Series B}, \textbf{76}, 349-372.

\textsc{Cai, T. , Liu, W. and Luo, X.} (2011). A constrained $l_1$ minimization approach to sparse precision matrix estimation. \textit{Journal of American Statistical Association}, \textbf{106}, 594-607.



\textsc{Donoho, D. and Jin, J.} (2004). Higher criticism for detecting sparse heterogeneous mixtures. \textit{The Annals of Statistics}, \textbf{32} 962-994.



\textsc{Efron, B.} (2007). Correlation and large-scale simultaneous significance testing. \textit{Journal of American Statistical Association}, \textbf{102} 93-103.

\textsc{Friedman, J. , Hastie, T. and Tibshirani R.} (2008). Sparse inverse covariance estimation with the graphical lasso. \textit{Biostatistics}, \textbf{9} 432-441.

\textsc{Frieze, A. and Molloy, M.} (1999). Splitting an expander graph. \textit{Journal of Algorithms}, \textbf{33}, 166-172.

\textsc{Genovese, C. and Wasserman, L.} (2002). Operating characteristics and extensions of the false discovery rate procedure. \textit{Journal of the Royal Statistical Society: Series B}, \textbf{64} 499-517.


\textsc{Hall, P. and Jin, J.} (2010). Innovated higher criticism for detecting sparse signals in correlated noise. \textit{The Annals of Statistics}, \textbf{38}, 1686-1732.

\textsc{Ji, P. and Jin, J.} (2012). UPS delivers optimal phase diagram in high-dimensional variable selection. \textit{The Annals of Statistics}, \textbf{40}, 73-103.

\textsc{Ji, P. and Zhao, Z.} (2014). Rate optimal multiple testing procedure in high-dimensional regression. \textit{Manuscript}.

\textsc{Klaus, B. and Strimmer, K.} (2013). Signal identification for rare and weak features: higher criticism or false discovery rates? \textit{Biostatistics}, \textbf{14}, 129-143.


\textsc{Qiu, X., Klebanov, L. and Yakovlev, A.} (2005). Correlation between gene expression levels and limitations of the empirical bayes methodology for finding differentially expressed genes. \textit{Statistical Applications in Genetics and Molecular Biology}, \textbf{4}, 34.

\textsc{Richardson, A., Wang, Z., Nicolo, A., Lu, X., Brown, M., Miron, A., Liao, X., Iglehart, J., Livingston, D. and Ganesan, S.} (2006). X chromosomal abnormalities in basal-like human breast cancer. \textit{Cancer Cell}, \textbf{9}, 121-132.


\textsc{Sun, W. and Cai, T.} (2007). Oracle and adaptive compound decision rules for false discovery rate control. \textit{Journal of American Statistical Association}, \textbf{102}, 901-912.

\textsc{Sun, W. and Cai, T.} (2009). Large-scale multiple testing under dependency. \textit{Journal of the Royal Statistical Society: Series B}, \textbf{71}, 393-424.


\textsc{Xie, J., Cai, T. and Li, H.} (2011). Sample size and power analysis for sparse signal recovery in genome-wide association studies. \textit{Biometrika}, \textbf{98}, 273-290.

\textsc{Xie, J., Cai, T., Maris, J. and Li, H.} (2011). Optimal false discovery rate control for dependent data. \textit{Statistics and Its Interface}, \textbf{4}, 417-430.

\textsc{Yuan, M. and Lin, Y.} (2007). Model selection and estimation in the Gaussian graphical model. \textit{Biometrika}, \textbf{94}, 19-35.

\textsc{Zhao, T. , Liu, H. , Roeder, K. , Lafferty, J. and Wasserman, L.} (2012). The huge package for high-dimensional undirected graph estimation in R. \textit{Journal of Machine Leaning Research}, \textbf{13}, 1059-1062.

\newpage

\begin{table}[t]
\centering
\small
\caption{The performance of $\mbox{DATE}_{\Omega}$ and $\mbox{DATE}_{\hat{\Omega}}$ in terms of mFDR and mFNR subject to different values of $s$ and $q$ chosen from two intervals separated by $\beta=0.6$ (mFDR and mFNR of $\mbox{DATE}_{\hat{\Omega}}$ are included in parenthesis). }
\label{case10}
 \begin{tabular}{ l|cccccc}
 \hline
 \backslashbox{$s$}{$q$} & 0.65 & 0.70& 0.75 & 0.80 & 0.85 & 0.90 \\
 \hline
 &   &   &  mFDR \\\hline
  0.25 & 0.045(0.041) & 0.047(0.042) & 0.038(0.033)& 0.053(0.046) & 0.036(0.029) & 0.038(0.030) \\
  0.30 & 0.038(0.039)  &  0.041(0.034)& 0.042(0.032)  & 0.043(0.030) & 0.050(0.035) & 0.032(0.020) \\
  0.35 & 0.041(0.037) & 0.033(0.040) & 0.025(0.025) & 0.048(0.041) & 0.033(0.028) & 0.048(0.041) \\
  0.40 & 0.046(0.051)  &  0.039(0.034)& 0.034(0.033)  & 0.031(0.034) & 0.044(0.041) & 0.043(0.035) \\  
  0.45 & 0.040(0.044) & 0.036(0.039) & 0.041(0.034) & 0.043(0.041) & 0.050(0.037) & 0.037(0.025) \\
  0.50 & 0.041(0.042)  &  0.031(0.030)& 0.038(0.033)  & 0.042(0.033) & 0.039(0.030) & 0.043(0.031) \\    
   \hline
 &   &   &  mFNR \\\hline
  0.25 & 0.005(0.006) & 0.005(0.006) & 0.006(0.007)& 0.005(0.007) & 0.006(0.007) & 0.006(0.007) \\
  0.30 & 0.006(0.007)  &  0.005(0.006)& 0.006(0.007)  & 0.005(0.006) & 0.005(0.006) & 0.005(0.007) \\
  0.35 & 0.005(0.006) & 0.006(0.007) & 0.005(0.006) & 0.006(0.007) & 0.005(0.007) & 0.005(0.007) \\
  0.40 & 0.006(0.006)  &  0.006(0.007)& 0.006(0.007)  & 0.006(0.007) & 0.006(0.007) & 0.005(0.007) \\  
  0.45 & 0.005(0.006) & 0.005(0.006) & 0.005(0.006) & 0.005(0.007) & 0.006(0.007) & 0.006(0.007) \\
  0.50 & 0.006(0.006)  &  0.005(0.006)& 0.005(0.007)  & 0.006(0.007) & 0.006(0.007) & 0.006(0.007) \\\hline   
  \end{tabular}
\end{table}

\begin{table*}[h]
\begin{center}
\caption{The number of differentially expressed genes identified by the BH, the DATE and both on chromosome X with the FDR controlled at the level of $\alpha=0.01, 0.005$ and $0.001$.}
\label{table6}
\small
\begin{tabular}{lccr}
\hline 
FDR-controlled level & BH & DATE & Both\\
 \hline
$0.01$ & $52$   & $56$  & $38$\\
$0.005$ & $43$   & $50$  & $33$ \\
$0.001$ & $27$   & $39$  & $22$ \\
\hline
\end{tabular}
\end{center}
\end{table*}

\begin{table}[bhtp]
\caption{The differentially expressed genes identified by the DATE not by the BH on chromosome X with the FDR controlled at level $0.001$.}
\small
\begin{center}
\begin{tabular}{lcr}
\hline
Gene symbol & Location & Description \\\hline
 PTCHD1   & Xp22.11 &  patched domain containing 1  \\
 DMD    & Xp21.2 &  dystrophin  \\
 SLC9A6    &  Xq26.3        & solute carrier family 9 (sodium/hydrogen exchanger), member 6  \\
KAL1   & Xp22.32 &  Kallmann syndrome 1 sequence  \\ 
TMSB15B     &   Xq22.2     & thymosin-like 8  \\ 
GPR64     &   Xp22.13     & G protein-coupled receptor 64 \\
ATP6AP1     & Xq28       & ATPase, H+ transporting, lysosomal accessory protein 1 \\ 
NXT2          & Xq23    & nuclear transport factor 2-like export factor 2\\
CLCN4         & Xp22.3     &   chloride channel 4  \\
VGLL1         & Xq26.3     & vestigial like 1 (Drosophila)\\
BEX1      & Xq22   & brain expressed, X-linked 1\\
SLC6A14      & Xq23   & solute carrier family 6 (amino acid transporter), member 14\\
BCOR         & Xp21.2-p11.4     & BCL6 corepressor  \\
BCORL1              & Xq25-q26.1    & BCL6 corepressor-like 1 \\
MUM1L1           &  Xq22.3   & melanoma associated antigen (mutated) 1-like 1\\
SYTL5         &  Xp21.1     & synaptotagmin-like 5\\
RLIM         &  Xq13-q21     & ring finger protein, LIM domain interacting\\
\hline
\end{tabular}
\end{center}
\label{table7}
\end{table}

\end{document}